\documentclass[journal, notitlepage,longbibliography]{revtex4-1}
\usepackage[T1]{fontenc}
\usepackage[latin9]{inputenc}
\usepackage{verbatim}
\usepackage{mathrsfs}
\usepackage{amsmath}
\usepackage{amssymb}
\usepackage{graphicx}
\usepackage[unicode=true,
 bookmarks=true,bookmarksnumbered=true,bookmarksopen=true,bookmarksopenlevel=1,
 breaklinks=false,pdfborder={0 0 1},backref=false,colorlinks=false]
 {hyperref}
\hypersetup{pdftitle={Your Title},
 pdfauthor={Your Name},
 pdfpagelayout=OneColumn, pdfnewwindow=true, pdfstartview=XYZ, plainpages=false}

\makeatletter

\providecommand{\tabularnewline}{\\}

\usepackage[caption=false,font=footnotesize]{subfig}
\usepackage{braket}

\usepackage{float}
\usepackage{mathtools}

\makeatother

\begin{document}
\title{Efficient quantum dot $\mathbf{k}\cdot\mathbf{p}$ in wurtzite systems
including spatially varying elastic and dielectric constants and smooth
alloy profile}
\author{Luc Robichaud\textsuperscript{1} and Jacob J. Krich\textsuperscript{1,2}}
\affiliation{\textsuperscript{1}Department of Physics, University of Ottawa, Ottawa,
Canada K1N 6N5}
\affiliation{\textsuperscript{2}School of Electrical Engineering and Computer
Science, University of Ottawa, Ottawa, Canada K1N 6N5}
\date{16 December 2020}
\begin{abstract}
We present Fourier-space based methods to calculate the electronic
structure of wurtzite quantum dot systems with continuous alloy profiles.
We incorporate spatially varying elastic and dielectric constants
in strain and piezoelectric potential calculations. A method to incorporate
smooth alloy profiles in all aspects of the calculations is presented.
We demonstrate our methodology for the case of a 1D InGaN quantum
dot array and show the importance of including these spatially varying
parameters in the modeling of devices. We demonstrate that convergence
of the lowest bound state energies is to good approximation determined
by the largest wave vector used in constructing the states. We also
present a novel approach of coupling strain into the $\mathbf{k}\cdot\mathbf{p}$
Hamiltonian, greatly reducing the computational cost of generating
the Hamiltonian.
\end{abstract}
\maketitle

\section{Introduction}

Given their large range of bandgaps, from 0.78 eV to 3.51 eV, InGaN
materials have attracted attention from various applications such
as LEDs, single-photon emitters, water splitting and solar cells \cite{Nguyen2011,Puchtler2016,Kibria2013,Sang2014,Cheriton2020}.
For any application, device performance depends on having an electronic
structure well tuned to its target application. Given that the electronic
structure of quantum dots can be drastically changed by varying their
size and composition, they can be quite attractive for applications.
The main problem in modeling complex structures such as quantum dots
is including all the necessary effects for the model to be accurate
while also keeping computational cost down.

Tight binding and $\mathbf{k}\cdot\mathbf{p}$ theory are standard
approaches for calculating single-particle electronic structures for
bulk materials and nanostructures \cite{Saito2002}. The $\mathbf{k}\cdot\mathbf{p}$
method gives a good balance between accuracy and computational requirements,
especially when considering large dots that contain large number of
atoms where the tight binding method becomes costly. $\mathbf{k}\cdot\mathbf{p}$
theory has been developed in both real space and Fourier space \cite{Winkelnkemper,Andreev2000}.
Following the Fourier-space method, symmetry adapted basis approaches
have been developed to reduce the required size of the Hamiltonian,
which block diagonalize the Hamiltonian, reducing the computational
cost of calculating the system's eigenstates \cite{Vukmirovic2005,Vukmirovic2006,Vukmirovc2008}.

InGaN materials are strongly piezoelectric, having both spontaneous
and strain-induced contributions to the piezoelectric polarization.
Strain calculations have been performed using valence force field
and Green's function based methods \cite{Stier1998,Andreev2000}.
The latter method has the advantage that it respects the symmetry
of the crystal lattice. From the strain, the piezoelectric potential
can be calculated from Maxwell's equations \cite{Andreev2000}. References
\cite{Andreev1999,Andreev2000,Vukmirovic2006} use the Green's function
method for calculating strain and calculated the piezoelectric potential
from Maxwell's equations. These works assume uniform elastic and dielectric
constants, which was justified for their respective InAs/GaAs and
GaN/AlN systems. However, in the case of InGaN, these constants vary
more significantly between dot and host. Additionally, InGaN devices
frequently do not have sharp interfaces between dot and barrier, with
indium diffusing over several nanometers. This smooth alloy profile
gives a spatial profile to every material parameter of the system,
effectively changing the confining potential seen by the electrons.

In this paper, we show the importance of including spatially varying
elastic and dielectric constants in strain and piezoelectric potential
calculations in the case InGaN systems. For strain calculations, we
implement a formalism previously presented for including spatially
varying elastic constants \cite{Andreev2000}. We present a new Fourier-space
formalism for the calculation of piezoelectric potentials with spatially
varying dielectric constants. We also present an approach to include
smooth indium profiles in the strain, piezoelectric potential and
electronic structure calculations, modeling the smooth alloy profiles
found in experimental devices. Considering smooth indium profiles
both increase the accuracy of the simulations and decrease their computational
cost by decreasing the number of plane waves required for convergence.

Strain plays an important role in the electronic structure properties
of quantum dots. In quantum dot $\mathbf{k}\cdot\mathbf{p}$, a single
real space unit cell is typically used when working in a Fourier-space
approach. However, strain decays more slowly than bound state wavefunctions.
When studying isolated dots, the difference in decay lengths makes
it computationally expensive to fully capture both the strain and
electronic structure using a single unit cell. Reference \cite{Vukmirovc2008}
presents an approach that implements two different unit cells; one
for the electronic structure and on for strain. This method allows
for the modeling of the electronic structure and strain, but introduces
some complexity in calculating the Hamiltonian, which requires the
calculation of multiple composed convolutions on different Fourier-space
meshes. These convolutions can be computationally costly depending
on the sizes of meshes needed for convergence. By fixing the strain
unit cell to be commensurate with the electronic unit cell, we present
an approach that reduces the number of needed convolutions, significantly
reducing the computational cost.

We demonstrate our methodology by calculating the electronic structure
for a 1D array of InGaN quantum dots, modeling devices grown as LEDs
and for water splitting \cite{Nguyen2011,Kibria2013}. In this example,
we show the importance of the inclusion of spatially varying elastic
and dielectric constants and smooth indium profiles for accurate electronic
structures. We also show that the most important criterion for convergence
of the lowest quantum dot electron and hole energies is the maximum
wave vector included in the Fourier-space sampling, which can be increased
with low computational cost by using a small unit cell.

Section \ref{sec: Non-uniform elastic and piezoelectric constants corrections}
contains strain and piezoelectric potential calculations using spatially
varying elastic and dielectric parameters. Section \ref{sec:Symmetry-adapted-basis k.p}
presents the $\mathbf{k}\cdot\mathbf{p}$ model used for electronic
structure calculations and our novel approach to efficiently include
strain through choices of unit cells. Section \ref{sec:Smooth-indium-profile}
introduces a method to use smooth indium profiles in all aspects of
our calculations. Section \ref{sec:Energy-shifts-from-corrections}
demonstrates our entire methodology for the case of a 1D quantum dot
array, such as quantum dots grown inside of nanowires \cite{Nguyen2011}.

\section{Spatially varying elastic and piezoelectric constants corrections\label{sec: Non-uniform elastic and piezoelectric constants corrections}}

We begin by considering quantum dot heterostructures with abrupt changes
in alloy fraction. Alloying the host material changes the local lattice
constants, leading to a lattice mismatch at the host and dot material
boundary. This lattice mismatch is a source of strain throughout the
QD system, affecting the electronic states of the system. For example,
InN has a larger lattice constant than GaN, so alloying GaN with indium
to form quantum dots induces change in the lattice constant. Additionally,
strain can generate strong piezoelectric potentials in materials such
as III-nitrides. The piezoelectric potential in III-nitrides is particularly
important along the c-axis and can be strong enough to spatially separate
electron and hole states through the quantum-confined Stark effect
\cite{Renard2009}.

In prior work, elastic and dielectric constants are largely assumed
to be spatially uniform in Fourier-based calculations of strain and
the piezoelectric potential. In fact, these material properties are
different in the dot and host materials, which can cause significant
errors when determining electronic structures. Here, we calculate
the strain and piezoelectric potential of a quantum dot superlattice
with elastic and dielectric constants that vary with alloy fraction,
while focusing on the changes brought on by spatially changing parameters.
In the case of the spatially varying elastic constants, we use a method
outlined in Ref.\ \cite{Andreev2000}. We present a version with
typos in Eqs. A3, A7 and A8 of Ref.\ \cite{Andreev2000} corrected
in Section \ref{subsec:strain }. For the piezoelectric potential,
we use a procedure similar to Ref.\ \cite{Andreev2000}, but we construct
a theory to include spatially varying dielectric constants. The strain
field and piezoelectric potential are coupled into a $\mathbf{k}\cdot\mathbf{p}$
model, presented in Section \ref{sec:Symmetry-adapted-basis k.p},
for electronic structure calculations.

\subsection{Quantum dot system\label{subsec:Quantum-dot-system}}

\begin{figure}[tbh]
\includegraphics[width=8.6cm]{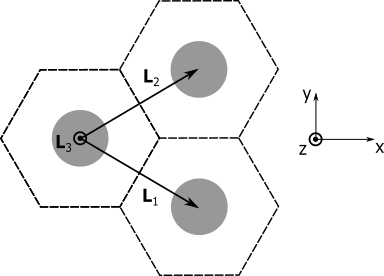}

\caption{Quantum dot superlattice and its basis vectors. White regions are
the host material and grey regions are the quantum dots. Dashed lines
show the unit cell boundaries of the quantum dot superlattice. Due
to the symmetry, we have $L_{12}\equiv\left|\mathbf{L}_{1}\right|=\left|\mathbf{L}_{2}\right|$.\label{fig:Quantum-dot-superlattice}}
\end{figure}

We consider a superlattice of cylindrical wurtzite quantum dots embedded
in a bulk host material, as shown in Fig.\ \ref{fig:Quantum-dot-superlattice}.
InGaN quantum dots such as those described in Ref.\ \cite{Nguyen2011}
have a lens-like shape and do not have a sharply defined boundary.
We approximate these quantum dots as being cylindrical. This choice
of dot geometry simplifies calculations, as described in Sec. \ref{subsec:strain },
and preserves the $C_{6v}$ symmetry of the material, which we take
advantage of in Section \ref{sec:Symmetry-adapted-basis k.p} for
electronic structure calculations. Hexagonal periodic boundary conditions
are used to also preserve the material's $C_{6v}$ symmetry. For single-dot
calculations, the superlattice unit cell must be large enough that
the choice of cell size does not affect results. For actual quantum
dot arrays, we consider only hexagonal superlattices in the plane.

In this periodic system, the real space quantum dot superlattice is
defined by the set of lattice vectors $\mathbf{L}_{i}$, as shown
in Fig.\ \ref{fig:Quantum-dot-superlattice}. We denote the real
space unit cell by $\Omega_{\mathsf{e}}$, its volume $V_{\mathsf{e}}$,
and the reciprocal-space unit cell by $\Omega_{\mathsf{e}}^{-1}$.
The index ``e'' indicates that these quantities relate to the electronic
cell, as opposed to the strain unit cell, which is introduced in Section
\ref{subsec:Including-strain-and-piezo-in-k.p}. Imposing periodic
conditions in real space implies a discrete reciprocal space with
wave vectors
\begin{equation}
\mathbf{q}=i_{1}\mathbf{b}_{1}+i_{2}\mathbf{b}_{2}+i_{3}\mathbf{b}_{3}\qquad i_{1},i_{2},i_{3}\in\mathbb{Z}\label{eq: reciprocal wave vectors-1}
\end{equation}
with the reciprocal basis vectors
\begin{equation}
\begin{gathered}\mathbf{b}_{1}=\frac{2\pi}{L_{1}}[1,\,-\frac{1}{\sqrt{3}},\,0]\\
\mathbf{b}_{2}=\frac{2\pi}{L_{2}}[1,\,\frac{1}{\sqrt{3}},\,0]\\
\mathbf{b}_{3}=\frac{2\pi}{L_{3}}[0,\,0,\,1]
\end{gathered}
\label{eq: reciprocal lattice vectors-1}
\end{equation}
Due to the symmetry of the system, we have $L_{1}=L_{2}$, which we
define as $L_{12}$. In our reciprocal-space calculations, we sample
on sets of the wave vectors $\mathbf{q}\in\Omega_{\mathsf{e}}^{-1}$.
We define $m_{12}$ and $m_{3}$ such that $i_{1},i_{2}=\left\{ -m_{12},\cdots,0,\cdots,m_{12}\right\} $
and $i_{3}=\left\{ -m_{3},\cdots,0,\cdots,m_{3}\right\} $. This sampling
produces a hexagonal mesh of size $N=N_{1}N_{2}N_{3}$ where $N_{i}=2m_{i}+1$.
To obtain a $C_{6}$ symmetric mesh, we remove points such that $\left|q_{x}\right|>m_{12}\frac{2\pi}{L_{1}}$,
leaving a mesh whose size we denote by $N_{\mathsf{e}}$.

By choosing the unit cell dimensions $L_{i}$ large enough, it is
possible to remove electronic coupling between neighboring dots. This
flexibility allows us to model 3D, 2D and 1D arrays of coupled dots.
The isolated dot case can also be obtained by choosing both $L_{12}$
and $L_{3}$ sufficiently large. Section \ref{subsec:strain } presents
a method that also uncouples dots in terms of strain, which is based
on calculating strain and the electronic structure using different
unit cells.

We illustrate the methods presented in this manuscript by modeling
a quantum dot system inspired by Ref.\ \cite{Nguyen2011}. That system
consists of InGaN dots grown in GaN nanowires. We approximate this
system as a 1D quantum dot array, by choosing $L_{3}$ to match the
measured dot-dot spacing and $L_{12}$ large enough to avoid dot-dot
coupling. We fix the dot indium alloy fraction, radius and height
based on the experimental device. System parameters are listed in
Table \ref{tab:System-parameters} and material parameters are in
Appendix \ref{sec:Bulk-k.p-parameters}.

\begin{table}[tbh]
\caption{Quantum dot superlattice system parameters used for calculations unless
specified otherwise. \label{tab:System-parameters}}

\begin{tabular}{cc}
 & \tabularnewline
\hline 
\hline 
Parameter & Value\tabularnewline
\hline 
$X_{0}$ & 0.45\tabularnewline
$h$ & 40 $\mathring{A}$\tabularnewline
$R$ & 200 $\mathring{A}$\tabularnewline
$L_{12}$ & 500 $\mathring{A}$\tabularnewline
$L_{3}$ & 70 $\mathring{A}$\tabularnewline
$m_{12}$ & 10\tabularnewline
$m_{3}$ & 4\tabularnewline
$n_{12}$ & 6\tabularnewline
$n_{3}$ & 1\tabularnewline
$\boldsymbol{\delta}$ & $[1.5,1.5,2.5]$ $\mathring{A}$\tabularnewline
\hline 
\end{tabular}
\end{table}

\subsection{Strain\label{subsec:strain }}

In this section, we present how we calculate strain with elastic constants
that depend on alloy fraction for 3D, 2D and 1D quantum dot superlattices
and isolated dots. Our method follows from Refs.\ \cite{Andreev2000,Vukmirovc2008}.

Materials such as InGaN have elastic constants that vary based on
the alloy fraction. Therefore, the spatial variation of the elastic
constants throughout the superlattice unit cell must be included for
accurate calculations of strain. We present a method, originally derived
in Ref.\ \cite{Andreev2000}, to include spatially varying elastic
constants in strain calculations. We calculate the strain produced
by a single isolated dot and construct the quantum dot superlattice
strain by linear superposition.

The calculated strain is to be coupled into the electronic structure
calculations. However, strain decays considerably slower than bound
electronic wavefunctions. In the case of isolated dots, the unit cell
must be large enough to accomodate the strain decay. Choosing a unit
cell large enough to capture the strain decay reduces the maximum
wave vector attainable when using a fixed number of plane waves. As
we demonstrate in Section \ref{sec:Energy-shifts-from-corrections},
accurately describing the electronic states requires using sufficiently
large wave vectors, and thus a large unit cell requires a large number
of plane waves. Following Ref.\ \cite{Vukmirovc2008}, we consider
that the electronic model and strain model each have their own real
space unit cells. This additional degree of freedom allows accurate
and computationally efficient determination of both electronic structure
of rapidly decaying confined quantum dot states and longer-range strain
effects in isolated dots. In the case of a quantum dot superlattice,
different real-space electronic and strain unit cells are not required.

\subsubsection{Isolated quantum dot strain\label{subsec:Isolated-quantum-dot}}

In prior work, lattice-mismatch-driven strain has been calculated
for a single dot using a continuum theory with a Green's function
approach while assuming spatially uniform elastic constants \cite{Andreev1999,Andreev2000,Nenashev2018}.
Here, we present the method outlined in Appendix A of \cite{Andreev2000}
to include spatially varying elastic constants. In this section, we
show how the spatially varying elastic constants modify strain and
how this modified strain changes the piezoelectric potential in Section
\ref{subsec:Piezoelectric-potential}. We show that the elastic constant
correction is necessary to obtain accurate strain and piezoelectric
potentials.

Consider a single InGaN QD in bulk GaN with spatially varying elastic
constants $\lambda_{ijmn}\left(\mathbf{r}\right)$ that depend on
the local alloy fraction,
\begin{equation}
\lambda_{ijmn}\left(\mathbf{r}\right)=\lambda_{ijmn}^{\mathsf{d}}+\lambda_{ijmn}^{\mathsf{h}}\left[1-\chi_{\mathsf{d}}\left(\mathbf{r}\right)\right],\label{eq: spatially varying elastic constants}
\end{equation}
where $\lambda_{ijmn}^{\mathsf{h}}$ and $\lambda_{ijmn}^{\mathsf{d}}$
are the host and dot's elastic constants, respectively. Assuming spatially
varying elastic constants, the Green's tensor $\tilde{G}_{in}$ for
the displacement field in an infinite anisotropic elastic medium must
satisfy
\begin{equation}
\frac{\partial}{\partial x_{k}}\lambda_{iklm}\left(\mathbf{r}\right)\frac{\partial}{\partial x_{m}}G\left(\mathbf{r},\mathbf{r}^{\prime}\right)=-\delta\left(\mathbf{r}-\mathbf{r}^{\prime}\right)\delta_{in}\label{eq: Diff eq. for Green's tensor}
\end{equation}
Taking the Fourier transform of Eq.\ \ref{eq: Diff eq. for Green's tensor},
we obtain
\[
\begin{aligned}\lambda_{iklm}^{\mathsf{h}}q_{k}q_{m} & \tilde{G}_{ln}\left(\mathbf{q},\mathbf{r}^{\prime}\right)\\
 & \phantom{\quad}+\Delta\lambda_{iklm}\sum_{\mathbf{q^{'}}}\tilde{\chi}_{\mathsf{d}}\left(\mathbf{q}-\mathbf{q}^{\prime}\right)q_{k}q_{m}^{\prime}\tilde{G}_{ln}\left(\mathbf{q}^{\prime},\mathbf{r}^{\prime}\right)\\
 & =\frac{1}{\left(2\pi\right)^{3}}\mathsf{e}^{i\mathbf{q}\cdot\mathbf{r}^{\prime}}\delta_{in}.
\end{aligned}
\]

The system strain is given by the superposition $\tilde{\epsilon}_{lm}\left(\mathbf{q}\right)=e_{lm}^{\mathsf{T}}\tilde{\chi}_{\mathsf{d}}\left(\mathbf{q}\right)+\tilde{\epsilon}_{lm}^{\mathsf{c}}\left(\mathbf{q}\right)$
where $e_{lm}^{\mathsf{T}}$ is the stress-free strain due to the
initial lattice mismatch and $\tilde{\epsilon}_{lm}^{\mathsf{c}}$
is the interface-driven strain \cite{Andreev1999,Andreev2000}. Reference
\cite{Andreev2000} showed that the Green's tensor can be related
to the strain $\tilde{\epsilon}_{lm}^{\mathsf{c}}\left(\mathbf{q}\right)$
to obtain
\begin{equation}
\begin{aligned}\lambda_{iklm}^{\mathsf{h}}q_{k} & \tilde{\epsilon}_{lm}^{c}\left(\mathbf{q}\right)\\
 & \phantom{\quad}+\Delta\lambda_{iklm}q_{k}\sum_{\mathbf{q^{'}}}\tilde{\chi}_{\mathsf{d}}\left(\mathbf{q}-\mathbf{q}^{'}\right)\tilde{\epsilon}_{lm}^{c}\left(\mathbf{q}^{'}\right)\\
 & =-\lambda_{ikpr}^{\mathsf{d}}\tilde{\epsilon}_{pr}^{T}q_{k}\tilde{\chi}^{d}\left(\mathbf{q}\right),
\end{aligned}
\label{eq:Strain system of equations}
\end{equation}
with typos fixed, where $\chi_{\mathsf{d}}$ is the characteristic
function of the dot, which is unity inside the dot and zero outside
(see Appendix \ref{sec:Characteristic-functions} for its Fourier
transform in our case of cylindrical dots), and $e_{pr}^{\mathsf{T}}$
is
\[
e_{pr}^{\mathsf{T}}=\varepsilon_{\mathsf{a}}\delta_{pr}+\varepsilon_{\mathsf{ca}}\delta_{p3}\delta_{r3}
\]
with $\varepsilon_{\mathsf{a}}=\left(a^{\mathsf{h}}-a^{\mathsf{d}}\right)/a^{\mathsf{d}}$,
$\varepsilon_{\mathsf{c}}=\left(c^{\mathsf{h}}-c^{\mathsf{d}}\right)/c^{\mathsf{d}}$
and $\varepsilon_{\mathsf{ca}}=\varepsilon_{\mathsf{c}}-\varepsilon_{\mathsf{a}}$.
Here, $a^{\mathsf{h}}$, $c^{\mathsf{h}}$ are the lattice constants
of the host material and, $a^{\mathsf{d}}$ and $c^{\mathsf{d}}$
are of dot material. More specifically, $a$ is the xy-plane lattice
constant and $c$ is the lattice constant along the z-axis. A solution
for $\tilde{\epsilon}_{lm}^{\mathsf{c}}\left(\mathbf{q}\right)$ can
be found by expanding $\tilde{\epsilon}_{lm}^{\mathsf{c}}\left(\mathbf{q}\right)$
in a power series,\textbf{
\begin{equation}
\tilde{\epsilon}_{lm}^{\mathsf{c}}\left(\mathbf{q}\right)=\tilde{\epsilon}_{lm}^{\left(0\right)}\left(\mathbf{q}\right)+\tilde{\epsilon}_{lm}^{\left(1\right)}\left(\mathbf{q}\right)+\tilde{\epsilon}_{lm}^{\left(2\right)}\left(\mathbf{q}\right)+\cdots,\label{eq:Strain power expansion}
\end{equation}
}where $\tilde{\epsilon}_{lm}^{\left(N\right)}\left(\mathbf{q}\right)\propto\left(\frac{\Delta\lambda}{\lambda}\right)^{N}$,
$\Delta\lambda_{ijmn}=\lambda_{ijmn}^{\mathsf{h}}-\lambda_{ijmn}^{\mathsf{d}}$
and the condition $\frac{\Delta\lambda}{\lambda}\ll1$ ensures convergence
of the series. The leading term $\tilde{\epsilon}_{lm}^{\left(0\right)}$
corresponds to uniform elastic constants of the dot with each subsequent
term being a correction to include spatial variations due to the alloy
profile. Using the Einstein summation convention, each term has the
form
\begin{equation}
\begin{aligned}\tilde{\epsilon}_{lm}^{\left(N\right)}\left(\mathbf{q}\right)=\frac{\left(2\pi\right)^{3}}{2}\left[F_{p}^{\left(N\right)}\left(\mathbf{q}\right)\right. & q_{l}\tilde{G}_{mp}^{\mathsf{h}}\left(\mathbf{q}\right)\\
 & \left.+F_{p}^{\left(N\right)}\left(\mathbf{q}\right)q_{m}\tilde{G}_{lp}^{\mathsf{h}}\left(\mathbf{q}\right)\right]
\end{aligned}
\label{eq: Strain eq.1}
\end{equation}
where
\begin{equation}
F_{i}^{\left(0\right)}\left(\mathbf{q}\right)=-\lambda_{ikpr}^{\mathsf{d}}\tilde{\epsilon}_{pr}^{T}q_{k}\tilde{\chi}_{\mathsf{d}}\left(\mathbf{q}\right)\label{eq: Strain eq.2}
\end{equation}
\begin{equation}
F_{i}^{\left(N\right)}\left(\mathbf{q}\right)=-\Delta\lambda_{iklm}q_{k}\frac{\left(2\pi\right)^{3}}{V}\sum_{\mathbf{q^{'}}}\tilde{\chi}_{\mathsf{d}}\left(\mathbf{q}-\mathbf{q}^{'}\right)\tilde{\epsilon}_{lm}^{\left(N-1\right)}\left(\mathbf{q}^{'}\right)\label{eq: Strain eq.3}
\end{equation}
where Eqs.\ \ref{eq: Strain eq.1} and \ref{eq: Strain eq.3} are
corrected from Ref.\ \cite{Andreev2000}. Here, $\tilde{G}_{in}^{\mathsf{h}}$
is the Green's tensor for the host material and is fully written out
in Appendix \ref{sec:Displacement-field-Green's}.

\begin{figure}[tbh]
\includegraphics[width=8.6cm]{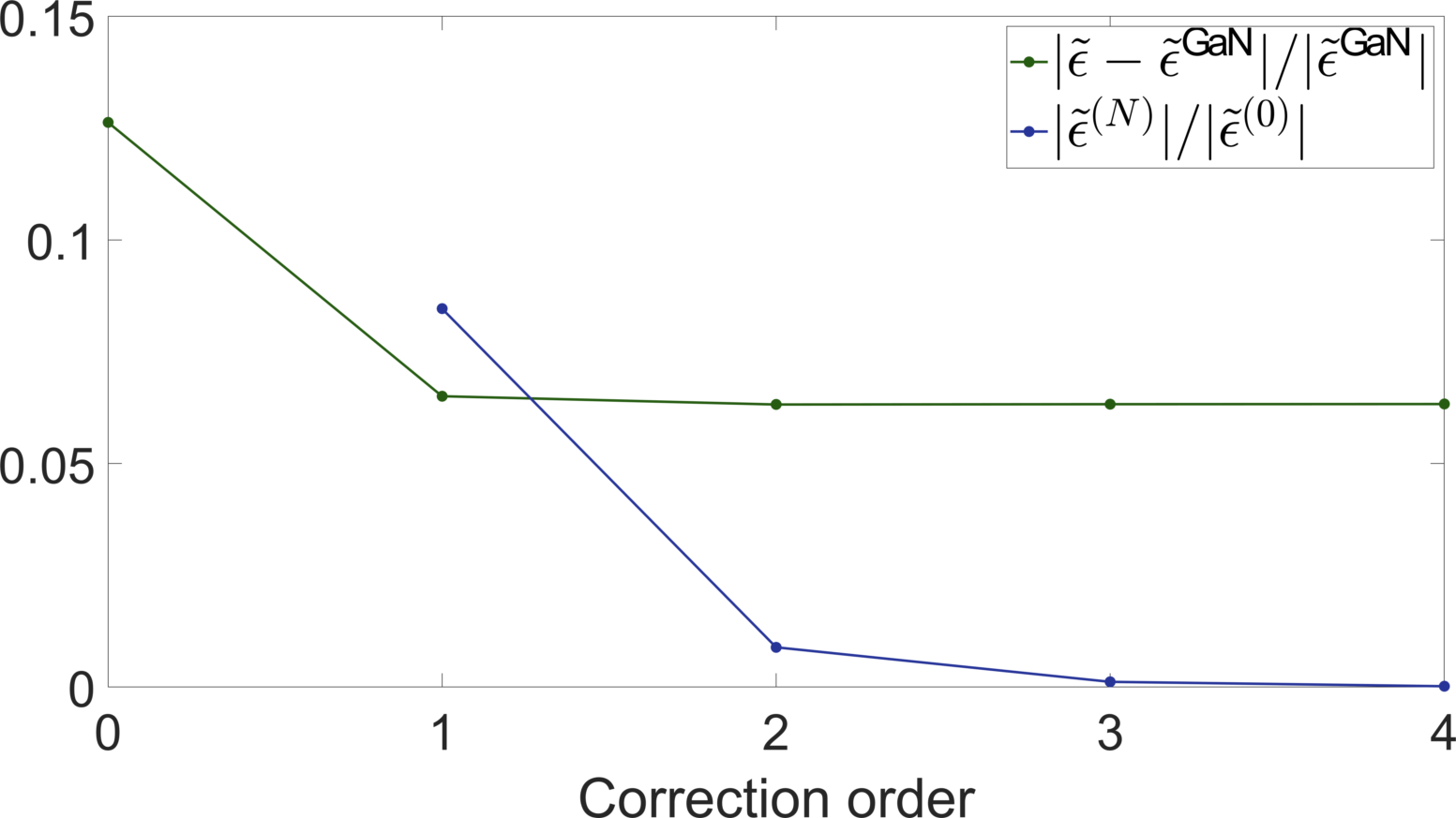}

\caption{Convergence of the strain with spatially varying elastic constants.
Green line is the relative difference between the corrected strain
and the case with uniform $\lambda$ of GaN for a 1D quantum dot array
as described in Table \ref{tab:System-parameters}. The zeroth term
is the case of uniform elastic constants of InGaN with alloy fraction
of the dot. Converged strain rests at a 6\% relative difference from
the case of uniform elastic constants of GaN, indicating that the
corrections are necessary for accurate strain fields. Blue line is
self-convergence of the power series in Eq.\ \ref{eq:Strain power expansion}
in respect to the zeroth term. Correction magnitudes are found to
be less than 1\% starting from 2nd order. We conclude that corrections
up to and including order 3 are sufficient for the calculation of
accurate strain fields. \label{fig:Convergence-of-strain}}
\end{figure}

\begin{figure}[tbh]
\includegraphics[width=8.6cm]{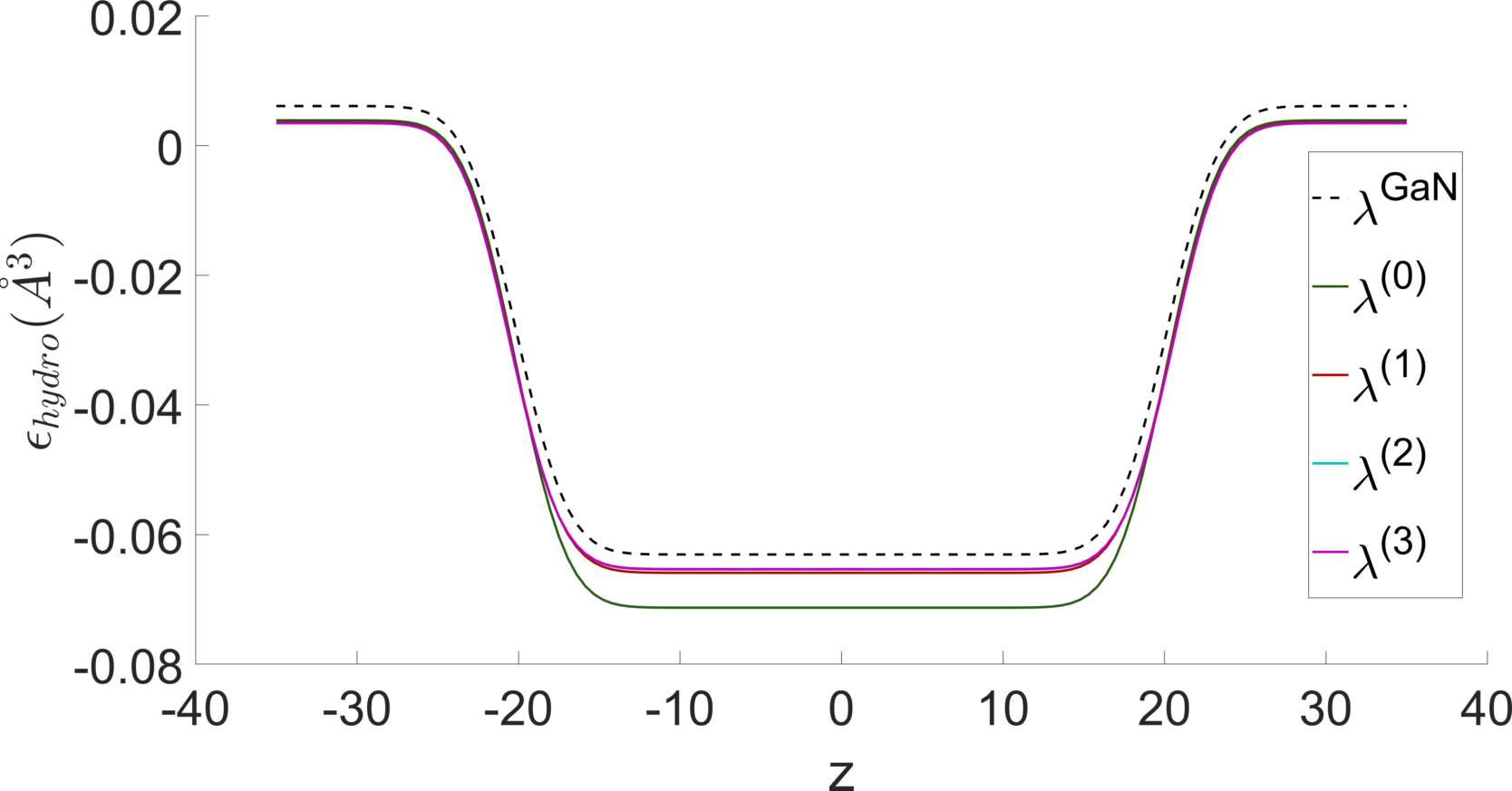}

\caption{Hydrostatic strain for increasing correction orders in the elastic
constants. Black dashed line is strain assuming uniform elastic constants
of the host material $\lambda^{\mathsf{GaN}}$ while solid lines are
for spatially varying elastic constants with correction order $\lambda^{\left(N\right)}$.
As the correction order $\lambda^{\left(N\right)}$ increases, the
strain moves toward the uniform case $\lambda^{\mathsf{GaN}}$. Note
that the lines for $\lambda^{\left(2\right)}$ and $\lambda^{\left(3\right)}$
are overlapping. However, the uniform case $\lambda^{\mathsf{GaN}}$
underestimates the strain in the dot. \label{fig:Hydrostatic-strain-comparison.}}

\end{figure}

It has been shown, when assuming uniform elastic constants, that using
the parameters for the host material gives more accurate results.
We compare the strain corrected at various orders according to Eq.\ \ref{eq:Strain power expansion}
to the usually considered case of uniform elastic constants of the
host material. Figure \ref{fig:Convergence-of-strain} shows the convergence
of the strain corrections for the 1D quantum dot array system described
in Section \ref{subsec:Quantum-dot-system}. We quantify convergence
with the following metric for the norm of the strain:
\begin{equation}
\left|\tilde{\epsilon}\right|=\sqrt{\sum_{m\geqslant l}\frac{V}{\left(2\pi\right)^{3}}\int\mathsf{d}\mathbf{q}^{3}\,\left|\tilde{\epsilon}_{lm}\left(\mathbf{q}\right)\right|^{2}}\label{eq:strain norm}
\end{equation}
where $m\geqslant l$ indicates the sum of the unique elements of
the strain tensor ($\tilde{\epsilon}_{11}$, $\tilde{\epsilon}_{22}$,
$\tilde{\epsilon}_{33}$, $\tilde{\epsilon}_{23}$, $\tilde{\epsilon}_{13}$,
$\tilde{\epsilon}_{12}$). The green line in Fig.\ \ref{fig:Convergence-of-strain}
compares the norm of the corrected strain $\tilde{\epsilon}$, calculated
from Eq.\ \ref{eq: Strain eq.1}, to the norm of the strain $\tilde{\epsilon}^{\mathsf{GaN}}$,
which is calculated assuming spatially uniform elastic constants of
GaN. Blue line shows the self-convergence of the power series in Eq.\ \ref{eq:Strain power expansion}.
From these results, we conclude that a 2nd order correction is sufficient
to have strain converged within 1\% in self-convergence and that this
converged strain differs from the uniform case by about 6\%, indicating
that the elastic constant corrections are important for accurate strain
fields in InGaN systems. In Sec.\ \ref{subsec:Piezoelectric-potential},
we show that the calculated piezoelectric potential remains essentially
unchanged from 3rd order corrections and up. Given that including
these corrections are not computationally costly, we have included
3rd order corrections in all of our calculations unless stated otherwise.
Figure \ref{fig:Hydrostatic-strain-comparison.} shows the hydrostatic
strain along a cut through the axis of the dot, showing relaxation
of strain inside the dot with each additional correction.


\subsubsection{Quantum dot superlattice strain\label{subsec:Quantum-dot-superlattice}}

Because the strain is linear in stress, the strain produced by the
QD superlattice can be obtained from linear superposition of the single-dot
strain. However, we want the ability to study dots that are completely
uncoupled, both electronically and from strains of the periodic array.
Reference \cite{Vukmirovc2008} proposed a method to allow simultaneous
treatment of large unit cell for the strain problem and small unit
cell for the electronic problem, which together allow isolated dots
to be considered in a computationally tractable manner. In this case
of two independent cells, the strain is calculated in its own real
space unit cell $\Omega_{\mathsf{s}}$ with volume $V_{\mathsf{s}}$.
We denote the strain reciprocal unit cell as $\Omega_{\mathsf{s}}^{-1}$
such that it contains the wave vectors $\mathbf{Q}$, which are defined
similarly to Eq.\ \ref{eq: reciprocal wave vectors-1} for the electronic
cell. Given that strain relaxes more slowly than bound electronic
wavefunctions, we only consider $V_{\mathsf{s}}\geq V_{\mathsf{e}}$.
In this two-unit-cells approximation, the Fourier transform of the
strain produced by the quantum dot array is
\begin{align}
\tilde{\epsilon}_{ij}^{\mathsf{a}}\left(\mathbf{q}\right) & =\frac{1}{V_{\mathsf{s}}}\sum_{\mathbf{Q}\in\Omega_{\mathsf{s}}^{-1}}\tilde{\epsilon}_{ij}\left(\mathbf{Q}\right)\tilde{\chi}_{\mathsf{e}}\left(\mathbf{q}-\mathbf{Q}\right)\nonumber \\
 & =\frac{1}{V_{\mathsf{s}}}\left(\tilde{\epsilon}_{ij}\ast\tilde{\chi}_{\mathsf{e}}\right)_{\mathsf{s}}\left(\mathbf{q}\right),\label{eq:QD superlattice strain}
\end{align}
where $\chi_{\mathsf{e}}$ is the characteristic function of the electronic
unit cell $\Omega_{\mathsf{e}}$ in $\Omega_{\mathsf{s}}$, which
is given for our case in Appendix \ref{sec:Characteristic-functions}.
Superscript ``$\mathsf{a}$'' indicates array. We follow the notation
that $\mathbf{q}\in\Omega_{\mathsf{e}}^{-1}$ and $\mathbf{Q}\in\Omega_{\mathsf{s}}^{-1}$.
$\left(\tilde{\epsilon}_{ij}\ast\tilde{\chi}_{\mathsf{e}}\right)_{\mathsf{s}}\left(\mathbf{q}\right)$
denotes a convolution where the subscript ``$\mathsf{s}$'' indicates
that the convolution is over the wave vectors $\mathbf{Q}\in\Omega_{\mathsf{s}}^{-1}$,
see Appendix \ref{sec:Conventions} for Fourier transform and convolution
definitions. We show in Sec.\ \ref{subsec:Including-strain-and-piezo-in-k.p}
that choosing the linear dimensions of $\Omega_{\mathsf{s}}$ to be
integer multiples of the linear dimensions of $\Omega_{\mathsf{e}}$
ensures that all vectors $\mathbf{q}\in\Omega_{\mathsf{e}}^{-1}$
are also in $\Omega_{\mathsf{s}}^{-1}$. This choice allows Eq.\ \ref{eq:QD superlattice strain}
to be evaluated efficiently.

\subsection{Piezoelectric potential\label{subsec:Piezoelectric-potential}}

III-nitride materials are strongly piezoelectric, having both spontaneous
and strain-driven polarizations \cite{Bernardini1997,Zoroddu2001}.
Calculation of the polarization from an electric field requires knowledge
of the static dielectric constant $\varepsilon$ of the material.
In prior work, all Fourier-space based approaches assumed a uniform
dielectric constant. We present a method to obtain the Fourier transform
of the scalar potential $\tilde{\varphi}\left(\mathbf{q}\right)$
assuming $\varepsilon\left(\mathbf{r}\right)$ changes with the local
alloy fraction. We find that correcting for the spatial dependence
of the dielectric function leads to important changes in the piezoelectric
potential. We also show in Sec.\ \ref{sec:Energy-shifts-from-corrections}
that this change in piezoelectric potential significantly shifts the
lowest quantum dot energy levels. We do not discuss metallic screening,
which can be important in highly doped materials \cite{Chichibu1998,Ibbetson2000,Kim2004}.

Generally, we can write the displacement field $\mathbf{D}\left(\mathbf{r}\right)$
as
\[
\mathbf{D}\left(\mathbf{r}\right)=\varepsilon_{0}\mathbf{E}\left(\mathbf{r}\right)+\mathbf{P}_{\mathsf{tot}}\left(\mathbf{r}\right),
\]
where $\mathbf{E}\left(\mathbf{r}\right)$ is the electric field,
$\varepsilon_{0}$ is the vacuum permittivity, and $\mathbf{P}_{\mathsf{tot}}$
is the total polarization. In the strained material, there are three
sources of polarization: bound charge, strain and spontaneous polarization,
\[
\mathbf{P}_{\mathsf{tot}}\left(\mathbf{r}\right)=\mathbf{P}_{\mathsf{bnd}}\left(\mathbf{r}\right)+\mathbf{P}_{\mathsf{st}}\left(\mathbf{r}\right)+\mathbf{P}_{\mathsf{sp}}\left(\mathbf{r}\right).
\]
Here, we assume no free charge screening and so an intrinsic material.
Assuming $\mathbf{P}_{\mathsf{bnd}}$ to be linear with the electric
field and incorporated into $\varepsilon\left(\mathbf{r}\right)$
as usual,
\begin{equation}
\mathbf{D}\left(\mathbf{r}\right)=\varepsilon\left(\mathbf{r}\right)\mathbf{E}\left(\mathbf{r}\right)+\mathbf{P}_{\mathsf{st}}\left(\mathbf{r}\right)+\mathbf{P}_{\mathsf{sp}}\left(\mathbf{r}\right),\label{eq:Displacement field}
\end{equation}
where $\mathbf{P}_{\mathsf{st}}\left(\mathbf{r}\right)+\mathbf{P}_{\mathsf{sp}}\left(\mathbf{r}\right)=\mathbf{P}\left(\mathbf{r}\right)$
is the residual polarization after electric-field induced bound charge
has been included in $\varepsilon\left(\mathbf{r}\right)$.

We take $\varepsilon\left(\mathbf{r}\right)$ to be $\varepsilon^{\mathsf{h}}$
in the host material and $\varepsilon^{\mathsf{d}}$ in the dot material,
so
\begin{equation}
\varepsilon\left(\mathbf{r}\right)=\varepsilon^{\mathsf{h}}+\left(\varepsilon^{\mathsf{d}}-\varepsilon^{\mathsf{h}}\right)\chi_{\mathsf{d}}\left(\mathbf{r}\right).\label{eq: dielectric constant}
\end{equation}
We obtain $\varepsilon^{\mathsf{d}}$ by linear interpolation of the
binary compounds' bulk dielectric constants. Taking the divergence
of Eq.\ \ref{eq:Displacement field}, using $\boldsymbol{\nabla}\cdot\mathbf{D}=0$,
taking the Fourier transform and solving for the electric field gives
\begin{equation}
E_{m}\left(\mathbf{r}\right)=-\frac{1}{\varepsilon\left(\mathbf{r}\right)}\mathscr{F}^{-1}\left\{ \frac{q_{n}}{q_{m}}\tilde{P}_{n}\left(\mathbf{q}\right)\right\} \label{eq: Electric field}
\end{equation}
where $\mathscr{F}^{-1}$ represents the inverse Fourier transform.
Using $E_{m}=-\partial_{m}\varphi$, where $\partial_{m}\equiv\frac{\partial}{\partial x_{m}}$
and $\varphi\left(\mathbf{r}\right)$ is the scalar potential, 
\begin{equation}
\tilde{\varphi}\left(\mathbf{q}\right)=-\frac{i}{q_{m}}\mathscr{F}\left\{ \frac{1}{\varepsilon\left(\mathbf{r}\right)}\mathscr{F}^{-1}\left\{ \frac{q_{n}}{q_{m}}\tilde{P}_{n}\left(\mathbf{q}\right)\right\} \left(\mathbf{r}\right)\right\} \left(\mathbf{q}\right).\label{eq: Piezoelectric potential 1}
\end{equation}
For the case of sharp alloy interfaces, $\chi_{\mathsf{d}}\left(\mathbf{r}\right)$
is either 1 or 0 and Eq.\ \ref{eq: dielectric constant} gives
\begin{equation}
\frac{1}{\varepsilon\left(\mathbf{r}\right)}=\frac{1}{\varepsilon^{\mathsf{h}}}+\left(\frac{1}{\varepsilon^{\mathsf{d}}}-\frac{1}{\varepsilon^{\mathsf{h}}}\right)\chi_{\mathsf{d}}\left(\mathbf{r}\right).\label{eq: inverse dielectric constant}
\end{equation}
We treat the case of smoothly varying alloy fraction in Sec.\ \ref{sec:Smooth-indium-profile}.
Putting this result in Eq.\ \ref{eq: Piezoelectric potential 1}
gives
\begin{align}
\tilde{\varphi}\left(\mathbf{q}\right) & =\tilde{\varphi}_{\mathsf{uni}}^{\mathsf{h}}\left(\mathbf{q}\right)+\Delta\tilde{\varphi}\left(\mathbf{q}\right)\label{eq: piezoelectric potential 2}
\end{align}
with
\begin{equation}
\tilde{\varphi}_{\mathsf{uni}}^{\mathsf{h}}\left(\mathbf{q}\right)=-\frac{i}{q_{m}}\frac{1}{\varepsilon^{\mathsf{h}}}\frac{q_{n}}{q_{m}}\tilde{P}_{n}\left(\mathbf{q}\right)\label{eq:uniform piezo}
\end{equation}
\begin{equation}
\Delta\tilde{\varphi}\left(\mathbf{q}\right)=-\frac{i}{q_{m}}\left(\frac{1}{\varepsilon^{\mathsf{d}}}-\frac{1}{\varepsilon^{\mathsf{h}}}\right)\mathscr{F}\left\{ \chi_{\mathsf{d}}\left(\mathbf{r}\right)\mathscr{F}^{-1}\left\{ \frac{q_{n}}{q_{m}}\tilde{P}_{n}\left(\mathbf{q}\right)\right\} \right\} \label{eq:delta piezo}
\end{equation}
Here, $\tilde{\varphi}_{\mathsf{uni}}^{\mathsf{h}}$ is the contribution
to $\varphi$ with $\varepsilon_{\mathsf{r}}\left(\mathbf{r}\right)=\varepsilon^{\mathsf{h}}$,
and $\Delta\tilde{\varphi}$ is the change in $\tilde{\varphi}$ due
to the dot material having a different dielectric constant. The polarization
fields $\tilde{P}_{n}\left(\mathbf{q}\right)$ for the wurtzite crystal
structure are given in terms of strain in Appendix \ref{sec:Polarization-fields}.

\begin{figure}[tbh]
\includegraphics[width=8.6cm]{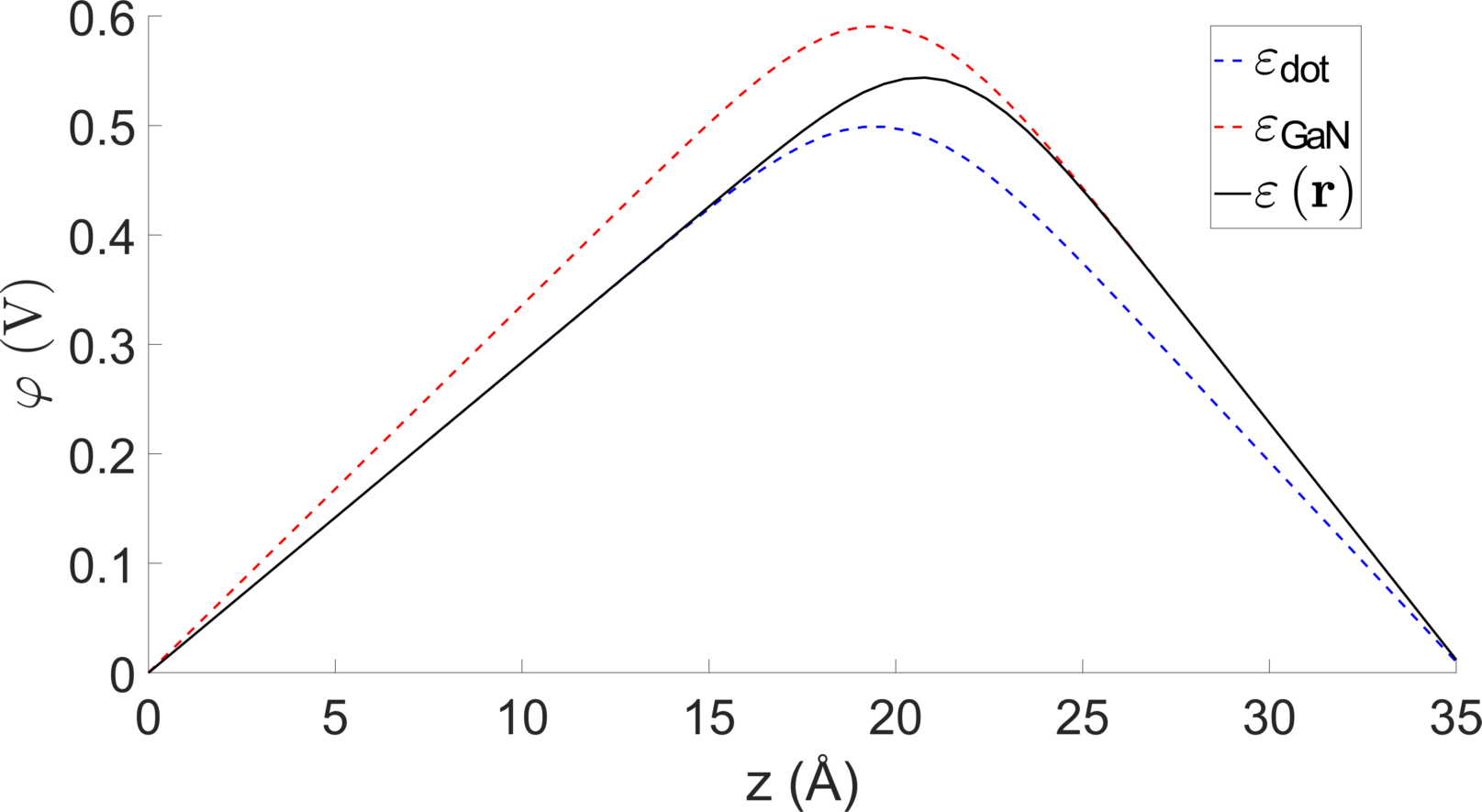}

\caption{Piezoelectric potential $\varphi\left(z\right)$ along the central
axis, beginning in the center of the dot, with parameters as in Table
\ref{tab:System-parameters}. Blue and red dashed lines show $\varphi$
calculated with uniform $\varepsilon\left(\mathbf{r}\right)=\varepsilon^{\mathsf{h}}$
and $\varepsilon\left(\mathbf{r}\right)=\varepsilon^{\mathsf{d}}$,
respectively. Black line shows the case with spatially varying $\varepsilon\left(\mathbf{r}\right)$.
Note that the $\varphi$ is antisymmetric in $z$. These results show
that the case of uniform $\varepsilon$ cannot capture $\varphi$
throughout the system. Strain calculations include third order corrections
for the nonuniform elastic constants. \label{fig:piezo and dielectric constant}}

\end{figure}

We now show the piezoelectric potentials that result from this formulation,
for our model system described in Sec.\ \ref{subsec:Quantum-dot-system}.
Figure \ref{fig:piezo and dielectric constant} shows $\varphi\left(z\right)$
along the central axis of the quantum dot calculated with constant
$\varepsilon$ of the dot and host and with Eq.\ \ref{eq: piezoelectric potential 2}.
The calculation with spatially varying $\varepsilon\left(\mathbf{r}\right)$
agrees with $\varphi_{\mathsf{uni}}^{\mathsf{d}}$ inside the dot
and also agrees with $\varphi_{\mathsf{uni}}^{\mathsf{h}}$ outside
the dot, with a transition near the boundary that is captured by neither
of the uniform cases.

\begin{figure}[tbh]
\includegraphics[width=8.6cm]{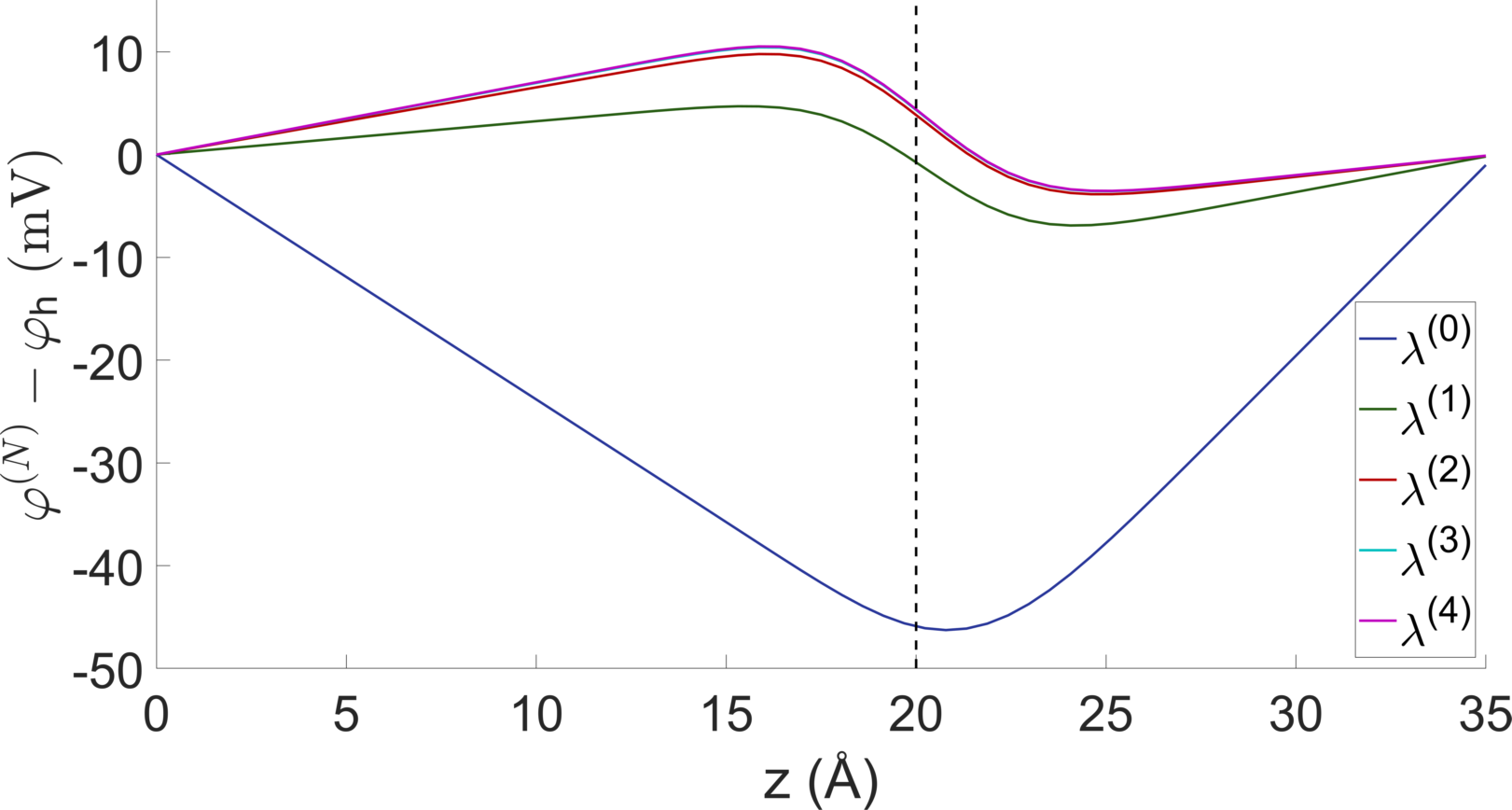}

\caption{Piezoelectric potential difference along the central axis of the dot
for various elastic constant corrections. Potential difference is
with respect to $\varphi_{\mathsf{h}}\left(\mathbf{r}\right)$, which
is calculated with a uniform $\lambda\left(\mathbf{r}\right)=\lambda^{\mathsf{GaN}}$.
Calculations implement an alloy smoothing of $\boldsymbol{\delta}=[1.5,1.5,2.5]\,\mathring{A}$,
which is described in Sec.\ \ref{sec:Smooth-indium-profile}, and
the vertical dashed line indicates the nominal material interface
without smoothing. These results are for the same system as Fig.\ \ref{fig:piezo and dielectric constant},
calculated using various orders of correction for the effects of nonuniform
elastic constants, described in Sec.\ \ref{subsec:Isolated-quantum-dot}.
Spatially varying dielectric constants are included. The zeroth order
case (blue line) shows $\lambda\left(\mathbf{r}\right)=\lambda^{\mathsf{d}}$.
\label{fig:piezo and elastic constants}}
\end{figure}

We showed in Fig.\ \ref{fig:Hydrostatic-strain-comparison.} how
spatially varying elastic constants change strain profiles. Figure
\ref{fig:piezo and elastic constants} shows how $\varphi$ changes
due to the elastic constants' correction propagates into the piezoelectric
potential. We find that the changes in piezoelectric potential, a
peak correction of 8 mV, are significant if looking to converge the
energy levels within a few meV's.


\section{Symmetry adapted basis $\mathbf{k}\cdot\mathbf{p}$ for wurtzite
quantum dots\label{sec:Symmetry-adapted-basis k.p}}

Here, we present the quantum dot $\mathbf{k}\cdot\mathbf{p}$ model
we use for electronic structure calculations. We first present a theory
for bulk materials and use it to construct a theory for quantum dots.
This quantum dot Hamiltonian is written in a symmetry adapted basis,
which reduces the computational cost for calculating and diagonalizing
the Hamiltonian. In this symmetry adapted basis, we show how the strain
produced by the quantum dots contributes to the Hamiltonian. We also
introduce strain effects using a different unit cell than that of
the electronic cell defined in Fig.\ \ref{fig:Quantum-dot-superlattice}.
In this section, our goal is to show our method of efficiently including
strain in the quantum dot $\mathbf{k}\cdot\mathbf{p}$ model, which
we do by choosing the strain unit cell's dimensions to be integer
multiples of the unit cell used for the electronic structure calculations.

\subsection{Bulk $\mathbf{k}\cdot\mathbf{p}$ model\label{subsec:Bulk k.p}}

To describe the electronic structure of bulk wurtzite systems, we
use an 8-band $\mathbf{k}\cdot\mathbf{p}$ model, which includes spin-orbit
coupling, crystal field splitting and strain. An 8-band $\mathbf{k}\cdot\mathbf{p}$
model for bulk wurtzite material has been presented by Ref. \cite{Winkelnkemper}
in the basis of $\Gamma$-point Bloch functions. References \cite{Winkelnkemper,Chuang}
presented a 6-band model using eigenfunctions of the angular momentum
operator $\hat{J}_{z}$. Since choosing $\hat{J}_{z}$ eigenfunctions
aids in the construction of a symmetry adapted basis, which is presented
in Sec.\ \ref{subsec:Quantum dot k.p}, we have used these two references
to construct an 8-band Hamiltonian in the $\hat{J}_{z}$ eigenfunctions
basis. More precisely, we have constructed the Hamiltonian using $S$,
$X$, $Y$ and $Z$ $\Gamma$-point Bloch functions as a basis and
then performed a basis transformation to obtain the $\hat{J}_{z}$
eigenfunctions basis. While $\mathbf{k}\cdot\mathbf{p}$ parameters
are usually obtained in $S$, $X$, $Y$ and $Z$ basis, recent work
has obtained $\mathbf{k}\cdot\mathbf{p}$ parameters directly in the
symmetry adapted basis using \textit{ab initio} calculations \cite{Jocic2020}.

We consider the time-independent Schr\"odinger equation for a single
electron
\begin{equation}
\hat{H}\ket{\psi}=E\ket{\psi}\label{eq: General Eigenvalue problem}
\end{equation}
where
\begin{equation}
\hat{H}=\hat{K}+\hat{V}+\hat{H}_{\mathsf{so}}+\hat{H}_{\mathsf{cr}}+\hat{H}_{\mathsf{st}}\label{eq: General Hamiltonian}
\end{equation}
Here, $\hat{K}$ is the kinetic term of the electrons, $\hat{V}$
the potential from the electron-ion interaction, $\hat{H}_{\mathsf{so}}$
is spin-orbit coupling, $\hat{H}_{\mathsf{cr}}$ is crystal field
splitting and $\hat{H}_{\mathsf{st}}$ is strain coupling. We expand
$\ket{\psi}$ in terms of the $\hat{J}_{z}$ eigenfunctions $\ket{u_{i}}$
\begin{equation}
\ket{\psi}=e^{i\mathbf{k}\cdot\mathbf{r}}\sum_{\alpha=1}^{8}C_{\alpha}\ket{u_{i}}\label{eq: Bulk eigenfunction expansion}
\end{equation}
where
\begin{align}
 & \ket{u_{1}}=\ket{iS,\uparrow} &  & \ket{u_{5}}=\ket{-iS,\downarrow}\nonumber \\
 & \ket{u_{2}}=\ket{-\frac{X+iY}{\sqrt{2}},\uparrow} &  & \ket{u_{6}}=\ket{\frac{X-iY}{\sqrt{2}},\downarrow}\nonumber \\
 & \ket{u_{3}}=\ket{\frac{X-iY}{\sqrt{2}},\uparrow} &  & \ket{u_{7}}=\ket{-\frac{X+iY}{\sqrt{2}},\downarrow}\nonumber \\
 & \ket{u_{4}}=\ket{Z,\uparrow} &  & \ket{u_{8}}=\ket{Z,\downarrow}\label{eq: bulk basis}
\end{align}
Here, $S$, $X$, $Y$ and $Z$ are $\Gamma$-point Bloch functions
with arrows indicating spin. The eigenvalues of the $\hat{J}_{z}$
eigenfunctions are
\[
J_{z}=\left\{ \frac{1}{2},\quad\frac{3}{2},\quad-\frac{1}{2},\quad\frac{1}{2},\quad-\frac{1}{2},\quad-\frac{3}{2},\quad\frac{1}{2},\quad-\frac{1}{2}\right\} ,
\]
respectively. Inserting Eq.\ \ref{eq: Bulk eigenfunction expansion}
into \ref{eq: General Eigenvalue problem}, the eigenvalue problem
can be written as
\begin{equation}
H_{\alpha^{\prime}\alpha}C_{\alpha}=EC_{\alpha^{\prime}}\label{eq: Bulk eigenvalue problem}
\end{equation}
Keeping terms only up to order $k^{2}$, the 8x8 $\mathbf{k}\cdot\mathbf{p}$
Hamiltonian is
\begin{align}
H & =\left[\begin{array}{cc}
g\left(\mathbf{k}\right) & \gamma\\
-\gamma^{*} & g^{*}\!\left(\mathbf{k}\right)
\end{array}\right]\label{eq: bulk Hamiltonian}
\end{align}
where
\begin{align*}
g\left(\mathbf{k}\right) & =g_{1}\left(\mathbf{k}\right)+g_{2}\left(\mathbf{k}\right)+g_{\mathsf{cr}}+g_{\mathsf{so}}+g_{\mathsf{st}}
\end{align*}

\begin{widetext}
\[
g_{1}\left(\mathbf{k}\right)=\left[\begin{array}{cccc}
E_{\mathsf{c}}^{\prime} & -\frac{P_{2}}{\sqrt{2}}k_{+} & \frac{P_{2}}{\sqrt{2}}k_{-} & P_{1}k_{z}\\
-\frac{P_{2}}{\sqrt{2}}k_{-} & E_{\mathsf{v}}^{\prime} & 0 & 0\\
\frac{P_{2}}{\sqrt{2}}k_{+} & 0 & E_{\mathsf{v}}^{\prime} & 0\\
P_{1}k_{z} & 0 & 0 & E_{\mathsf{v}}^{\prime}
\end{array}\right]
\]
\[
g_{2}\left(\mathbf{k}\right)=\left[\begin{array}{cccc}
A_{2}^{\prime}\left(k_{x}^{2}+k_{y}^{2}\right)+A_{1}^{\prime}k_{z}^{2} & 0 & 0 & 0\\
0 & \left(\frac{L_{1}^{\prime}+M_{1}}{2}\right)\left(k_{x}^{2}+k_{y}^{2}\right)+M_{2}k_{z}^{2} & -\frac{1}{2}N_{1}^{\prime}k_{-}^{2} & -\frac{1}{\sqrt{2}}N_{2}^{\prime}k_{-}k_{z}\\
0 & -\frac{1}{2}N_{1}^{\prime}k_{+}^{2} & \left(\frac{L_{1}^{\prime}+M_{1}}{2}\right)\left(k_{x}^{2}+k_{y}^{2}\right)+M_{2}k_{z}^{2} & \frac{1}{\sqrt{2}}N_{2}^{\prime}k_{+}k_{z}\\
0 & -\frac{1}{\sqrt{2}}N_{2}^{\prime}k_{+}k_{z} & \frac{1}{\sqrt{2}}N_{2}^{\prime}k_{-}k_{z} & M_{3}\left(k_{x}^{2}+k_{y}^{2}\right)+L_{2}^{\prime}k_{z}^{2}
\end{array}\right]
\]
\begin{equation}
g_{\mathsf{st}}=\left[\begin{array}{cccc}
a_{2}\left(\epsilon_{xx}+\epsilon_{yy}\right)+a_{1}\epsilon_{zz} & 0 & 0 & 0\\
0 & \frac{1}{2}\left(l_{1}+m_{1}\right)\left(\epsilon_{xx}+\epsilon_{yy}\right)+m_{2}\epsilon_{zz} & -\frac{1}{2}\left(l_{1}-m_{1}\right)\left(\epsilon_{xx}-\epsilon_{yy}\right)+in_{1}\epsilon_{xy} & -\frac{n_{2}\left(\epsilon_{xz}-i\epsilon_{yz}\right)}{\sqrt{2}}\\
0 & -\frac{1}{2}\left(l_{1}-m_{1}\right)\left(\epsilon_{xx}-\epsilon_{yy}\right)-in_{1}\epsilon_{xy} & \frac{1}{2}\left(l_{1}+m_{1}\right)\left(\epsilon_{xx}+\epsilon_{yy}\right)+m_{2}\epsilon_{zz} & \frac{n_{2}\left(\epsilon_{xz}+i\epsilon_{yz}\right)}{\sqrt{2}}\\
0 & -\frac{n_{2}\left(\epsilon_{xz}+i\epsilon_{yz}\right)}{\sqrt{2}} & \frac{n_{2}\left(\epsilon_{xz}-i\epsilon_{yz}\right)}{\sqrt{2}} & m_{3}\left(\epsilon_{xx}+\epsilon_{yy}\right)+l_{2}\epsilon_{zz}
\end{array}\right]\label{eq: Bulk strain Hamiltonian}
\end{equation}

\end{widetext}
\[
g_{\mathsf{cr}}=\Delta_{\mathsf{cr}}\left[\begin{array}{cccc}
0 & 0 & 0 & 0\\
0 & 1 & 0 & 0\\
0 & 0 & 1 & 0\\
0 & 0 & 0 & 0
\end{array}\right]
\]
\[
g_{\mathsf{so}}=\frac{\Delta_{\mathsf{so}}}{3}\left[\begin{array}{cccc}
0 & 0 & 0 & 0\\
0 & 1 & 0 & 0\\
0 & 0 & -1 & 0\\
0 & 0 & 0 & 0
\end{array}\right]
\]
\begin{align*}
\gamma & =\frac{\sqrt{2}\Delta_{\mathsf{so}}}{3}\left[\begin{array}{cccc}
0 & 0 & 0 & 0\\
0 & 0 & 0 & 0\\
0 & 0 & 0 & 1\\
0 & 0 & -1 & 0
\end{array}\right]
\end{align*}
Here, $\Delta_{\mathsf{cr}}$ and $\Delta_{\mathsf{so}}$ are the
crystal field splitting and spin-orbit coupling, respectively. The
band edges are $E_{\mathsf{c}}^{\prime}=E_{\mathsf{v}}+E_{\mathsf{g}}+\Delta_{\mathsf{cr}}+\frac{\Delta_{\mathsf{so}}}{3}+\varphi$
and $E_{\mathsf{v}}^{\prime}=E_{\mathsf{v}}+\varphi$ where $\varphi$
is any additional scalar potential such as the piezoelectric potential.
The $A_{i}^{\prime}$ parameters are related to the Kane parameters
$P_{i}$ and $L_{i}^{\prime}$, $M_{i}$, $N_{i}^{\prime}$ to the
Luttinger-like parameters $A_{i}$, all of which are shown in Appendix
\ref{sec:Bulk-k.p-parameters}. $g_{\mathsf{st}}$ is the contribution
to the Hamiltonian due to strain $\epsilon_{ij}$. The parameters
$a_{i}$, $l_{i}$. $m_{i}$ and $n_{i}$ for the strain contribution
are given in Appendix \ref{sec:Bulk-k.p-parameters} in terms of deformation
potentials.

In example calculations, alloy parameters have been obtained by linearly
interpolating between bulk GaN and InN parameters, which are given
in Appendix \ref{sec:Bulk-k.p-parameters}, except the band gap, which
has bowing included.

\subsection{Quantum dot $\mathbf{k}\cdot\mathbf{p}$ \label{subsec:Quantum dot k.p}}

For the quantum dot system, we construct the Hamiltonian from the
bulk system described in Section \ref{subsec:Bulk k.p}. We use slowly
varying envelope functions and apply a spatial dependence to the bulk
Hamiltonian. The problem is expressed in a symmetry adapted basis
to obtain a block diagonal Hamiltonian from which we calculate the
eigenstates of the quantum dot.

We start from Eq.\ \ref{eq: General Eigenvalue problem}, but expand
$\ket{\psi}$ in terms of envelope functions $F_{\alpha}\left(\mathbf{r}\right)$
that are slowly varying compared to the lattice constant \cite{Andreev2000,Tomic2006,Vukmirovic2005,Vukmirovc2008},
\[
\ket{\psi}=\sum_{\alpha=1}^{8}\ket{F,\alpha}
\]
\begin{equation}
\braket{\mathbf{r}|F,\alpha}=F_{\alpha}\left(\mathbf{r}\right)u_{\alpha}\left(\mathbf{r}\right)\label{eq:envelope function definiton}
\end{equation}
where the $u_{\alpha}\left(\mathbf{r}\right)$ are defined by Eq.\ \ref{eq: bulk basis}
and are periodic with the crystal lattice. Analogous to Eq.\ \ref{eq: Bulk eigenvalue problem},
this envelope function expansion leads to
\begin{equation}
\sum_{\alpha=1}^{8}H_{\alpha^{'}\alpha}F_{\beta}\left(\mathbf{r}\right)=EF_{\alpha^{'}}\left(\mathbf{r}\right)\label{eq: Quantum dot eigenvalue problem 1}
\end{equation}
where $H_{\alpha^{'}\alpha}$ are the bulk Hamiltonian matrix elements
from Eq.\ \ref{eq: bulk Hamiltonian}. Due to the broken translation
symmetry in the quantum dot system, we apply the substitution
\begin{equation}
k_{j}\rightarrow-i\frac{\partial}{\partial x_{j}}\label{eq: k to der substitution}
\end{equation}
to the bulk Hamiltonian in Eq.\ \ref{eq: bulk Hamiltonian}. Each
parameter in the bulk Hamiltonian also has a spatial dependence based
on the alloy distribution,
\begin{equation}
f\left(\mathbf{r}\right)=f^{\mathsf{d}}\chi_{\mathsf{d}}\left(\mathbf{r}\right)+f^{\mathsf{h}}\left[1-\chi_{\mathsf{d}}\left(\mathbf{r}\right)\right]\label{eq: Spatial dependance of k.p parameters}
\end{equation}
Here, $f$ stands for any of the parameters in the bulk Hamiltonian
that are material dependent. $f^{\mathsf{h}}$ and $f^{\mathsf{d}}$
are the parameter values of the host and alloyed dot material, respectively.
Applying the substitution in Eq.\ \ref{eq: k to der substitution}
to Eq.\ \ref{eq: bulk Hamiltonian}, the Hamiltonian consists of
terms of the form $f\left(\mathbf{r}\right)$, $f\left(\mathbf{r}\right)\frac{\partial}{\partial x_{j}}$
and $f\left(\mathbf{r}\right)\frac{\partial^{2}}{\partial x_{i}\partial x_{j}}$.
To preserve Hermiticity, we symmetrize the derivatives \cite{Morrow1984,Vukmirovic2005,Tomic2006}:
\begin{equation}
f\left(\mathbf{r}\right)\frac{\partial}{\partial x_{j}}\rightarrow\frac{1}{2}\left(f\left(\mathbf{r}\right)\frac{\partial}{\partial x_{j}}+\frac{\partial}{\partial x_{j}}f\left(\mathbf{r}\right)\right)\label{eq:2nd type matrix element symmetrisation}
\end{equation}
\begin{equation}
f\left(\mathbf{r}\right)\frac{\partial^{2}}{\partial x_{i}\partial x_{j}}\rightarrow\frac{1}{2}\left(\frac{\partial}{\partial x_{i}}f\left(\mathbf{r}\right)\frac{\partial}{\partial x_{j}}+\frac{\partial}{\partial x_{j}}f\left(\mathbf{r}\right)\frac{\partial}{\partial x_{i}}\right)\label{eq:3rd type of matrix element symmetrisation}
\end{equation}

\begin{figure}[tbh]
\includegraphics[width=8.6cm]{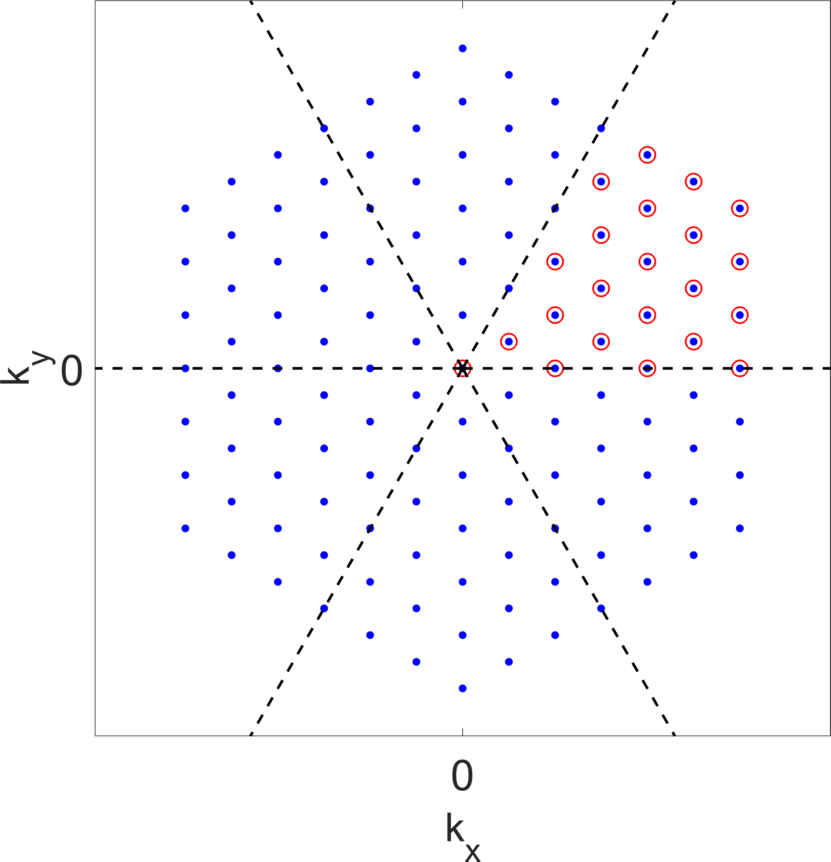}

\caption{Fourier space sampling used for the symmetry adapted basis. Red circles
are the Fourier space points $\mathbf{q}$ used in the symmetry adapted
basis. Blue dots are the full Fourier space sampled by rotations $\protect\overleftrightarrow{\mathbf{R}}_{\!l}\mathbf{q}$.
Dashed lines highlight the 6 sextants.\label{fig:Example-of-SAB-sampling}}
\end{figure}

The envelope functions $F_{\alpha}\left(\mathbf{r}\right)$ are periodic
with the superlattice and can be expanded in Fourier domain using
the superlattice reciprocal wave vectors $\mathbf{q}$ defined in
Eq.\ \ref{eq: reciprocal wave vectors-1}. Writing the envelope functions
$F_{\alpha}\left(\mathbf{r}\right)$ in terms of plane waves leads
to a non-sparse Hamiltonian \cite{Andreev2000,Tomic2006}. For computational
efficiency, we use a symmetry adapted basis, which takes advantage
of the $C_{6}$ symmetry of the wurtzite crystal structure by block
diagonalizing the Hamiltonian. Symmetry adapted bases have been fully
described for both zincblende and wurtzite systems \cite{Vukmirovic2005,Vukmirovic2006}.
We use a symmetry-adapted basis with elements $\ket{m_{f},\alpha,\mathbf{q}}$
where $\mathbf{q}=\left(q_{x},q_{y},q_{z}\right)$ are chosen within
a single sextant, so $0\leq q_{y}\leq\tan\left(\frac{2\pi}{6}\right)q_{x}$,
and $m_{f}=\left\{ -5/2,-3/2,-1/2,1/2,3/2,5/2\right\} $ can be interpreted
as a total quasi angular momentum \cite{Vukmirovic2006,Vukmirovc2008}.
This basis consists of the basis functions of the irreducible representations
of the double group $\bar{C}_{6}$. Using this basis reduces the Fourier
space sampling to a single sextant of the full space and block diagonalizes
the Hamiltonian into 6 blocks, which are labeled by $m_{f}$. This
block diagonalization greatly reduces the computational cost to diagonalize
the Hamiltonian. Figure \ref{fig:Example-of-SAB-sampling} shows an
example of the Fourier space sampling used in the symmetry adapted
basis. Written out, the basis states are
\begin{equation}
\ket{m_{f},\alpha,\mathbf{q}}=\Lambda\left(m_{f},\alpha,\mathbf{q},\mathbf{r}\right)\ket{u_{\alpha}}\label{eq:Symmetry adapted basis}
\end{equation}
\begin{widetext}
\begin{equation}
\Lambda\left(m_{f},\alpha,\mathbf{q},\mathbf{r}\right)=\begin{cases}
\frac{1}{\sqrt{6}}\sum_{l=0}^{5}e^{i\mbox{\ensuremath{\left(\overleftrightarrow{\mathbf{R}}_{\!l}\mathbf{q}\right)}}\cdot\mathbf{r}}e^{il\frac{2\pi}{6}\left[m_{f}-J_{z}\left(\alpha\right)\right]} & q_{x}\neq0\,\mathsf{or}\,q_{y}\neq0\\
e^{i\mathbf{q}\cdot\mathbf{r}} & q_{x}=q_{y}=0\qquad J_{z}\left(\alpha\right)=m_{f}
\end{cases}\label{eq:SAB defintion}
\end{equation}
\end{widetext} where$\overleftrightarrow{\mathbf{R}}_{l}$ is the
$l\frac{2\pi}{6}$ rotation around the z-axis. Equation \ref{eq:SAB defintion}
distinguishes wave vectors that are purely along the z-axis from those
that have an xy-component, which we denote by $\mathbf{q}_{z}$ and
$\mathbf{q}$, respectively. These two cases differ because a z-axis
rotation leaves $\mathbf{q}_{z}$ invariant while sending $\mathbf{q}$
to a new wave vector $\overleftrightarrow{\mathbf{R}}_{\!l}\mathbf{q}$.
The case of $\ket{m_{f},\alpha,\mathbf{q}_{z}}$ with $J_{z}\left(\alpha\right)\neq m_{f}$
does not exist in the basis set. Using the symmetry adapted basis,
the eigenstates can be written
\begin{equation}
\ket{\psi_{i,m_{f}}}=\sum\limits _{\alpha=1}^{8}\sum_{\mathbf{q}}A_{im_{f}}^{\alpha}\left(\mathbf{q}\right)\ket{m_{f},\alpha,\mathbf{q}}\label{eq:Solutions in SAB}
\end{equation}
where the $\mathbf{q}$ summation is restricted to the sextant, shown
in Fig.\ \ref{fig:Example-of-SAB-sampling}.

Writing the envelope functions in the symmetry adapted basis, the
eigenvalue problem in Eq.\ \ref{eq: Quantum dot eigenvalue problem 1}
can then be written
\begin{equation}
\sum_{\alpha=1}^{8}\sum_{\mathbf{q}}\mathcal{H}_{m_{f}\alpha^{\prime}\alpha}\left(\mathbf{q}^{\prime},\mathbf{q}\right)A_{im_{f}}^{\alpha}\left(\mathbf{q}\right)=E_{i}A_{im_{f}}^{\alpha^{\prime}}\left(\mathbf{q}^{\prime}\right)\label{eq:Eigenvalue problem in basis}
\end{equation}
with
\begin{align}
\mathcal{H}_{m_{f}\alpha^{\prime}\alpha}\left(\mathbf{q}^{\prime},\mathbf{q}\right) & \equiv\braket{m_{f},\alpha^{\prime},\mathbf{q}^{\prime}|\hat{\mathcal{H}}|m_{f},\alpha,\mathbf{q}}\nonumber \\
 & =\frac{1}{V_{\mathsf{e}}}\int\limits _{V_{\mathsf{e}}}\mathsf{d}^{3}\mathbf{r}\,\Lambda^{*}\left(\mathbf{r}\right)H_{\alpha^{\prime}\alpha}\Lambda\left(\mathbf{r}\right)\label{eq:General Hamiltonian matrix element}
\end{align}
where $H_{\alpha^{\prime}\alpha}$ are the bulk Hamiltonian matrix
elements presented in Sec.\ \ref{subsec:Bulk k.p}. Expressions for
$\mathcal{H}_{m_{f}\alpha^{\prime}\alpha}\left(\mathbf{q}^{\prime},\mathbf{q}\right)$
are fully written out in Appendix \ref{sec:QD-k.p-Hamiltonian} in
terms of the bulk Hamiltonian matrix elements and quantum dot characteristic
function.

\subsection{Including strain and piezoelectric effects\label{subsec:Including-strain-and-piezo-in-k.p}}

Deformation potentials and piezoelectric effects, which are both strain-driven,
are important for accurate calculations of electronic structure in
III-N materials. However, including deformation potentials can be
computationally costly for the case of isolated dots. The two-unit
cell approach presented in Sec.\ \ref{subsec:Quantum-dot-superlattice}
allows for the study of isolated dots, but at the cost of computationally
expensive convolutions. Additionally, another layer of convolutions
appears in the Hamiltonian matrix elements, leading to composed convolutions.
Here, we present the matrix elements due to strain and show our computationally
efficient approach of dealing with these composed convolutions by
choosing the linear dimensions of the real-space strain cell $\Omega_{\mathsf{s}}$
to be integer multiples of those of the electronic cell $\Omega_{\mathsf{e}}$.

The bulk strain Hamiltonian matrix elements in Eqs.\ \ref{eq: bulk Hamiltonian}
and \ref{eq: Bulk strain Hamiltonian} can be written as
\[
H_{\alpha^{\prime}\alpha}=\sum_{ij}f_{\alpha^{\prime}\alpha}^{ij}\epsilon_{ij}\left(\mathbf{r}\right)
\]
where the $f_{\alpha^{\prime}\alpha}^{ij}$ consist of $\mathbf{k}\cdot\mathbf{p}$
parameters ($a_{i}$, $l_{i}$, $m_{i}$ and $n_{i}$). Using the
prescription of Sec.\ \ref{subsec:Quantum dot k.p}, the strain contributions
to the quantum dot Hamiltonian are \begin{widetext}
\[
\mathcal{H}_{m_{f}\alpha^{\prime}\alpha}^{ij,\mathsf{st}}\left(\mathbf{q}^{\prime},\mathbf{q}\right)=\frac{1}{6}\sum_{l^{\prime}=0}^{5}\sum_{l=0}^{5}\mathsf{e}^{i\frac{2\pi}{6}\left\{ l\left[m_{f}-J_{z}\left(\alpha\right)\right]-l^{\prime}\left[m_{f}-J_{z}\left(\alpha^{\prime}\right)\right]\right\} }h_{\alpha^{\prime}\alpha}^{ij,\mathsf{st}}\left(\overleftrightarrow{\mathbf{R}}_{\!l^{\prime}}\mathbf{q}^{\prime},\overleftrightarrow{\mathbf{R}}_{\!l}\mathbf{q}\right)
\]
\[
\mathcal{H}_{m_{f}\alpha^{\prime}\alpha}^{ij,\mathsf{st}}\left(\mathbf{q}^{\prime},\mathbf{q}_{z}\right)=\frac{1}{\sqrt{6}}\sum_{l'=0}^{5}\mathsf{e}^{-il^{\prime}\phi\left[m_{f}-J_{z}\left(\alpha^{\prime}\right)\right]}h_{\alpha^{\prime}\alpha}^{ij,\mathsf{st}}\left(\overleftrightarrow{\mathbf{R}}_{\!l^{\prime}}\mathbf{q}^{\prime},\mathbf{q}_{z}\right)
\]
\[
\mathcal{H}_{m_{f}\alpha^{\prime}\alpha}^{ij,\mathsf{st}}\left(\mathbf{q}_{z}^{\prime},\mathbf{q}_{z}\right)=h_{\alpha^{\prime}\alpha}^{ij,\mathsf{st}}\left(\mathbf{q}_{z}^{\prime},\mathbf{q}_{z}\right)
\]
where
\begin{align}
h_{\alpha^{\prime}\alpha}^{ij,\mathsf{st}}\left(\mathbf{q}^{\prime},\mathbf{q}\right) & =\frac{\left(2\pi\right)^{3}f_{\alpha^{\prime}\alpha}^{ij,\mathsf{h}}}{V_{\mathsf{e}}}\tilde{\epsilon}_{ij}^{\mathsf{a}}\left(\mathbf{q}^{\prime}-\mathbf{q}\right)+\frac{\left(2\pi\right)^{6}\left(f_{\alpha^{\prime}\alpha}^{ij,\mathsf{d}}-f_{\alpha^{\prime}\alpha}^{ij,\mathsf{h}}\right)}{V_{\mathsf{e}}^{2}}\left(\tilde{\chi}_{\mathsf{d}}\ast\tilde{\epsilon}_{ij}^{\mathsf{a}}\right)_{\mathsf{e}}\left(\mathbf{q}^{\prime}-\mathbf{q}\right).\label{eq:Strain matrix element}
\end{align}
\end{widetext}Here, $f_{\alpha^{\prime}\alpha}^{ij,\mathsf{h}}$
and $f_{\alpha^{\prime}\alpha}^{ij,\mathsf{d}}$ are the $\mathbf{k}\cdot\mathbf{p}$
parameters for bulk host and dot materials, respectively. $\tilde{\epsilon}_{ij}^{\mathsf{a}}$
is the strain produced by the quantum dot array calculated in Sec.\ \ref{subsec:strain }.
The subscript ``e'' in $\left(\tilde{\chi}_{\mathsf{d}}\ast\tilde{\epsilon}_{ij}^{\mathsf{a}}\right)_{\mathsf{e}}$
indicates that the convolution is over the wave vectors $\mathbf{q}\in\Omega_{\mathsf{e}}^{-1}$.
Inserting the superlattice strain from Eq.\ \ref{eq:QD superlattice strain}
into the $\mathbf{k}\cdot\mathbf{p}$ strain matrix elements from
Eq.\ \ref{eq:Strain matrix element} leads to composed convolutions,
\begin{align}
\left(\tilde{\chi}_{\mathsf{d}}\ast\tilde{\epsilon}_{ij}^{\mathsf{a}}\right)_{\mathsf{e}} & \left(\mathbf{q}\right)=\sum_{\mathbf{q}^{\prime}\in\Omega_{\mathsf{e}}^{-1}}\tilde{\chi}_{\mathsf{d}}\left(\mathbf{\mathbf{q^{\prime}}}\right)\tilde{\epsilon}_{ij}^{\mathsf{a}}\left(\mathbf{q}-\mathbf{q^{\prime}}\right)\label{eq:Composed convolutions 1}\\
 & =\frac{1}{V_{\mathsf{s}}}\sum_{\mathbf{q}^{\prime}\in\Omega_{\mathsf{e}}^{-1}}\tilde{\chi}_{\mathsf{d}}\left(\mathbf{\mathbf{q^{\prime}}}\right)\sum_{\mathbf{Q}\in\Omega_{\mathsf{s}}^{-1}}\tilde{\epsilon}_{ij}\left(\mathbf{Q}\right)\tilde{\chi}_{\mathsf{e}}\left(\mathbf{q}-\mathbf{q^{\prime}}-\mathbf{Q}\right)\label{eq:Composed convolutions 2}
\end{align}
which can be computationally demanding depending on the number of
wave vectors used. The original proposal of using a large strain cell
with a smaller electronic cell imposed no relationship between their
sizes \cite{Vukmirovc2008}. Equation \ref{eq:Composed convolutions 2}
then requires evaluating $\tilde{\chi}_{\mathsf{e}}$ at points $\mathbf{q}-\mathbf{q^{\prime}}-\mathbf{Q}$,
which are contained on neither the electronic nor strain meshes, requiring
a unique convolution be calculated for every $\mathbf{q^{\prime}}$.
It is well known that using the convolution theorem to compute a convolution
between two vectors of length $N$ has a computational cost that scales
as $N\log\left(N\right)$. Similarly, the computational cost for a
convolution on a 3D $N\times N\times N$ mesh scales as $N^{3}\log\left(N\right)$.
Computing the composed convolutions in Eq.\ \ref{eq:Composed convolutions 2}
would then scale as $N_{\mathsf{e}}^{3}\log\left(N_{\mathsf{e}}\right)N_{\mathsf{s}}^{3}\log\left(N_{\mathsf{s}}\right)$
since a convolution in $\mathbf{Q}$ has to be calculated for each
individual $\mathbf{\mathbf{q}^{\prime}}$. Note that the convolutions
from Eqs.\ \ref{eq:Composed convolutions 1}-\ref{eq:Composed convolutions 2}
are linear convolutions, which implies that the arrays of function
values must be padded with zeros before using the convolution theorem
as detailed in Appendix \ref{sec:Conventions}. This zero padding
increases both $N_{\mathsf{e}}$ and $N_{\mathsf{s}}$. We show that
choosing a strain unit cell to be a supercell of the electronic unit
cell reduces the number of convolutions to compute, leading to an
improved scaling of $N_{\mathsf{e}}^{3}\log\left(N_{\mathsf{e}}\right)+N_{\mathsf{s}}^{3}\log\left(N_{\mathsf{s}}\right)$.

Choosing the strain unit cell linear dimensions to be multiples of
the electronic cell, we have
\begin{equation}
L_{i}^{\mathsf{s}}=n_{i}L_{i}^{\mathsf{e}}\qquad i=12,3\label{eq: MeshFact definition}
\end{equation}
where the $n_{i}$ take positive integer values. This choice of real-space
unit cells leads to the electronic Fourier-space mesh being contained
in the strain mesh $\mathbf{\Omega}_{\mathsf{e}}^{-1}\subset\mathbf{\Omega}_{\mathsf{s}}^{-1}$.
The wave vectors $\mathbf{Q}$ then have a spacing that is a fraction
of the spacing of the electronic wave vectors $\mathbf{q}$,
\begin{equation}
\Delta Q_{i}=\frac{\Delta q_{i}}{n_{i}}\label{eq: Overlapping q meshes}
\end{equation}
Note that from Eq.\ \ref{eq:Composed convolutions 1}, $\tilde{\epsilon}_{ij}^{\mathsf{a}}$
is only sampled at points $\Delta\mathbf{q}=\mathbf{q}-\mathbf{q^{\prime}}$,
which belong to the electronic mesh. Our procedure starts with using
the convolution theorem (see Appendix \ref{sec:Conventions}) to efficiently
calculate the inner convolution $\left(\tilde{\epsilon}_{ij}\ast\tilde{\chi}_{\mathsf{e}}\right)_{\mathsf{s}}\left(\mathbf{Q}\right)$
on the strain mesh to obtain $\tilde{\epsilon}_{ij}^{\mathsf{a}}\left(\mathbf{Q}\right)$.
Since the wave vectors $\mathbf{Q}$ also contain the wave vectors
$\mathbf{q}$, we can then extract the points that lie on the electronic
mesh to obtain $\tilde{\epsilon}_{ij}^{\mathsf{a}}\left(\mathbf{q}\right)$.
Lastly, we perform the second convolution $\left(\tilde{\chi}_{\mathsf{d}}\ast\tilde{\epsilon}_{ij}^{\mathsf{a}}\right)_{\mathsf{e}}\left(\mathbf{q}\right)$,
again utilizing the convolution theorem. This workflow is shown in
Fig.\ \ref{fig: Strain and piezo workflows}(a). In our method, we
compute only two 3D convolutions and so get a complexity scaling of
$N_{\mathsf{e}}^{3}\log\left(N_{\mathsf{e}}\right)+N_{\mathsf{s}}^{3}\log\left(N_{\mathsf{s}}\right)$,
which is a considerable improvement compared to the non-overlapping
case. Note that $N_{\mathsf{s}}$ is generally much larger than $N_{\mathsf{e}}$
to obtain appropriate convergence, so the computational cost is dominated
by the convolutions on $\Omega_{\mathsf{s}}^{-1}$.

\begin{figure}[tbh]
\includegraphics[width=8.6cm]{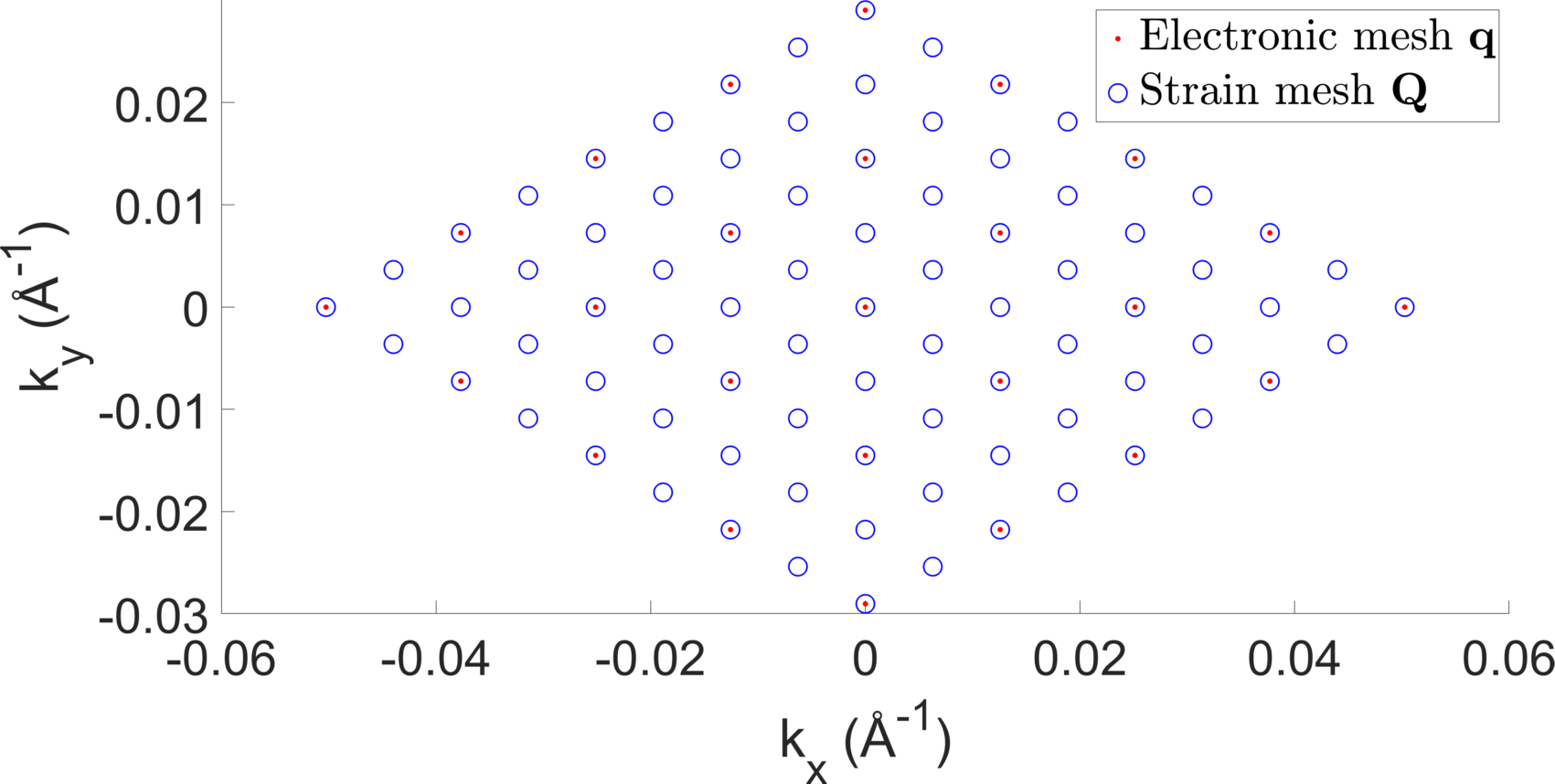}

\caption{Example of Fourier space meshes where $n_{12}=2$, leading to $\Delta Q_{i}=\frac{\Delta q_{i}}{2}$,
which implies the vectors $\mathbf{Q}$ contain all the vectors $\mathbf{q}$.}
\end{figure}

The piezoelectric potential brings no additional complexity, and the
workflow for calculating the piezoelectric potential is shown in Fig.\ \ref{fig: Strain and piezo workflows}(b).
The potential is initially calculated on the strain mesh, and the
electronic mesh portion is extracted to calculate the piezoelectric
potential contributions to the Hamiltonian, which are written out
in Appendix \ref{sec:QD-k.p-Hamiltonian}.

\begin{figure}[tbh]
(a)
\[
\begin{array}{ccccc}
 & \text{Extract}\\
 & \text{ to e-mesh}\\
\underbrace{\tilde{\epsilon}_{ij}^{\mathsf{a}}\left(\mathbf{Q}\right)} & \longrightarrow & \tilde{\epsilon}_{ij}^{\mathsf{a}}\left(\mathbf{q}\right) & \longrightarrow & \underbrace{\left(\tilde{\chi}_{\mathsf{d}}\ast\tilde{\epsilon}_{ij}^{\mathsf{a}}\right)_{\mathsf{e}}\left(\mathbf{q}\right)}\\
\text{Array strain } &  &  &  & \text{Calculate on e-mesh}\\
\text{on s-mesh}
\end{array}
\]

(b)
\[
\begin{array}{ccccc}
\underbrace{\tilde{\epsilon}^{\mathsf{arr}}\left(\mathbf{Q}\right)\longrightarrow\tilde{\varphi}\left(\mathbf{Q}\right)} & \longrightarrow & \underbrace{\tilde{\varphi}^{\mathsf{arr}}\left(\mathbf{Q}\right)=\left(\tilde{\varphi}\ast\tilde{\chi}_{e}\right)_{\mathsf{s}}\left(\mathbf{Q}\right)} & \longrightarrow & \underbrace{\tilde{\varphi}^{\mathsf{arr}}\left(\mathbf{q}\right)}\\
\text{Strain gives potential } &  & \text{Truncate to } &  & \text{Extract }\\
\text{in strain box} &  & \text{electronic unit cell} &  & \text{on e-mesh}
\end{array}
\]

\caption{(a) Workflow for calculation of composed convolutions of the form
shown in Eq.\ \ref{eq:Composed convolutions 1}. (b) Workflow used
to obtain piezoelectric potential on the electronic mesh from the
strain on the strain mesh. e-mesh and s-mesh signify the Fourier space
electronic and strain meshes, respectively. \label{fig: Strain and piezo workflows}}
\end{figure}

\section{Smooth alloy profile\label{sec:Smooth-indium-profile}}

\begin{figure}[tbh]
\includegraphics[width=8.6cm]{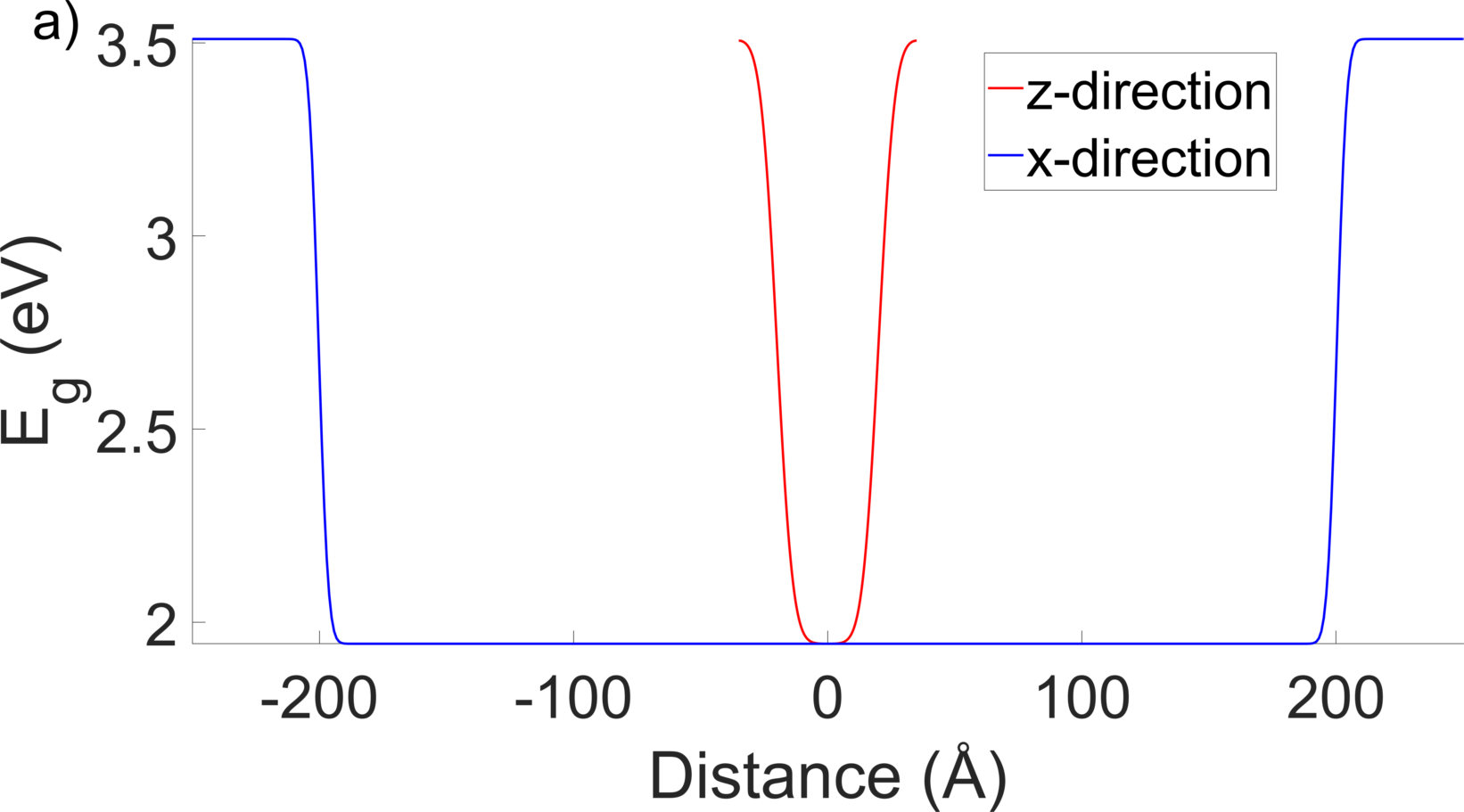}

\includegraphics[width=8.6cm]{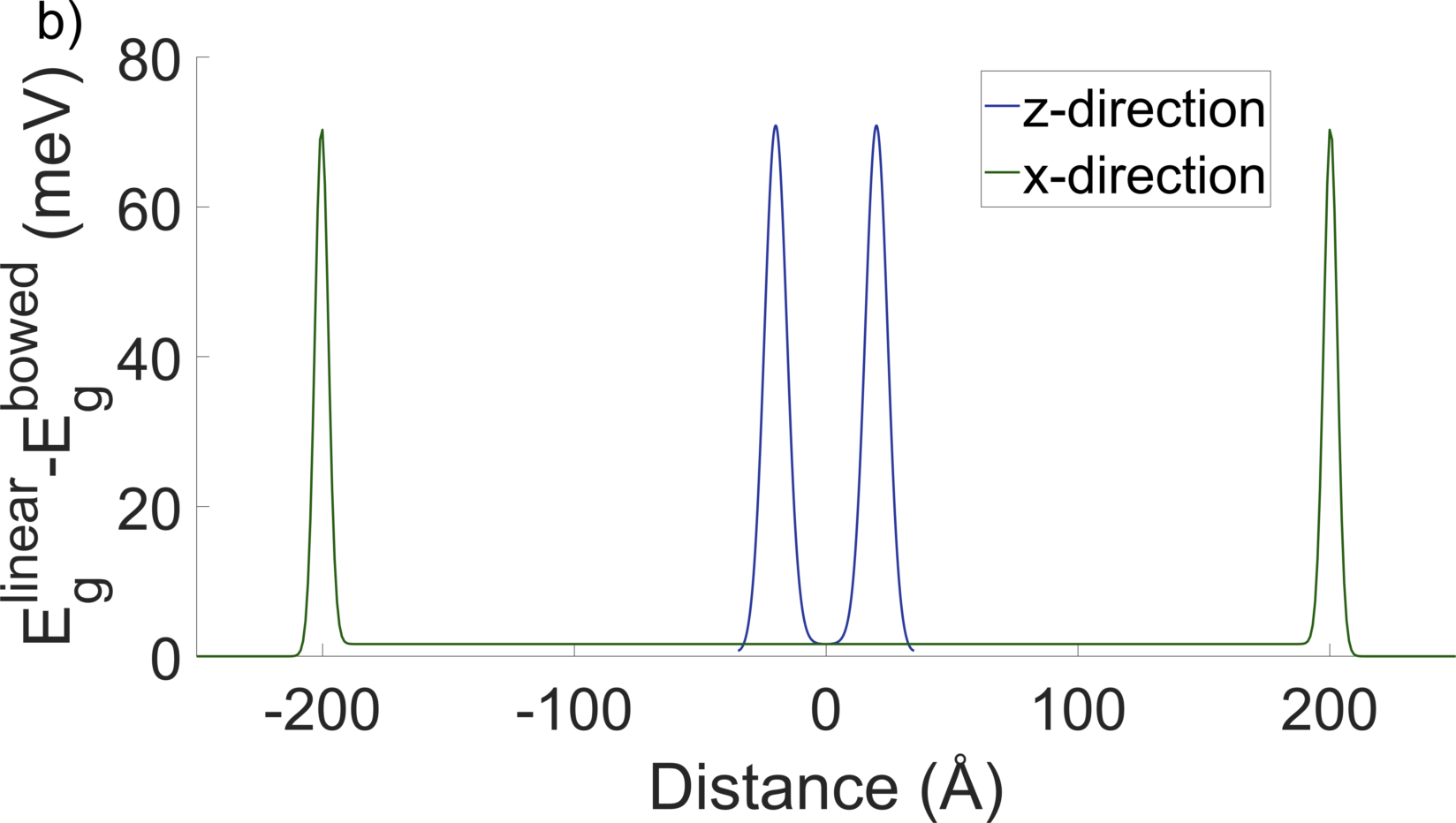}

\caption{(a) Band gap in a quantum dot system obtained by bowed interpolation
using Eq.\ \ref{eq: General smoothed bandgap} along the z and y
axes through the center of the dot. Band gap obtained from linear
interpolation using Eq.\ \ref{eq: Approx band gap} is visually indistinguishable.
(b) Difference of the bowed and linearly interpolated band gaps along
the z and x axes. Quantum dot parameters are listed in Table \ref{tab:System-parameters}.
However, for computational simplicity and smooth curves, these results
were obtained using a rectangular real-space unit cell with dimensions
$L_{\mathsf{x}}=L_{\mathsf{y}}=500\mathring{A}$ and $L_{\mathsf{z}}=70\mathring{A}$
and a smoothing of $\delta=[3,3,5]\mathring{A}$. \label{fig:(a) smoothing and bandgap}}
\end{figure}

\begin{figure}[tbh]
\includegraphics[width=8.6cm]{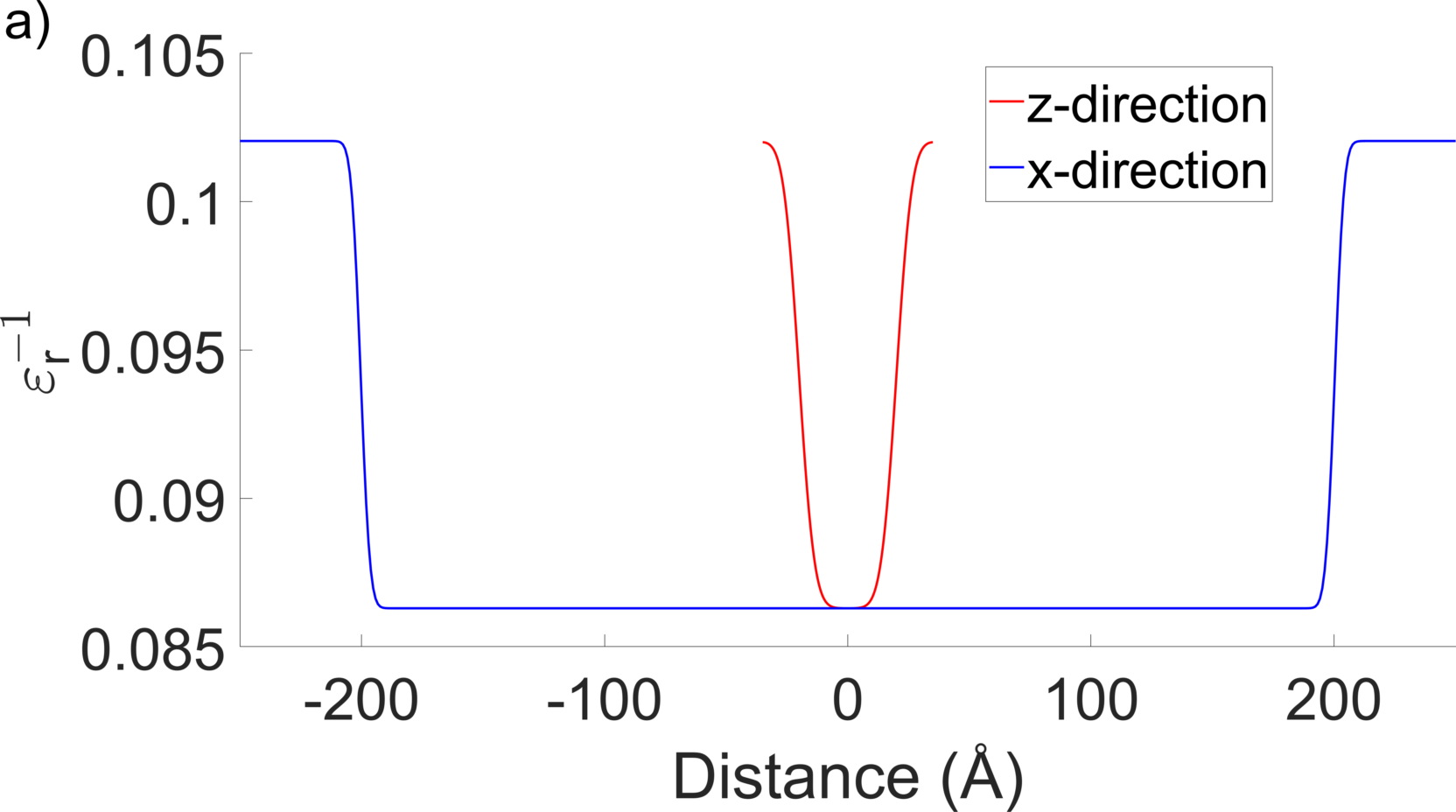}

\includegraphics[width=8.6cm]{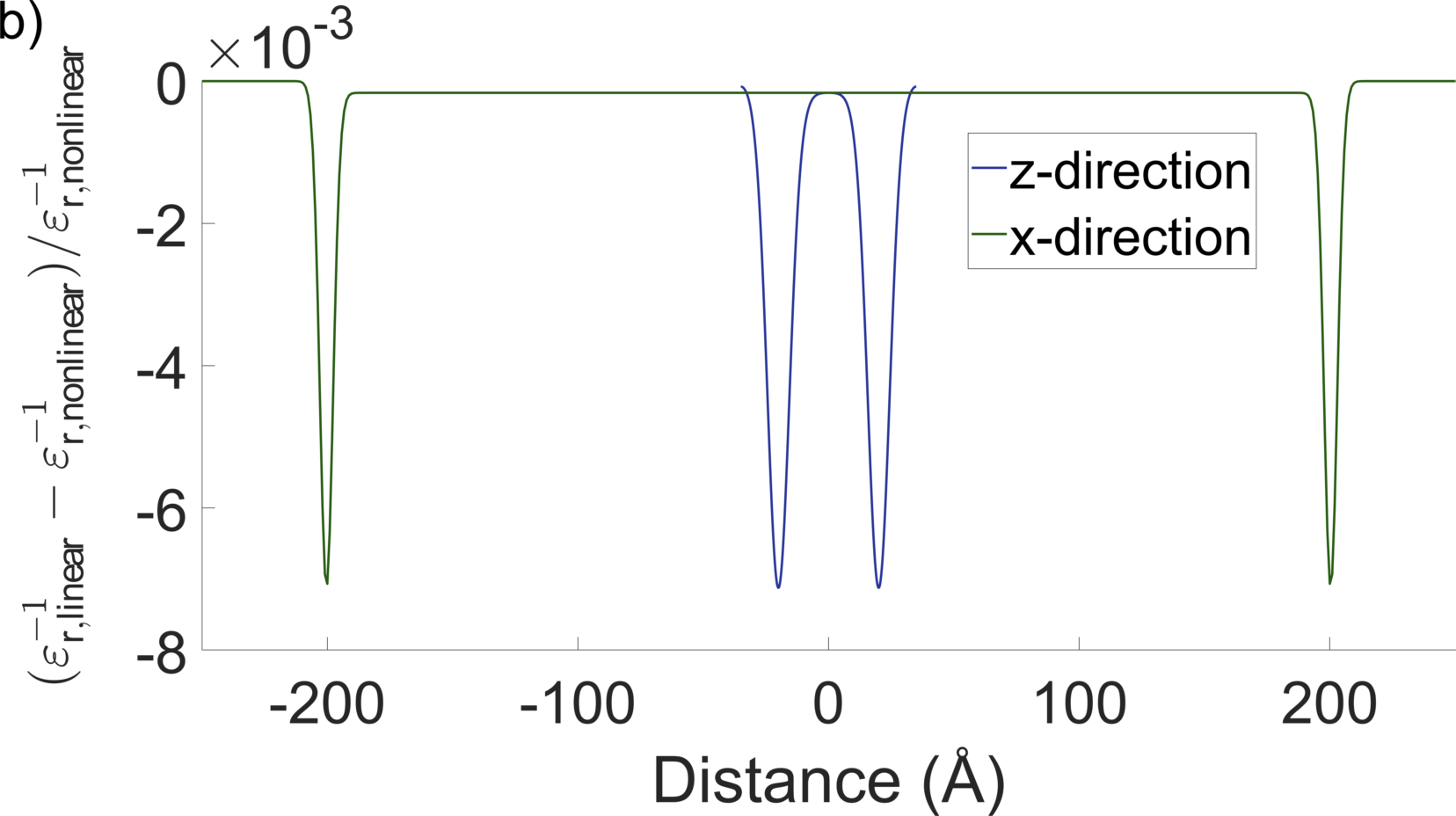}

\caption{(a) Inverse dielectric constant from Eq.\ \ref{eq: Exact smooth dielectric constant}
along the z and x axes through the center of the dot. Linearly interpolated
inverse from Eq.\ \ref{eq: Approx smoothed inverse dielectric constant}
is visually indistinguishable. (b) Relative difference of the linear
and nonlinear dielectric constants along the z and x axes. Quantum
dot parameters are listed in Table \ref{tab:System-parameters}. However,
for computational simplicity and smooth curves, these results were
obtained using a rectangular real-space unit cell with dimensions
$L_{\mathsf{x}}=L_{\mathsf{y}}=500\mathring{A}$ and $L_{\mathsf{z}}=70\mathring{A}$
and a smoothing of $\delta=[3,3,5]\mathring{A}$. \label{fig: Smoothing dielectric constant}}
\end{figure}

When InGaN devices are grown by molecular beam epitaxy (MBE), indium
diffuses between layers \cite{Nguyen2011}. While most studies of
MBE-grown materials simulate abrupt junctions, this diffusion leads
to smoothing of the material interfaces, producing a continuously
varying alloy fraction, which changes the local band properties and
lattice constant, which in turn change strain and polarization fields.
This smooth alloy profile must be included for accurate modeling.
Smooth indium profiles also provide a computational benefit, since
sharp features of the confining potentials are removed, so fewer wave
vectors are required to attain convergence. In this section, we present
a method to include alloy diffusion effects by effectively smoothing
the characteristic function of the dot. We focus on indium alloying
here for the examples, but the methods are general for all $\mathbf{k}\cdot\mathbf{p}$
calculations of alloy structures.

\subsubsection{Smoothing method}

In the case of a sharp material interface, the local alloy fraction
$X\left(\mathbf{r}\right)$ can be defined by the characteristic function
of the dot
\[
X\left(\mathbf{r}\right)=X_{0}\chi_{\mathsf{d}}\left(\mathbf{r}\right),
\]
where the characteristic function $\chi_{\mathsf{d}}\left(\mathbf{r}\right)$
defines the geometry of the dot with indium fraction $X_{0}$. By
convolving with a Gaussian $G\left(\mathbf{r},\boldsymbol{\delta}\right)=\frac{1}{\sqrt{2\pi}\delta_{x}\delta_{y}\delta_{z}}\mathsf{e}^{-\frac{1}{2}\left(\frac{x^{2}}{\delta_{x}^{2}}+\frac{y^{2}}{\delta_{y}^{2}}+\frac{z^{2}}{\delta_{z}^{2}}\right)}$
or other kernel, we can obtain a smooth version of the characteristic
function
\begin{align*}
X_{\mathsf{sm}}\left(\mathbf{r}\right) & =\left(X_{0}\chi_{\mathsf{d}}\ast G\right)\left(\mathbf{r}\right)\\
 & =X_{0}\chi_{\mathsf{sm}}\left(\mathbf{r}\right),
\end{align*}
where $\boldsymbol{\delta}=\left[\delta_{x},\delta_{y},\delta_{z}\right]$
controls the radius of smoothing and needs to be chosen to model the
desired alloy diffusion. $G\left(\mathbf{r},\boldsymbol{\delta}\right)$
is normalized to preserve the total amount of alloying element, and
$\chi_{\mathsf{sm}}\left(\mathbf{r}\right)$ is a smoothed characteristic
function. Using the convolution theorem, the smoothed characteristic
function satisfies
\begin{align*}
\tilde{\chi}_{\mathsf{sm}}\left(\mathbf{q}\right) & =\tilde{\chi}_{\mathsf{d}}\left(\mathbf{q}\right)\mathsf{e}^{-\frac{\left(\delta_{x}^{2}q_{x}^{2}+\delta_{y}^{2}q_{y}^{2}+\delta_{z}^{2}q_{z}^{2}\right)}{2}}.
\end{align*}
Note that $\chi_{\mathsf{sm}}\left(\mathbf{r}\right)$ is no longer
strictly a characteristic function, as it takes values continuously
between 0 and 1. We now show that it can be inserted in place of the
characteristic function in the previous sections to give $\mathbf{k}\cdot\mathbf{p}$
parameters, strain and piezoelectric fields accurately with a smooth
alloy profile.

\subsubsection{Material parameters}

We now focus on the case of InGaN to illustrate the interpolation
of material parameters. In the case of sharp material interfaces,
the host and dot regions each consist of uniform material. The host
material is a binary material and has well-defined parameters. The
dot region consists of alloyed InGaN, and its parameters are obtained
by either linear or bowed interpolation of bulk GaN and InN parameters,
which are listed in Appendix \ref{sec:Bulk-k.p-parameters}.

In the case of a smooth alloy profile, the dot and host regions are
no longer uniform, giving the material parameters a smooth spatial
dependence. Parameters that were linearly interpolated in the sharp
interface case can still be obtained from a simple linear interpolation
based on the local alloy fraction $X\left(\mathbf{r}\right)$. The
band gap $E_{\mathsf{g}}$ is nonlinear in the alloy fraction due
to a bowing factor. This nonlinearity prevents us from using the convolution
theorem in calculating the Hamiltonian matrix elements. However, we
show that neglecting the bowing parameters in the alloy-smoothing
region can still give computationally efficient and accurate smoothed
profiles when the alloy fraction is not too large.

The local value for any of the linearly interpolated material parameters
depends on the local alloy fraction
\begin{equation}
f\left(\mathbf{r}\right)=f^{\mathsf{B}}X\left(\mathbf{r}\right)+\left[1-X\left(\mathbf{r},X_{0}\right)\right]f^{\mathsf{A}}\label{eq: Definition of linear parameter}
\end{equation}
where $f$ can be a parameter such as lattice constant, and subscripts
A and B stand for the two binary materials, GaN and InN for example.
For this case of linearly interpolated quantities, smoothed parameters
can be written:%
\begin{equation}
f\left(\mathbf{r}\right)=f^{\mathsf{d}}\left(X_{0}\right)\chi_{\mathsf{sm}}\left(\mathbf{r}\right)+\left[1-\chi_{\mathsf{sm}}\left(\mathbf{r}\right)\right]f^{\mathsf{A}}\label{eq: smoothed linear parameters}
\end{equation}
where $f^{\mathsf{d}}\left(X_{0}\right)$ is the linearly interpolated
material parameter at the nominal alloy fraction $X_{0}$ of the quantum
dot.

Band gaps do not vary linearly with alloy fraction and are generally
well described with a bowing term, as
\begin{equation}
E_{g}\left(\mathbf{r}\right)=E_{g}^{\mathsf{B}}X\left(\mathbf{r}\right)+\left[1-X\left(\mathbf{r}\right)\right]E_{g}^{\mathsf{A}}-X\left(\mathbf{r}\right)\left[1-X\left(\mathbf{r},X_{0}\right)\right]C\label{eq: Definition of bowed parameter}
\end{equation}
where $C$ is a bowing constant. Following the same procedure as in
Eq.\ \ref{eq: smoothed linear parameters}, a smoothed version can
be written:
\begin{align}
E_{\mathsf{g}}\left(\mathbf{r}\right) & =X_{0}\chi_{\mathsf{sm}}\left(\mathbf{r}\right)E_{\mathsf{g}}^{\mathsf{B}}+\left[1-X_{0}\chi_{\mathsf{sm}}\left(\mathbf{r}\right)\right]E_{\mathsf{g}}^{\mathsf{A}}-CX_{0}\chi_{\mathsf{sm}}\left(\mathbf{r}\right)\left[1-X_{0}\chi_{\mathsf{sm}}\left(\mathbf{r}\right)\right]\nonumber \\
 & =E_{\mathsf{g}}^{\mathsf{A}}+\left[E_{\mathsf{g}}^{\mathsf{B}}-E_{\mathsf{g}}^{\mathsf{A}}\right]E_{\mathsf{g}}^{\mathsf{A}}X_{0}\chi_{\mathsf{sm}}\left(\mathbf{r}\right)-CX_{0}\chi_{\mathsf{sm}}\left(\mathbf{r}\right)\left[1-X_{0}\chi_{\mathsf{sm}}\left(\mathbf{r}\right)\right]\label{eq: General smoothed bandgap}
\end{align}
where the first two terms are the linear interpolation and the last
term is the bowing. This bowing term brings additional complexity
when performing $\mathbf{k}\cdot\mathbf{p}$ calculations due to the
nonlinearlity in $\chi_{\mathsf{sm}}\left(\mathbf{r}\right)$. We
approximate the band gap by a linear interpolation between the host
and dot band gaps,
\begin{equation}
E_{g}\left(\mathbf{r}\right)\approx E_{g}^{\mathsf{d}}\left(X_{0}\right)\chi_{\mathsf{sm}}\left(\mathbf{r}\right)+\left[1-\chi_{\mathsf{sm}}\left(\mathbf{r}\right)\right]E_{g}^{\mathsf{h}}.\label{eq: Approx band gap}
\end{equation}
Here, $E_{g}^{\mathsf{d}}$ is the bulk band gap at an alloy fraction
of $X_{0}$ and $E_{g}^{\mathsf{h}}$ is the bulk band gap of the
host material. This linear interpolation gives a good approximation
for the band gap for most regions and as well for moderate indium
fractions, as shown in Fig.\ \ref{fig:(a) smoothing and bandgap}.
The regions with largest deviation are in the same locations where
$E_{\mathsf{g}}$ changes over 1.5 eV, so we expect the slight shift
of position where each band gap value occurs to have minimal effect.
The neglect of the $\chi_{\mathsf{sm}}^{2}$ term allows the theory
to stay linear and therefore efficiently calculated with the convolution
theorem.

\subsubsection{Strain and the piezoelectric potential}

Here we show how smoothing is included in the strain and piezoelectric
potential calculations. Once calculated, those strains and piezoelectric
potentials can be included in the $\mathbf{k}\cdot\mathbf{p}$ model
exactly as shown in Sec.\ \ref{subsec:Including-strain-and-piezo-in-k.p}.

Following the derivations from Refs.\ \cite{Andreev1999,Andreev2000},
it is not obvious how smoothing is to be implemented in strain calculations
since they begin from the stress of the sharp interface dot/barrier
interface. However, Ref.\ \cite{Nenashev2018} presents an alternative
derivation for the same strain calculation indicating that $\chi\left(\mathbf{r}\right)$
in the strain expressions can be exchanged for the smoothed version
$\chi_{\mathsf{sm}}\left(\mathbf{r}\right)$ without any further changes.

For the piezoelectric potential, Eq.\ \ref{eq: inverse dielectric constant}
for the spatially varying inverse dielectric constant assumed sharp
interfaces. In the case of a smooth indium profile, we use Eq.\ \ref{eq: Definition of linear parameter}
to write
\begin{equation}
\varepsilon\left(\mathbf{r}\right)=\varepsilon^{\mathsf{B}}X\left(\mathbf{r}\right)+\left[1-X\left(\mathbf{r}\right)\right]\varepsilon^{\mathsf{A}}\label{eq: Exact smooth dielectric constant}
\end{equation}

In the scenario where $X\left(\mathbf{r}\right)$ is spatially varying,
Eq.\ \ref{eq: inverse dielectric constant} can no longer be applied,
because the inverse of the dielectric constant is not a linear function
of indium. However, similar to the band gap, we find that
\begin{equation}
\frac{1}{\varepsilon\left(\mathbf{r}\right)}\approx\frac{1}{\varepsilon^{\mathsf{d}}\left(X_{0}\right)}\chi_{\mathsf{sm}}\left(\mathbf{r}\right)+\left[1-\chi_{\mathsf{sm}}\left(\mathbf{r}\right)\right]\frac{1}{\varepsilon^{\mathsf{A}}}\label{eq: Approx smoothed inverse dielectric constant}
\end{equation}
still gives an accurate representation of $\varepsilon^{-1}\left(\mathbf{r}\right)$.
Figure \ref{fig: Smoothing dielectric constant} shows a disagreement
of less than 1 percent between the inverse dielectric from Eq.\ \ref{eq: Exact smooth dielectric constant}
and the linear interpolation in Eq.\ \ref{eq: Approx smoothed inverse dielectric constant}.
The form of Eq.\ \ref{eq: Approx smoothed inverse dielectric constant}
allows us to use Eqs.\ \ref{eq: piezoelectric potential 2}-\ref{eq:delta piezo}
for the piezoelectric potential with a simple substitution of $\chi\left(\mathbf{r}\right)$
by $\chi_{\mathsf{sm}}\left(\mathbf{r}\right)$.

\begin{figure}[tbh]
\includegraphics[width=8.6cm]{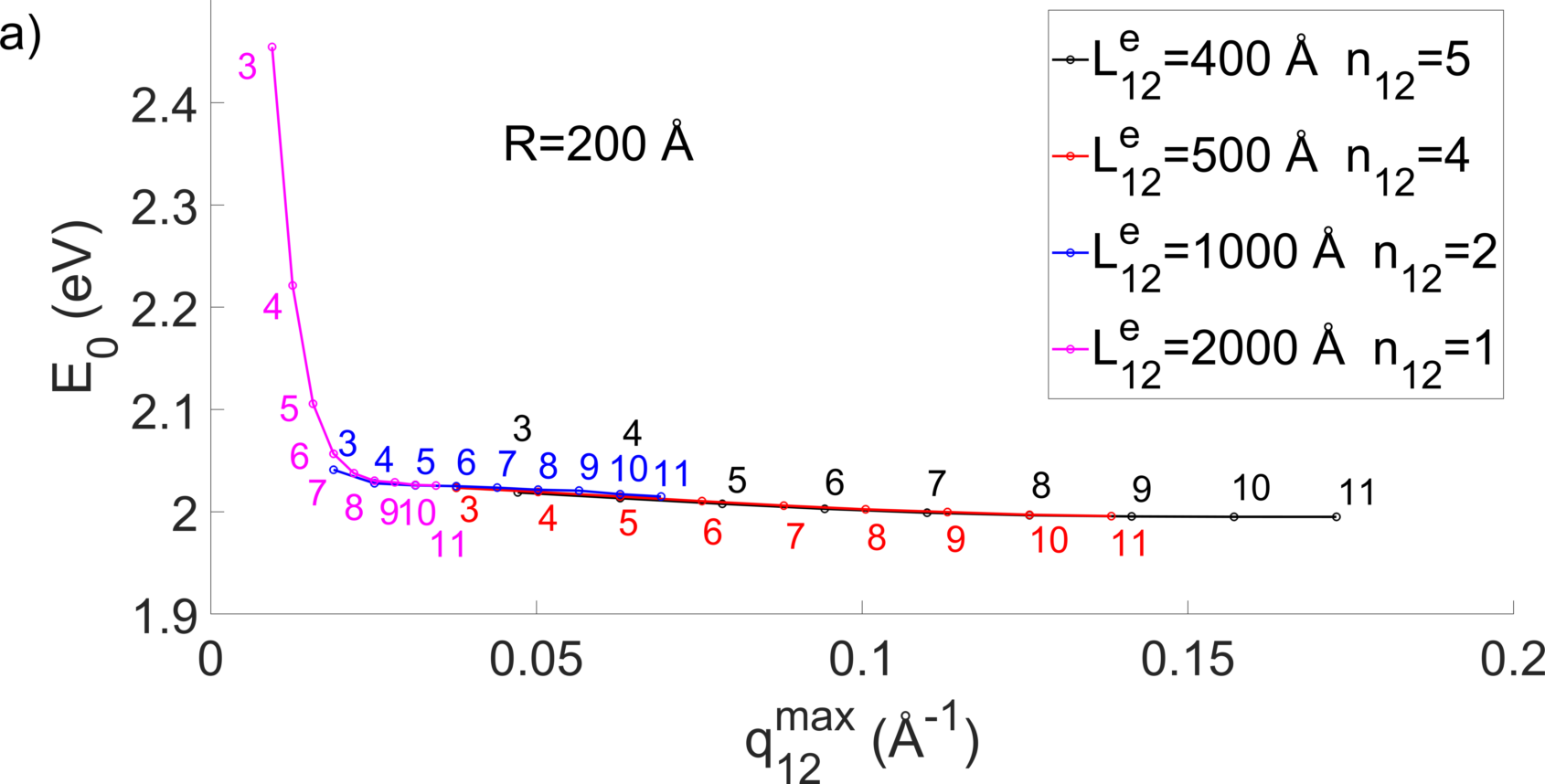}

\includegraphics[width=8.6cm]{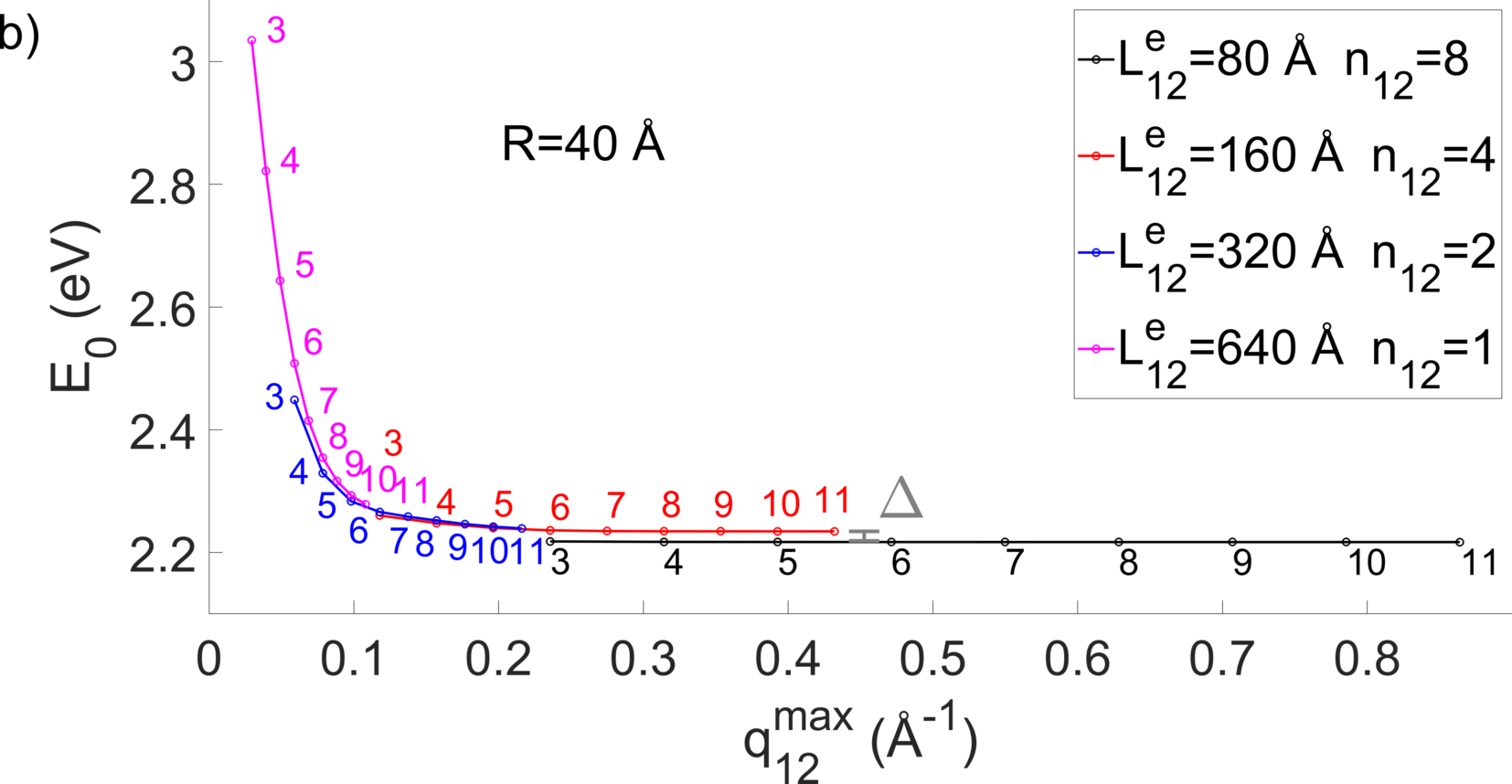}

\caption{Convergence of the fundamental gap of the quantum dot $E_{\mathsf{0}}$
for a 1D array of quantum dots with the maximum magnitude of $q_{12}$
included in the $\mathbf{k}\cdot\mathbf{p}$ calculations. $L_{12}^{\mathsf{e}}$
and $n_{12}$ vary while (a) has $R=200\mathring{A}$ with strain
box size held constant at $L_{12}^{\mathsf{s}}=10R$ and (b) has $R=40\mathring{A}$
with $L_{12}^{\mathsf{s}}=16R$. Plane wave sampling $m_{12}$ from
3 to 11 are shown for each choice of $L_{12}^{\mathsf{e}}$; the number
next to each point indicates $m_{12}$. Different values of $m_{12}$,
$L_{12}^{\mathsf{e}}$ that produce the same $q_{12}^{\mathsf{max}}$
can be seen to produce approximately the same $E_{\mathsf{0}}$, showing
that $q_{12}^{\mathsf{max}}$ is a useful metric for convergence of
these states. Since $q_{12}^{\mathsf{max}}=m_{12}\pi/L_{12}^{\mathsf{e}}$
and computational cost scales with $m_{12}$, smaller $L_{12}^{\mathsf{e}}$
allows easier access to large $q_{12}^{\mathsf{max}}$. In both panels,
the black curves have $L_{12}^{\mathsf{e}}=2R$, so the dots touch
each other at the 6 edges of the hexagonal unit cell. For larger dot,
there is no visible deviation of $E_{\mathsf{0}}$ from the trend
with separated dots. With the smaller dot, tunneling of the wavefunctions
into neighboring dots causes $E_{\mathsf{0}}$ to have a significant
change, labeled $\Delta$. \label{fig: kmax study}}
\end{figure}

\begin{figure}[tbh]
\includegraphics[width=8.6cm]{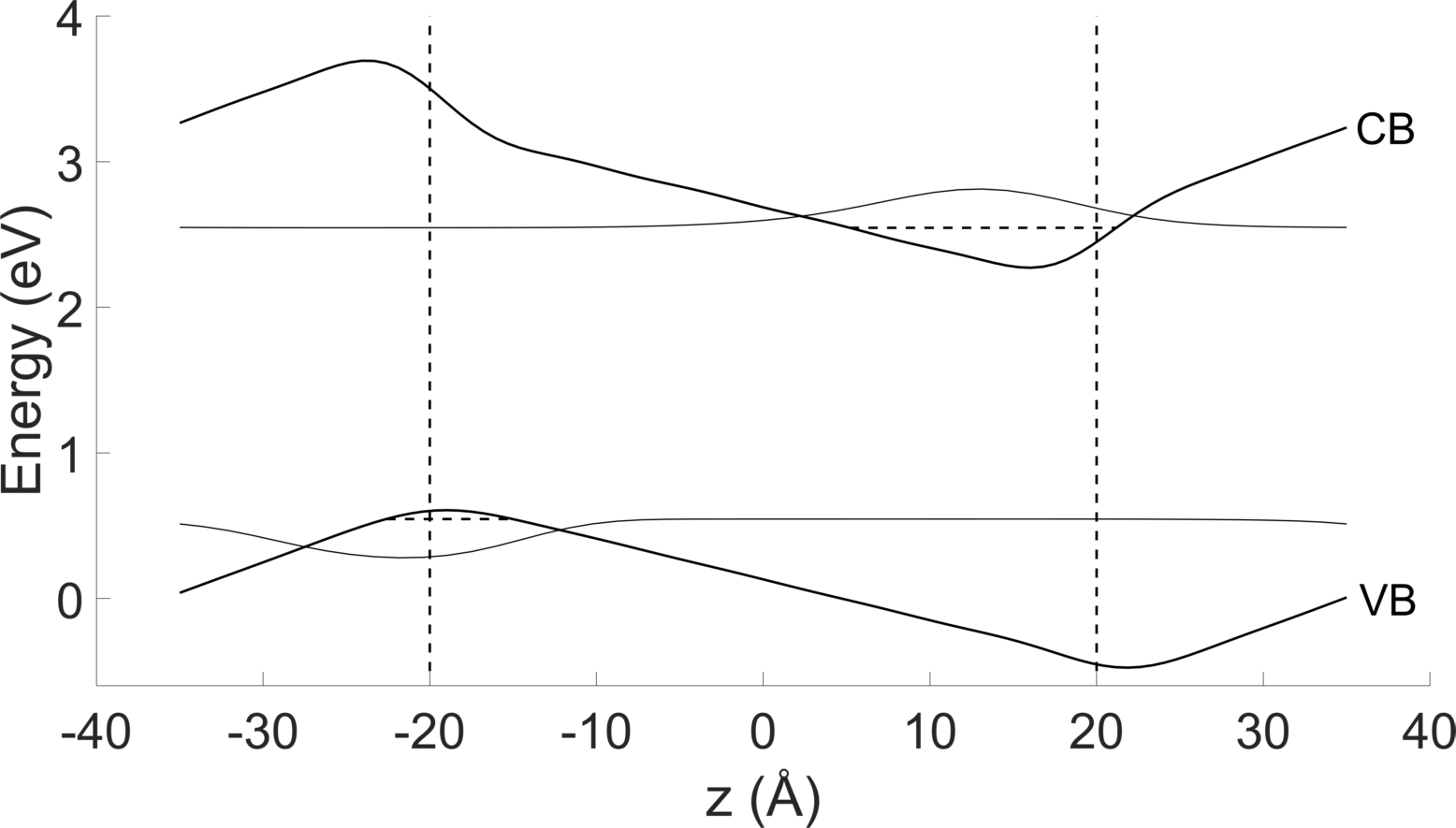}

\caption{Electronic structure of the QD superlattice system along the central
axis of the dot for $\boldsymbol{\delta}=[1.5,1.5,2.5]\mathring{A}$,
corresponding to $s=1$ in Fig.\ \ref{fig: Smoothing and edge states}(a).
Other system parameters are in Table \ref{tab:System-parameters}.
Lowest bound electron and hole states energies are shown by the horizontal
dashed lines. Thick solid lines are the bulk band edges under the
influence of the piezoelectric field and strain. Thin solid lines
are z-axis projections of the probability distributions obtained from
the envelope functions. Dashed vertical lines are the nominal material
interfaces before smoothing. These calculations include spatially
varying elastic and dielectric constants. \label{fig: electronic structure}}
\end{figure}

\begin{table}[h]
\caption{Energy shifts due to spatially varying $\lambda\left(\mathbf{r}\right)$
and $\varepsilon\left(\mathbf{r}\right)$ relative to the case with
uniform constants of the host material $\lambda^{\mathsf{GaN}}$ and
$\varepsilon^{\mathsf{GaN}}$, respectively. System parameters in
Table \ref{tab:System-parameters}. \label{tab: Energy shifts}}

\begin{tabular}{cccc}
 &  &  & \tabularnewline
\hline 
\hline 
Energy shifts & $\quad\lambda\left(\mathbf{r}\right)$ \& $\varepsilon^{\mathsf{GaN}}$ & $\quad\lambda^{\mathsf{GaN}}$ \& $\varepsilon\left(\mathbf{r}\right)$ & $\quad\lambda\left(\mathbf{r}\right)$ \& $\varepsilon\left(\mathbf{r}\right)$\tabularnewline
\hline 
$\Delta E_{\mathsf{c}}$ (meV) & 16.7 & 46.7 & 64.7\tabularnewline
$\Delta E_{\mathsf{v}}$ (meV) & -5.1 & -30.6 & -37.4\tabularnewline
$\Delta E_{\mathsf{0}}$ (meV) & 21.7 & 77.4 & 102.1\tabularnewline
\hline 
\end{tabular}
\end{table}

\begin{figure}[tbh]
\includegraphics[width=8.6cm]{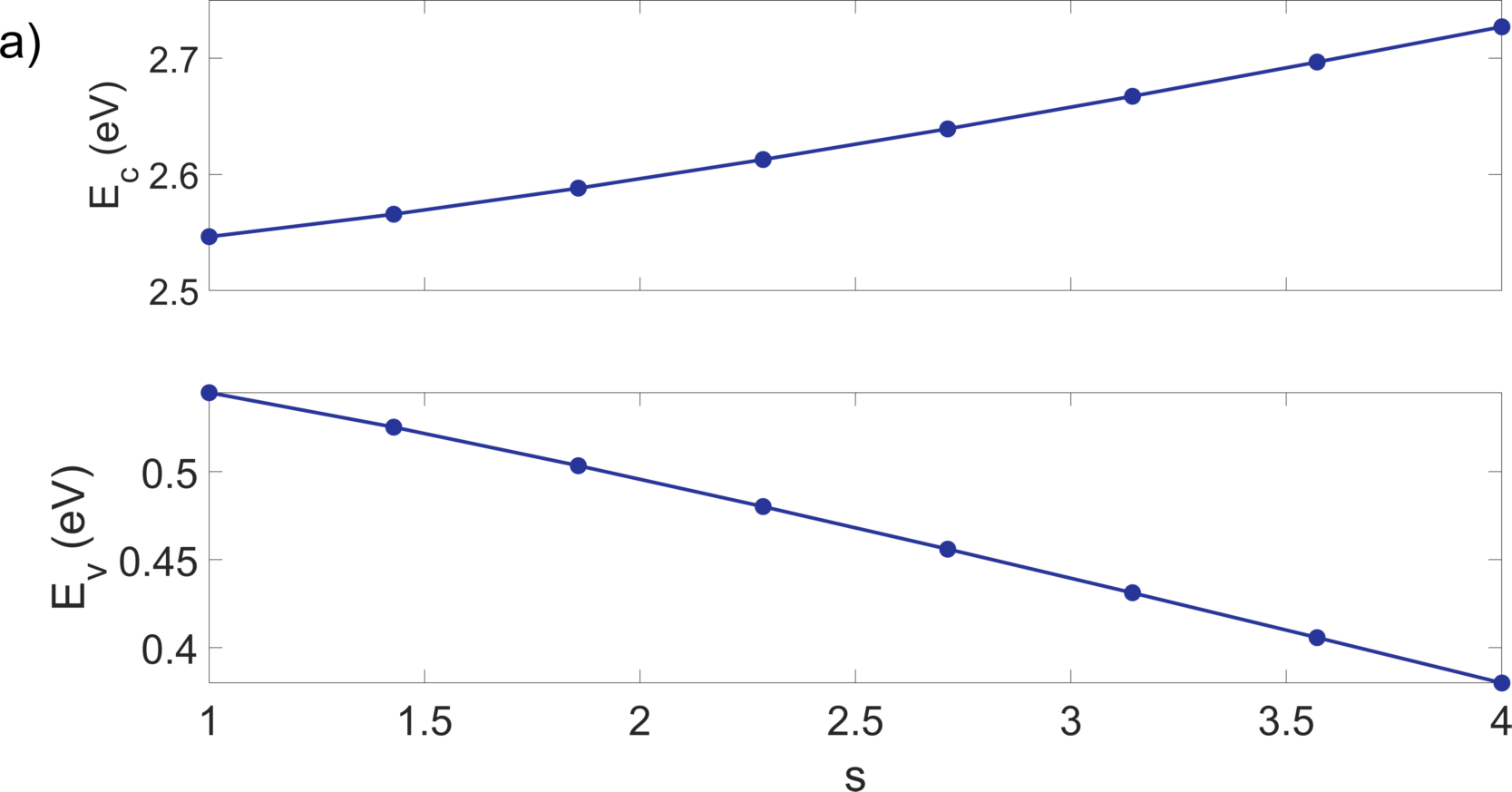}

\includegraphics[width=8.6cm]{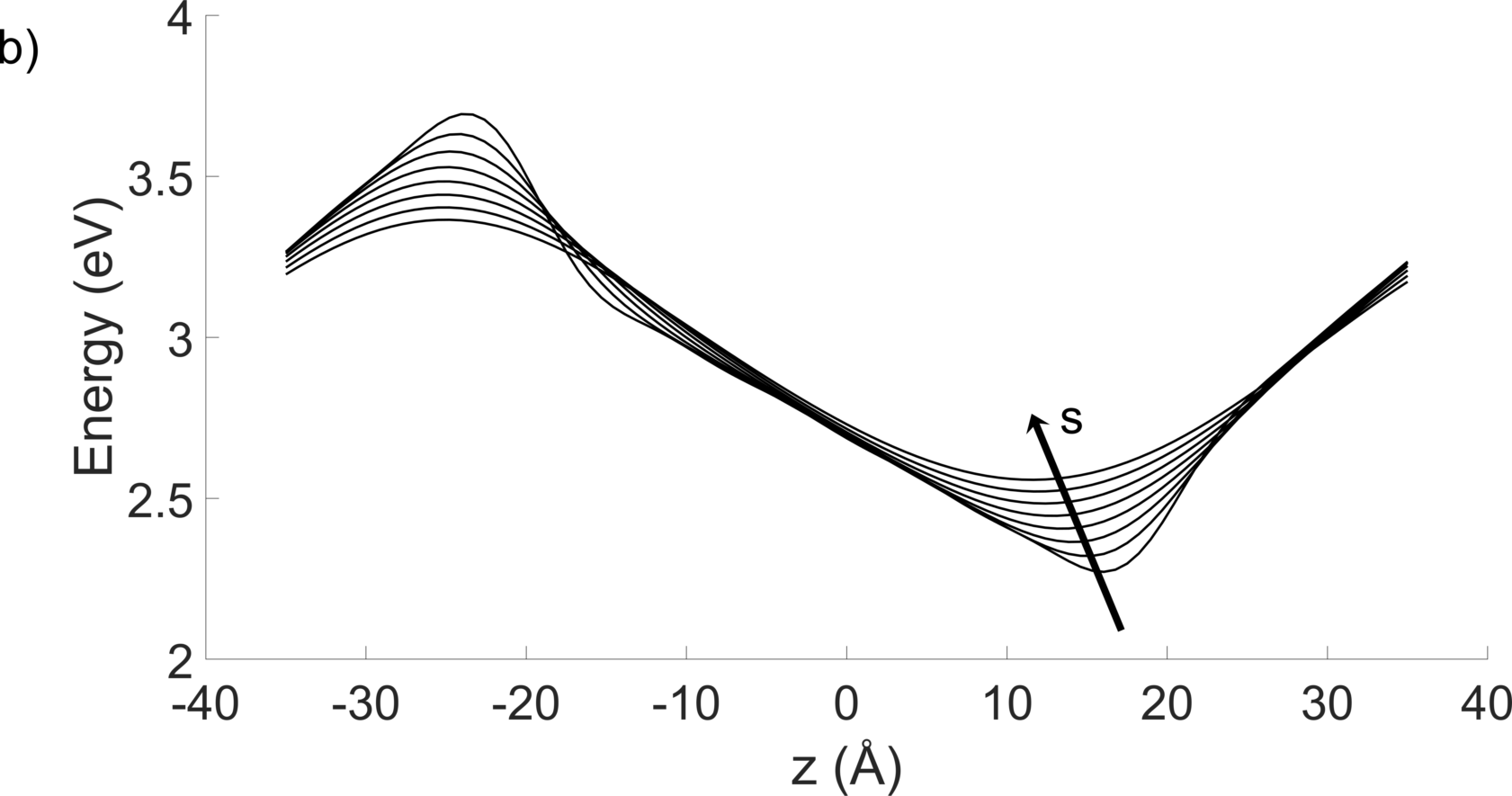}

\caption{Effects of indium diffusion, given by $\boldsymbol{\delta}=s[1.5,1.5,2.5]\mathring{A}$,
for (a) $E_{\mathsf{c}}$ and $E_{\mathsf{v}}$ and (b) bulk conduction
band edge. Remaining parameters are as listed in Table \ref{tab:System-parameters}.
Increase in indium diffusion leads to less confinement, which pushes
the two states apart in energy, widening the electronic gap $E_{\mathsf{0}}$.
Electronic structure for $s=1$ is shown in Fig.\ \ref{fig: electronic structure}.
\label{fig: Smoothing and edge states}}
\end{figure}

\section{Impacts of corrections\label{sec:Energy-shifts-from-corrections}}

In this section, we apply our methodology to study the case of a 1D
array of quantum dots, such as described in Ref.~\cite{Nguyen2011},
though we do not consider the boundaries of the nanowire. We achieve
this 1D array by taking $n_{3}=1$ to fully couple the dots in the
z-direction and $n_{12}>1$ to avoid strain effects from neighboring
dots in the xy-plane. In this section, we investigate convergence
of the lowest electron and hole state energies $E_{\mathsf{c}}$ and
$E_{\mathsf{v}}$, which define the fundamental gap of the dot $E_{\mathsf{0}}=E_{\mathsf{c}}-E_{\mathsf{v}}$.
More specifically, we show that the largest wave vector sampled plays
a dominant role in convergence. We also show the energy shifts experienced
by these two states when using uniform or spatially varying material
parameters and when including alloy smoothing.

We model an infinite 1D quantum dot array with parameters listed in
Table \ref{tab:System-parameters}. The 1D dot array has an experimentally
well characterized dot-dot spacing in z, which fixes $L_{3}^{\mathsf{e}}=L_{3}^{\mathsf{s}}=L_{3}$,
leaving $L_{12}$ and $n_{12}$ to be fixed. These quantum dots have
a rather large radius, so the smallest spatial feature that we need
to resolve is the decay of the bound wavefunctions into the classically
forbidden region. Given that bound wavefunctions decay faster than
strain, we need wave vectors that are relatively large to be able
to resolve the wavefunctions. Increasing $m_{12}$ increases the maximum
wave vector contained in the mesh, but we can also sample at larger
wave vectors by using a smaller $L_{12}^{\mathsf{e}}$. However, if
the electronic cell is chosen too small, then there can be electronic
wavefunction overlap between states of neighboring dots. We must then
choose $L_{12}^{\mathsf{e}}$ as small as possible while also avoiding
dot-dot interactions. As for strain, in order to study a 1D array,
we must choose $n_{12}$ sufficiently large to have $L_{12}^{\mathsf{s}}=n_{12}L_{12}^{\mathsf{e}}$
large enough that the strain of the quantum dot superlattice does
not extend across neighboring strain unit cells in the xy-plane.

With this intuition, we turn to the convergence of $E_{0}$ in terms
of $m_{12}$, $L_{12}^{e}$ and $n_{12}$. Figure \ref{fig: kmax study}(a)
shows the importance of the largest $\mathbf{q}$ in the electronic
mesh, $q_{12}^{\mathsf{max}}$, for convergence of $E_{0}$. In this
study, $L_{12}^{\mathsf{e}}$ and $n_{12}$ are chosen to keep a constant
$L_{12}^{\mathsf{s}}=L_{12}^{\mathsf{e}}n_{12}=2000\mathring{A}$.
We observe that $E_{0}$ is to good approximation a function of only
$q_{12}^{\mathsf{max}}$ and not of $m_{12}$ and $L_{12}^{\mathsf{e}}$,
converging towards the same value for all choices of $L_{12}^{\mathsf{e}}$.
We also observe that the smallest $L_{12}^{\mathsf{e}}$ with highest
$m_{12}$ gives the most converged $E_{0}$, since $q_{12}^{\mathsf{max}}=m_{12}\pi/L_{12}^{\mathsf{e}}$.
The black line in Fig.\ \ref{fig: kmax study}(a) represents the
case of dots touching in the xy plane and, interestingly, does not
break the convergence trend. However, we do find a break in the convergence
trend for smaller dots in Fig.\ \ref{fig: kmax study}(b). This difference
in convergence is due to the larger quantum dots having better confined
states compared to the smaller dots. Smaller dots have wavefunctions
that extend further outside the dot region, which makes them more
able to tunnel to a neighboring dot. Consequently, care has to be
taken in choosing the unit cell dimensions for small quantum dots.

The lowest quantum dot confined electron and hole energies, $E_{\mathsf{c}}$
and $E_{\mathsf{v}}$, have respectively been converged within $5$
meV by choosing $m_{12}$, $m_{3}$ and $n_{12}$ sufficiently large,
see Table \ref{tab:System-parameters}. Material parameters are listed
in Appendix \ref{sec:Bulk-k.p-parameters}. Band edges and lowest-energy
confined states are shown in Fig.\ \ref{fig: electronic structure}.
The thick black solid lines represent the bulk band edges modified
by the piezoelectric potential and strain. To include strain effects
in the bulk band edges, we have used the $(1,1)$ matrix element from
Eq.\ \ref{eq: Bulk strain Hamiltonian} to modify the conduction
band edge and a third of the trace of the $3\times3$ valence band
block for the valence band edge.

The modifications in both the strain and piezoelectric potential due
to spatially varying elastic and dielectric constants also have effects
on the electronic structure. Table \ref{tab: Energy shifts} shows
how much $E_{\mathsf{v}}$ and $E_{\mathsf{c}}$ shift due to the
corrections. We find that both corrections push the states apart,
leading to an energy gap 100 meV larger than from simpler calculations
with uniform $\varepsilon$ and $\lambda$, a significant change that
shows the importance of accurate modeling of dielectric and elastic
parameters.

Figure \ref{fig: Smoothing and edge states}(a) shows that indium
diffusion pushes the lowest electron and hole states apart, which
is due to changes in the confining potentials. From Fig.\ \ref{fig: Smoothing and edge states}(b),
we see that indium diffusion reduces the depth of the confining potential.
We have observed similar behavior for the hole state confining potential,
leading to $E_{\mathsf{v}}$ being pushed down in energy. Consequently,
the gap $E_{0}$ increases in energy as indium diffusion is increased.

In the case of sharp material interfaces, large wave vectors are needed
to resolve the discontinuous parameter profiles. Smoothing removes
the sharp interfaces and yields smoothly varying material parameters.
Consequently, the required $q_{12}^{\mathsf{max}}$ for the same degree
of convergence is smaller, which means that smaller $m_{12}$ and
$m_{3}$ and therefore reduced computational cost are needed with
increasing $\boldsymbol{\delta}$. In Figs.\ \ref{fig: Smoothing and edge states}
(a) and (b), we converged for the case of $s=1$, guaranteeing convergence
for the rest of the sweep.

\section{Conclusions}

We have demonstrated techniques for and results of four modifications
of standard quantum dot $\mathbf{k}\cdot\mathbf{p}$ theory. We have
included spatially varying elastic and dielectric constants as alloy
fraction changes in strain and piezoelectric potential calculations.
The effects of the spatially varying parameters are non-negligible
on the strain and piezoelectric potential and also produce important
shifts of the lowest electron and hole states, significantly changing
the calculated gap of the quantum dot. We have also presented a method
to include smoothly varying alloy profiles in Fourier-based strain,
piezoelectric potential and $\mathbf{k}\cdot\mathbf{p}$ calculations.
This smoothing has to be chosen to represent the device of interest,
such as for indium diffusion in InGaN systems. For the case of $\mathbf{k}\cdot\mathbf{p}$
theory for isolated dots, we have presented a new methodology of overlapping
electronic and strain meshes to facilitate the coupling of strain
into the $\mathbf{k}\cdot\mathbf{p}$ Hamiltonian, greatly reducing
the computational cost of calculating the Hamiltonian matrix elements.
Lastly, we have shown that the maximum wave vector contained in the
electronic sampling mesh is the most important criterion for determining
convergence of quantum dot levels.

\appendix
\begin{acknowledgments}
We acknowledge useful conversation with Stanko Tomi\'{c} about Fourier-space
$\mathbf{k}\cdot\mathbf{p}$ methods. We acknowledge funding from
the Ontario Early Researcher Award and NSERC CREATE TOP-SET program,
Award number 497981.
\end{acknowledgments}

\section{Bulk $\mathbf{k}\cdot\mathbf{p}$ parameters \label{sec:Bulk-k.p-parameters}}

From Ref.\ \cite{Winkelnkemper}, the $\mathbf{k}\cdot\mathbf{p}$
parameters $A_{i}^{\prime}$ are related to the Kane parameters $P_{i}$
as
\[
\begin{aligned}A_{1}^{\prime} & =\frac{\hbar^{2}}{2m_{\mathsf{e}}^{\parallel}}-\frac{P_{1}^{2}}{E_{\mathsf{g}}}\\
A_{2}^{\prime} & =\frac{\hbar^{2}}{2m_{\mathsf{e}}^{\perp}}-\frac{P_{2}^{2}}{E_{\mathsf{g}}}
\end{aligned}
\]

where
\[
\begin{aligned}P_{1}^{2} & =\frac{\hbar^{2}}{2m_{0}}\left(\frac{m_{0}}{m_{\mathsf{e}}^{\parallel}}-1\right)\frac{3E_{\mathsf{g}}\left(\Delta_{\mathsf{so}}+E_{\mathsf{g}}\right)+\Delta_{\mathsf{cr}}\left(2\Delta_{\mathsf{so}}+3E_{\mathsf{g}}\right)}{2\Delta_{\mathsf{so}}+3E_{\mathsf{g}}}\\
P_{2}^{2} & =\frac{\hbar^{2}}{2m_{0}}\left(\frac{m_{0}}{m_{\mathsf{e}}^{\perp}}-1\right)\frac{E_{\mathsf{g}}\left[3E_{\mathsf{g}}\left(\Delta_{\mathsf{so}}+E_{\mathsf{g}}\right)\right]+\Delta_{\mathsf{cr}}\left(2\Delta_{\mathsf{so}}+3E_{\mathsf{g}}\right)}{\Delta_{\mathsf{cr}}\Delta_{\mathsf{so}}+3\Delta_{\mathsf{cr}}E_{\mathsf{g}}+2\Delta_{\mathsf{so}}E_{\mathsf{g}}+E_{\mathsf{g}}^{2}}.
\end{aligned}
\]
Here, $m_{\mathsf{e}}^{\parallel}$ and $m_{\mathsf{e}}^{\perp}$
are the electron effective masses along the z-axis and in the xy-plane,
respectively. $\Delta_{\mathsf{cr}}$ and $\Delta_{\mathsf{so}}$
are the crystal field splitting and spin-orbit coupling, respectively.
The Luttinger-like parameters $L_{i}^{\prime}$, $M_{i}$ and $N_{i}^{\prime}$
are related to the $A_{i}$ parameters by
\[
\begin{aligned}L_{1}^{\prime} & =\frac{\hbar^{2}}{2m_{0}}\left(A_{2}+A_{4}+A_{5}\right)+\frac{P_{2}^{2}}{E_{g}}\\
L_{2}^{\prime} & =\frac{\hbar^{2}}{2m_{0}}A_{1}+\frac{P_{1}^{2}}{E_{g}}\\
M_{1} & =\frac{\hbar^{2}}{2m_{0}}\left(A_{2}+A_{4}-A_{5}\right)\\
M_{2} & =\frac{\hbar^{2}}{2m_{0}}\left(A_{1}+A_{3}\right)\\
M_{3} & =\frac{\hbar^{2}}{2m_{0}}A_{2}\\
N_{1}^{\prime} & =\frac{\hbar^{2}}{2m_{0}}2A_{5}+\frac{P_{2}^{2}}{E_{g}}\\
N_{2}^{\prime} & =\frac{\hbar^{2}}{2m_{0}}\sqrt{2}A_{6}+\frac{P_{1}P_{2}}{E_{g}}
\end{aligned}
\]

Note there is an error in the relations for $L_{1}^{\prime}$, $L_{2}^{\prime}$
and $N_{1}^{\prime}$ in Ref.\ \cite{Winkelnkemper}, which we have
corrected in agreement with Appendix B of Ref.\ \cite{Berkowicz2000}.
The parameters $A_{i}$, $P_{i}$ and $E_{g}$ used in our numerical
study of InGaN systems are given in Table \ref{tab:Material-parameters}.

Similarly to the $\mathbf{k}\cdot\mathbf{p}$ parameters $L_{i}^{\prime}$,
$M_{i}$ and $N_{i}^{\prime}$, the strain parameters are
\[
\begin{aligned}l_{1} & =D_{2}+D_{4}+D_{5}\\
l_{2} & =D_{1}\\
m_{1} & =D_{2}+D_{4}-D_{5}\\
m_{2} & =D_{1}+D_{3}\\
m_{2} & =D_{1}+D_{3}\\
m_{3} & =D_{2}\\
n_{1} & =2D_{5}\\
n_{2} & =\sqrt{2}D_{6}
\end{aligned}
\]

where the deformation potentials $D_{i}$ are listed in Table \ref{tab:Material-parameters}.

\begin{table}[tbh]
\caption{Material parameters used to model the InGaN system. Parameters were
taken from Ref.~\cite{Winkelnkemper}. \label{tab:Material-parameters}}

\begin{center}

\begin{tabular}{ccc}
 &  & \tabularnewline
\hline 
\hline 
Parameters & GaN & InN\tabularnewline
\hline 
$a$ ($\mathsf{\mathring{A}}$) & 3.189 & 3.545\tabularnewline
$c$ ($\mathsf{\mathring{A}}$) & 5.185 & 5.703\tabularnewline
$C_{11}$ (GPa) & 390 & 223\tabularnewline
$C_{12}$ (GPa) & 145 & 115\tabularnewline
$C_{13}$ (GPa) & 106 & 92\tabularnewline
$C_{33}$ (GPa) & 398 & 224\tabularnewline
$C_{44}$ (GPa) & 105 & 48\tabularnewline
$e_{15}$ ($\mathsf{C/m^{2}}$) & 0.326 & 0.264\tabularnewline
$e_{31}$ ($\mathsf{C/m^{2}}$) & -0.527 & -0.484\tabularnewline
$e_{33}$ ($\mathsf{C/m^{2}}$) & 0.895 & 1.06\tabularnewline
$P_{\mathsf{sp}}$ ($\mathsf{C/m^{2}}$) & -0.034 & -0.042\tabularnewline
$\varepsilon_{\mathsf{r}}$ & 9.8 & 13.8\tabularnewline
$E_{\mathsf{g}}$ (eV) & 3.51 & 0.78\tabularnewline
$E_{\mathsf{v}}$ (eV) & 0 & 0.5\tabularnewline
$\Delta_{\mathsf{cr}}$ (eV) & 0.010 & 0.040\tabularnewline
$\Delta_{\mathsf{so}}$ (eV) & 0.017 & 0.005\tabularnewline
$m_{\parallel}/m_{0}$ & 0.20 & 0.07\tabularnewline
$m_{\perp}/m_{0}$ & 0.20 & 0.07\tabularnewline
$A_{1}$ & -7.21 & -8.21\tabularnewline
$A_{2}$ & -0.44 & -0.68\tabularnewline
$A_{3}$ & 6.68 & 7.57\tabularnewline
$A_{4}$ & -3.46 & -5.23\tabularnewline
$A_{5}$ & -3.40 & -5.11\tabularnewline
$A_{6}$ & -4.90 & -5.96\tabularnewline
$a_{1}$ (eV) & -4.9 & -3.5\tabularnewline
$a_{2}$ (eV) & -11.3 & -3.5\tabularnewline
$D_{1}$ (eV) & -3.7 & -3.7\tabularnewline
$D_{2}$ (eV) & 4.5 & 4.5\tabularnewline
$D_{3}$ (eV) & 8.2 & 8.2\tabularnewline
$D_{4}$ (eV) & -4.1 & -4.1\tabularnewline
$D_{5}$ (eV) & -4.0 & -4.0\tabularnewline
$D_{6}$ (eV) & -5.5 & -5.5\tabularnewline
\hline 
\end{tabular}

\end{center}
\end{table}

\section{Characteristic functions\label{sec:Characteristic-functions}}

The characteristic function of a single dot is unity inside the dot
and zero outside,
\[
\chi_{\mathsf{d}}\left(\mathbf{r}\right)=\begin{cases}
1 & \mathbf{r}\in\Omega_{\mathsf{d}}\\
0 & \text{otherwise}
\end{cases}
\]
where $\Omega_{\mathsf{d}}$ is the space inside the dot. For a cylindrical
dot centered on the origin with radius $R$ and height $h$ along
the z-axis, the Fourier transform of $\chi_{\mathsf{d}}$ is
\begin{align*}
\tilde{\chi}_{\mathsf{d}}\left(\mathbf{q}\right) & =\frac{1}{\left(2\pi\right)^{3}}\frac{4\pi R}{q_{3}\sqrt{q_{x}^{2}+q_{y}^{2}}}\sin\left(\frac{h}{2}q_{3}\right)J_{1}\left(R\sqrt{q_{x}^{2}+q_{y}^{2}}\right).
\end{align*}
The characteristic function of the electronic cell, the hexagonal
prism shown in Fig.\ \ref{fig:Quantum-dot-superlattice}, is defined
as
\[
\chi_{\mathsf{e}}\left(\mathbf{r}\right)=\begin{cases}
1 & \mathbf{r}\in\Omega_{\mathsf{e}}\\
0 & \text{otherwise}
\end{cases}
\]
and its Fourier transform is
\begin{align*}
\tilde{\chi}_{\mathsf{e}}\left(\mathbf{q}\right) & =L_{3}\text{sinc}\left(q_{3}\frac{L_{3}}{2}\right)\frac{q_{1}\cos\left(\frac{L_{12}q_{1}}{2}\right)+q_{2}\cos\left(\frac{L_{12}q_{1}}{2}\right)-\left(q_{1}+q_{2}\right)\cos\left(L_{12}\frac{q_{1}+q_{2}}{2}\right)}{\sqrt{3}q_{1}q_{2}\left(q_{1}+q_{2}\right)}.
\end{align*}

\section{Fourier and Convolution Conventions\label{sec:Conventions}}

We define the Fourier forward and inverse transforms of a function
$g\left(\mathbf{r}\right)$ as
\[
\mathscr{F}\left\{ g\left(\mathbf{r}\right)\right\} \left(\mathbf{q}\right)\equiv\tilde{g}\left(\mathbf{q}\right)=\frac{1}{\left(2\pi\right)^{3}}\int\limits _{-\infty}^{\infty}\mathsf{d}^{3}\mathbf{r}\,g\left(\mathbf{r}\right)\mathsf{e}^{-i\mathbf{q}\cdot\mathbf{r}}
\]
\[
\mathscr{F}^{-1}\left\{ \tilde{g}\left(\mathbf{q}\right)\right\} \left(\mathbf{r}\right)=g\left(\mathbf{r}\right)=\int\limits _{-\infty}^{\infty}\mathsf{d}^{3}\mathbf{q}\,\tilde{g}\left(\mathbf{q}\right)\mathsf{e}^{i\mathbf{q}\cdot\mathbf{r}}
\]
A convolution is denoted by
\[
\left(f\ast g\right)\left(\mathbf{r}\right)=\int\limits _{-\infty}^{\infty}\,f\left(\mathbf{r}^{\prime}\right)g\left(\mathbf{r}-\mathbf{r}^{\prime}\right)\mathsf{d}\mathbf{r}^{\prime}
\]
The convolution theorem states
\[
\left(\tilde{f}\ast\tilde{g}\right)\left(\mathbf{q}\right)=\mathscr{F}\left\{ f\left(\mathbf{r}\right)g\left(\mathbf{r}\right)\right\} \left(\mathbf{q}\right).
\]

For a system that is periodic in real space with a unit cell $\Omega$
of volume $V$, we define
\[
\tilde{g}\left(\mathbf{q}\right)=\frac{1}{V}\int\limits _{\Omega}\mathsf{d}^{3}\mathbf{r}\,g\left(\mathbf{r}\right)\mathsf{e}^{-i\mathbf{q}\cdot\mathbf{r}}
\]
\[
g\left(\mathbf{r}\right)=\sum_{\mathbf{q}\in\Omega^{-1}}\tilde{g}\left(\mathbf{q}\right)\mathsf{e}^{i\mathbf{q}\cdot\mathbf{r}}
\]
where $\Omega^{-1}$ is the reciprocal space to the unit cell $\Omega$.
Defining the Fourier space convolution as
\begin{equation}
\left(\tilde{f}\ast\tilde{g}\right)\left(\mathbf{q}\right)=\sum\limits _{\mathbf{q}^{\prime}\in\Omega^{-1}}\tilde{f}\left(\mathbf{q}^{\prime}\right)\tilde{g}\left(\mathbf{q}-\mathbf{q}^{\prime}\right),\label{eq:Finite cell convolution}
\end{equation}
the convolution theorem is
\[
\left(\tilde{f}\ast\tilde{g}\right)\left(\mathbf{q}\right)=\mathscr{F}\left\{ f\left(\mathbf{r}\right)g\left(\mathbf{r}\right)\right\} \left(\mathbf{q}\right).
\]
Since we consider different superlattice unit cells for the electronic
and strain properties, $\left(\tilde{f}\ast\tilde{g}\right)_{\mathsf{e}}$
indicates a convolution on the electronic space $\Omega_{\mathsf{e}}^{-1}$
and $\left(\tilde{f}\ast\tilde{g}\right)_{\mathsf{s}}$ on the strain
space $\Omega_{\mathsf{s}}^{-1}$.

The reciprocal space $\Omega^{-1}$ contains a discrete infinity of
wave vectors on which to evaluate $\tilde{f}$ and $\tilde{g}$. The
convolution in Eq.\ \ref{eq:Finite cell convolution} then sums over
the infinite number of wave vectors, with $\tilde{f}$ and $\tilde{g}$
being functions that decay at large wave vectors. For the calculations
in this manuscript, we choose a finite number of wave vectors. By
choosing this mesh to contain wave vectors sufficiently large to capture
the decay of $\tilde{f}$ and $\tilde{g}$, we can calculate the linear
convolution in Eq.\ \ref{eq:Finite cell convolution} to good approximation
by padding the $\tilde{f}$ and $\tilde{g}$ arrays with zeros and
performing a circular convolution, defined below.

We denote the finite Fourier-space mesh by $\mathbf{q}_{i_{1}i_{2}i_{3}}$
with $-m_{j}<i_{j}<m_{j}$ such that $N_{j}=2m_{j}+1$ is the dimension
of the mesh in each direction. For simplicity, we use the mapping
$p_{j}=i_{j}+m_{j}+1$ to start indexing from 1. Evaluating a function
$\tilde{f}$ on the mesh $\mathbf{q}_{p_{1}p_{2}p_{3}}$ gives the
array $\tilde{f}_{p_{1}p_{2}p_{3}}$. The Fourier-space array $\tilde{f}_{p_{1}p_{2}p_{3}}$
and its real-space counterpart $f_{v_{1}v_{2}v_{3}}$ are then related
through the discrete Fourier transform and its inverse,
\begin{equation}
\tilde{f}_{p_{1}p_{2}p_{3}}=\mathscr{F}\left\{ f\right\} _{p_{1}p_{2}p_{3}}=\frac{1}{N}\sum_{v_{1},v_{2},v_{3}=1}^{N_{1},N_{2},N_{3}}f_{v_{1}v_{2}v_{3}}e^{-i2\pi\left(p_{1}v_{1}/N_{1}+p_{2}v_{2}/N_{2}+p_{3}v_{3}/N_{3}\right)}\label{eq:DFT}
\end{equation}
\begin{align}
f_{v_{1}v_{2}v_{3}} & =\mathscr{F}^{-1}\left\{ \tilde{f}\right\} _{v_{1}v_{2}v_{3}}=\sum_{p_{1},p_{2},p_{3}=1}^{N_{1},N_{2},N_{3}}\tilde{f}_{p_{1}p_{2}p_{3}}e^{i2\pi\left(p_{1}v_{1}/N_{1}+p_{2}v_{2}/N_{2}+p_{3}v_{3}/N_{3}\right)}.\label{eq:IDFT}
\end{align}
where $N=N_{1}N_{2}N_{3}$. The circular convolution is defined as
\[
\left(\tilde{f}\bullet\tilde{g}\right)_{p_{1}p_{2}p_{3}}=\sum_{p_{1}^{\prime},p_{2}^{\prime},p_{3}^{\prime}=1}^{N_{1},N_{2},N_{3}}\tilde{f}_{p_{1}^{\prime}p_{2}^{\prime}p_{3}^{\prime}}\tilde{g}_{\left(p_{1}-p_{1}^{\prime}\right),\left(p_{2}-p_{2}^{\prime}\right),\left(p_{3}-p_{3}^{\prime}\right)}
\]
where $\tilde{f}_{\left(p_{1}+N_{1}\right),\left(p_{2}+N_{2}\right),\left(p_{3}+N_{3}\right)}=\tilde{f}_{p_{1}p_{2}p_{3}}$
and likewise for $\tilde{g}$. The convolution theorem is then
\[
\left(\tilde{f}\bullet\tilde{g}\right)_{p_{1}p_{2}p_{3}}=\mathscr{F}\left\{ fg\right\} _{p_{1}p_{2}p_{3}}.
\]
To perform a linear convolution, we pad the arrays with zeros, which
increases the dimensions of the mesh to $N_{j}^{\mathsf{p}}=N_{j}+2m_{j}$
and yields the padded array $\tilde{f}_{p_{1}p_{2}p_{3}}^{\mathsf{p}}$
containing $N^{\mathsf{p}}=N_{1}^{\mathsf{p}}N_{2}^{\mathsf{p}}N_{3}^{\mathsf{p}}$
elements. The linear convolution is then 
\begin{align}
\left(\tilde{f}\ast\tilde{g}\right)_{p_{1}p_{2}p_{3}} & =\left(\tilde{f}^{\mathsf{p}}\bullet\tilde{g}^{\mathsf{p}}\right)_{p_{1}p_{2}p_{3}}\label{eq:DConv theo.}
\end{align}
This result is independent of the basis used to generate the mesh
and is valid in the case of hexagonal meshes.

\section{Displacement field Green's tensor \label{sec:Displacement-field-Green's}}

The Green's function for the displacement field for spatially varying
elastic constants must satisfy \cite{Andreev2000}
\[
\frac{\partial}{\partial x_{k}}\lambda_{iklm}\left(\mathbf{r}\right)\frac{\partial}{\partial x_{m}}G_{ln}\left(\mathbf{r}-\mathbf{r}^{\prime}\right)=-\delta\left(\mathbf{r}-\mathbf{r}^{\prime}\right)\delta_{in}
\]
From Ref.\ \cite{Andreev2000}, the Green's tensor when $\lambda_{jklm}$
is spatially invariant is
\[
\tilde{G}_{in}^{\mathsf{h}}=\frac{1}{\left(2\pi\right)^{3}}\frac{1}{\beta q^{2}+\rho q_{3}^{2}}\left\{ \delta_{in}-\left[\left(\rho q^{2}+\gamma q_{3}^{2}\right)\delta_{i3}+\left(\kappa+\rho\right)q_{i}q_{3}\right]\frac{F\delta_{3n}-I}{FP-IQ}q_{n}-\left[\left(\alpha+\beta\right)q_{i}+\left(\kappa+\rho\right)\delta_{i3}q_{3}\right]\frac{P-Q\delta_{3n}}{FP-IQ}q_{n}\right\} ,
\]
where 
\[
\begin{aligned}F\left(\mathbf{q}\right) & =\left(C_{13}+2C_{44}-C_{11}\right)q_{3}^{2}+C_{11}q^{2}\\
Q\left(\mathbf{q}\right) & =\left(C_{33}-2C_{13}-4C_{44}+C_{11}\right)q_{3}^{2}+\left(C_{13}+2C_{44}-C_{11}\right)q^{2}\\
P\left(\mathbf{q}\right) & =C_{44}q^{2}+\left(C_{33}-C_{13}-2C_{44}\right)q_{3}^{2}\\
I\left(\mathbf{q}\right) & =\left(C_{13}+C_{44}\right)q_{3}^{2}\\
\alpha & =C_{12}\\
\beta & =\frac{1}{2}\left(C_{11}-C_{12}\right)\\
\gamma & =C_{33}-2C_{13}-4C_{44}+C_{11}\\
\kappa & =C_{13}-C_{12}\\
\rho & =C_{44}+\frac{C_{12}-C_{11}}{2}
\end{aligned}
\]
with $C_{ij}$ being the elastic constants.

\section{Polarization fields \label{sec:Polarization-fields}}

There are two contributions to the piezoelectric polarization fields:
strain driven and spontaneous. The strain-driven polarization can
be written \cite{Vukmirovic2006}
\[
\left[\begin{array}{c}
P_{1}^{\mathsf{st}}\\
P_{2}^{\mathsf{st}}\\
P_{3}^{\mathsf{st}}
\end{array}\right]=\left[\begin{array}{c}
2e_{15}\epsilon_{13}\\
2e_{15}\epsilon_{23}\\
e_{31}\left(\epsilon_{11}+\epsilon_{22}\right)+e_{33}\epsilon_{33}
\end{array}\right],
\]
where $e_{ij}$ are the piezoelectric constants and $\epsilon_{ij}$
the strain fields. We write the piezoelectric constants using the
characteristic function of the quantum dot
\[
e_{ij}\left(\mathbf{r}\right)=e_{ij}^{\mathsf{d}}\chi_{\mathsf{d}}\left(\mathbf{r}\right)+e_{ij}^{\mathsf{h}}\left[1-\chi_{\mathsf{d}}\left(\mathbf{r}\right)\right].
\]
The Fourier transform of the polarization fields is then \begin{widetext}
\[
\left[\begin{array}{c}
\tilde{P}_{1}^{\mathsf{st}}\\
\tilde{P}_{2}^{\mathsf{st}}\\
\tilde{P}_{3}^{\mathsf{st}}
\end{array}\right]=\left[\begin{array}{c}
2e_{15}^{\mathsf{h}}\tilde{\epsilon}_{13}\left(\mathbf{q}\right)+\frac{2\left(2\pi\right)^{3}\left(e_{15}^{\mathsf{d}}-e_{15}^{\mathsf{h}}\right)}{V}\left(\tilde{\chi}_{d}\ast\tilde{\epsilon}_{13}\right)_{e}\left(\mathbf{q}\right)\\
2e_{15}^{\mathsf{h}}\tilde{\epsilon}_{23}\left(\mathbf{q}\right)+\frac{2\left(2\pi\right)^{3}\left(e_{15}^{\mathsf{d}}-e_{15}^{\mathsf{h}}\right)}{V}\left(\tilde{\chi}_{d}\ast\tilde{\epsilon}_{23}\right)_{e}\left(\mathbf{q}\right)\\
e_{31}^{\mathsf{h}}\left[\tilde{\epsilon}_{11}\left(\mathbf{q}\right)+\tilde{\epsilon}_{22}\left(\mathbf{q}\right)\right]+\frac{\left(2\pi\right)^{3}\left(e_{31}^{\mathsf{d}}-e_{31}^{\mathsf{h}}\right)}{V}\left[\tilde{\chi}_{\mathsf{d}}\ast\left(\tilde{\epsilon}_{11}+\tilde{\epsilon}_{22}\right)\right]_{e}\left(\mathbf{q}\right)+e_{33}^{\mathsf{h}}\tilde{\epsilon}_{33}\left(\mathbf{q}\right)+\frac{\left(2\pi\right)^{3}\left(e_{33}^{\mathsf{d}}-e_{33}^{\mathsf{h}}\right)}{V}\left(\tilde{\chi}_{\mathsf{d}}\ast\tilde{\epsilon}_{33}\right)_{e}\left(\mathbf{q}\right)
\end{array}\right]
\]
\end{widetext}

In a bulk wurtzite material, the spontaneous polarization is along
the c-axis and uniform throughout the material
\[
\mathbf{P}^{\mathsf{sp}}\left(\mathbf{r}\right)=\left[\begin{array}{c}
0\\
0\\
P^{\mathsf{sp}}
\end{array}\right]
\]
Then
\[
\tilde{P}_{3}^{\mathsf{sp}}\left(\mathbf{q}\right)=P^{\mathsf{sp,h}}\delta_{\mathbf{q},\mathbf{0}}+\left(P^{\mathsf{sp,d}}-P^{\mathsf{sp,h}}\right)\tilde{\chi}\left(\mathbf{q}\right)
\]
where $P^{\mathsf{sp,d}}$ and $P^{\mathsf{sp,h}}$ are the spontaneous
polarizations in the dot and host materials, respectively.

\section{QD $\mathbf{k\cdot\mathbf{p}}$ Hamiltonian \label{sec:QD-k.p-Hamiltonian}}

For the quantum dot superlattice system, the $\mathbf{k\cdot\mathbf{p}}$
Hamiltonian matrix elements of Eq.\ \ref{eq:General Hamiltonian matrix element}
in the symmetry adapted basis are given in terms of the parameters
$f_{\alpha'\alpha}^{\mathsf{d}}$ and $f_{\alpha'\alpha}^{\mathsf{h}}$
that make up the bulk $\mathbf{k\cdot\mathbf{p}}$ Hamiltonian matrix
elements $H_{\alpha^{\prime}\alpha}$, presented in Sec.\ \ref{sec:Symmetry-adapted-basis k.p}.
We take the convention where superscript ``$\left(0\right)$'' indicates
a bulk Hamiltonian matrix element containing no wave vector, ``$\left(i\right)$''
indicates a single wave vector $k_{i}$ and ``$\left(i,j\right)$''
two wave vectors $k_{i}$ and $k_{j}$. By defining $\phi=\frac{2\pi}{6}$
and 
\[
S_{\alpha^{\prime},\alpha}^{ll^{\prime}m_{f}}=e^{i\phi\left\{ l\left[m_{f}-J_{z}\left(\alpha\right)\right]-l^{\prime}\left[m_{f}-J_{z}\left(\alpha^{\prime}\right)\right]\right\} }
\]
\[
S_{\alpha^{\prime}}^{l^{\prime}m_{f}}=e^{-il^{\prime}\phi\left[m_{f}-J_{z}\left(\alpha^{\prime}\right)\right]},
\]
the quantum dot Hamiltonian matrix elements are\begin{widetext}
\begin{eqnarray*}
\mathcal{H}_{m_{f}\alpha^{\prime}\alpha}^{\left(0\right)}\left(\mathbf{q}^{\prime},\mathbf{q}\right) & = & \frac{1}{6}\sum_{l^{\prime}=0}^{5}\sum_{l=0}^{5}S_{\alpha^{\prime},\alpha}^{ll^{\prime}m_{f}}h_{\alpha^{\prime}\alpha}\left(\overleftrightarrow{\mathbf{R}}_{\!l^{\prime}}\mathbf{q}^{\prime},\overleftrightarrow{\mathbf{R}}_{\!l}\mathbf{q}\right)
\end{eqnarray*}
\[
\mathcal{H}_{m_{f}\alpha^{\prime}\alpha}^{\left(i\right)}\left(\mathbf{q}^{\prime},\mathbf{q}\right)=\frac{1}{6}\sum_{l^{\prime}=0}^{5}\sum_{l=0}^{5}S_{\alpha^{\prime},\alpha}^{ll^{\prime}m_{f}}\frac{\left(\overleftrightarrow{\mathbf{R}}_{\!l^{'}}\mathbf{q}^{\prime}\right)_{i}+\left(\overleftrightarrow{\mathbf{R}}_{\!l}\mathbf{q}\right)_{i}}{2}h_{\alpha^{\prime}\alpha}\left(\overleftrightarrow{\mathbf{R}}_{\!l^{\prime}}\mathbf{q}^{\prime},\overleftrightarrow{\mathbf{R}}_{\!l}\mathbf{q}\right)
\]
\[
\mathcal{H}_{m_{f}\alpha^{\prime}\alpha}^{\left(i,j\right)}\left(\mathbf{q}^{\prime},\mathbf{q}\right)=\frac{1}{6}\sum_{l^{\prime}=0}^{5}\sum_{l=0}^{5}S_{\alpha^{\prime},\alpha}^{ll^{\prime}m_{f}}\frac{\left(\overleftrightarrow{\mathbf{R}}_{\!l}\mathbf{q}\right)_{j}\left(\overleftrightarrow{\mathbf{R}}_{\!l^{'}}\mathbf{q}^{\prime}\right)_{i}+\left(\overleftrightarrow{\mathbf{R}}_{\!l}\mathbf{q}\right)_{i}\left(\overleftrightarrow{\mathbf{R}}_{\!l^{'}}\mathbf{q}^{\prime}\right)_{j}}{2}h_{\alpha^{\prime}\alpha}\left(\overleftrightarrow{\mathbf{R}}_{\!l^{\prime}}\mathbf{q}^{\prime},\overleftrightarrow{\mathbf{R}}_{\!l}\mathbf{q}\right)
\]

\textcompwordmark
\begin{eqnarray*}
\mathcal{H}_{m_{f}\alpha^{\prime}\alpha}^{\left(0\right)}\left(\mathbf{q}^{\prime},\mathbf{q}_{z}\right) & = & \frac{1}{\sqrt{6}}\sum_{l^{\prime}=0}^{5}S_{\alpha^{\prime}}^{l^{\prime}m_{f}}h_{\alpha^{\prime}\alpha}\left(\overleftrightarrow{\mathbf{R}}_{\!l^{\prime}}\mathbf{q}^{\prime},\mathbf{q}_{z}\right)
\end{eqnarray*}
\[
\mathcal{H}_{m_{f}\alpha^{\prime}\alpha}^{\left(i\right)}\left(\mathbf{q}^{\prime},\mathbf{q}_{z}\right)=\frac{1}{\sqrt{6}}\sum_{l^{\prime}=0}^{5}S_{\alpha^{\prime}}^{l^{\prime}m_{f}}\frac{\left(\overleftrightarrow{\mathbf{R}}_{\!l^{'}}\mathbf{q}^{\prime}\right)_{i}+\left(\mathbf{q}_{z}\right)_{i}}{2}h_{\alpha^{\prime}\alpha}\left(\overleftrightarrow{\mathbf{R}}_{\!l^{\prime}}\mathbf{q}^{\prime},\mathbf{q}_{z}\right)
\]
\[
\mathcal{H}_{m_{f}\alpha^{\prime}\alpha}^{\left(i,j\right)}\left(\mathbf{q}^{\prime},\mathbf{q}_{z}\right)=\frac{1}{\sqrt{6}}\sum_{l^{\prime}=0}^{5}S_{\alpha^{\prime}}^{l^{\prime}m_{f}}\frac{\left(\mathbf{q}_{z}\right)_{j}\left(\overleftrightarrow{\mathbf{R}}_{\!l^{'}}\mathbf{q}^{\prime}\right)_{i}+\left(\mathbf{q}_{z}\right)_{i}\left(\overleftrightarrow{\mathbf{R}}_{\!l^{'}}\mathbf{q}^{\prime}\right)_{j}}{2}h_{\alpha^{\prime}\alpha}\left(\overleftrightarrow{\mathbf{R}}_{\!l^{\prime}}\mathbf{q}^{\prime},\mathbf{q}_{z}\right)
\]
\[
\mathcal{H}_{m_{f}\alpha^{\prime}\alpha}^{\left(0\right)}\left(\mathbf{q}_{z}^{\prime},\mathbf{q}_{z}\right)=h_{\alpha'\alpha}\left(\mathbf{q}_{z}^{\prime},\mathbf{q}_{z}\right)
\]
\[
\mathcal{H}_{m_{f}\alpha^{\prime}\alpha}^{\left(i\right)}\left(\mathbf{q}_{z}^{\prime},\mathbf{q}_{z}\right)=\frac{\left(\mathbf{q}_{z}^{\prime}\right)_{i}+\left(\mathbf{q}_{z}\right)_{i}}{2}h_{\alpha'\alpha}\left(\mathbf{q}_{z}^{\prime},\mathbf{q}_{z}\right)
\]
\[
\mathcal{H}_{m_{f}\alpha^{\prime}\alpha}^{\left(i,j\right)}\left(\mathbf{q}_{z}^{\prime},\mathbf{q}_{z}\right)=\frac{\left(\mathbf{q}_{z}\right)_{j}\left(\mathbf{q}_{z}^{\prime}\right)_{i}+\left(\mathbf{q}_{z}\right)_{i}\left(0,0,q_{z}^{\prime}\right)_{j}}{2}h_{\alpha'\alpha}\left(\mathbf{q}_{z}^{\prime},\mathbf{q}_{z}\right)
\]
 where

\[
h_{\alpha^{\prime}\alpha}\left(\overleftrightarrow{\mathbf{R}}_{\!l^{\prime}}\mathbf{q}^{\prime},\overleftrightarrow{\mathbf{R}}_{\!l}\mathbf{q}\right)=f_{\alpha'\alpha}^{\mathsf{h}}\delta_{l,l^{\prime}}\delta_{q,q^{\prime}}+\frac{\left(2\pi\right)^{3}\left(f_{\alpha^{\prime}\alpha}^{\mathsf{d}}-f_{\alpha^{\prime}\alpha}^{\mathsf{h}}\right)}{V}\tilde{\chi}_{\mathsf{d}}\left(\overleftrightarrow{\mathbf{R}}_{\!l^{\prime}}\mathbf{q}^{\prime}-\overleftrightarrow{\mathbf{R}}_{\!l}\mathbf{q}\right)
\]

\end{widetext}

The contributions of the piezoelectric potential to the Hamiltonian
are\begin{widetext}
\[
\mathcal{H}_{m_{f}\alpha^{\prime}\alpha}^{\mathsf{pz}}\left(\mathbf{q}^{\prime},\mathbf{q}\right)=-\delta_{\alpha^{\prime},\alpha}\frac{\left(2\pi\right)^{3}e_{\mathsf{c}}}{6V}\sum_{l^{\prime}=0}^{5}\sum_{l=0}^{5}e^{i\phi\left(l-l^{\prime}\right)\left[m_{f}-J_{z}\left(\alpha\right)\right]}\tilde{\varphi}\left(\overleftrightarrow{\mathbf{R}}_{\!l^{\prime}}\mathbf{q}^{\prime}-\overleftrightarrow{\mathbf{R}}_{\!l}\mathbf{q}\right)
\]
\[
\mathcal{H}_{m_{f}\alpha^{\prime}\alpha}^{\mathsf{pz}}\left(\mathbf{q}^{\prime},\mathbf{q}_{z}\right)=-\delta_{\alpha^{\prime},\alpha}\frac{\left(2\pi\right)^{3}e_{\mathsf{c}}}{V\sqrt{6}}\sum_{l^{\prime}=0}^{5}e^{-il^{\prime}\phi\left[m_{f}-J_{z}\left(\alpha^{\prime}\right)\right]}\tilde{\varphi}\left(\overleftrightarrow{\mathbf{R}}_{\!l^{\prime}}\mathbf{q}^{\prime}-\mathbf{q}_{z}\right)
\]
\[
\mathcal{H}_{m_{f}\alpha^{\prime}\alpha}^{\mathsf{pz}}\left(\mathbf{q}_{z}^{\prime},\mathbf{q}_{z}\right)=-\delta_{\alpha^{\prime},\alpha}\frac{\left(2\pi\right)^{3}e_{\mathsf{c}}}{V}\tilde{\varphi}\left(\mathbf{q}_{z}^{\prime}-\mathbf{q}_{z}\right)
\]
\end{widetext} where $\tilde{\varphi}$ is described in Secs\@.\ \ref{subsec:Piezoelectric-potential}
and \ref{subsec:Including-strain-and-piezo-in-k.p} and $e_{\mathsf{c}}$
is the electric charge.

\bibliography{References}

\begin{thebibliography}{25}%
\makeatletter
\providecommand \@ifxundefined [1]{%
 \@ifx{#1\undefined}
}%
\providecommand \@ifnum [1]{%
 \ifnum #1\expandafter \@firstoftwo
 \else \expandafter \@secondoftwo
 \fi
}%
\providecommand \@ifx [1]{%
 \ifx #1\expandafter \@firstoftwo
 \else \expandafter \@secondoftwo
 \fi
}%
\providecommand \natexlab [1]{#1}%
\providecommand \enquote  [1]{``#1''}%
\providecommand \bibnamefont  [1]{#1}%
\providecommand \bibfnamefont [1]{#1}%
\providecommand \citenamefont [1]{#1}%
\providecommand \href@noop [0]{\@secondoftwo}%
\providecommand \href [0]{\begingroup \@sanitize@url \@href}%
\providecommand \@href[1]{\@@startlink{#1}\@@href}%
\providecommand \@@href[1]{\endgroup#1\@@endlink}%
\providecommand \@sanitize@url [0]{\catcode `\\12\catcode `\$12\catcode
  `\&12\catcode `\#12\catcode `\^12\catcode `\_12\catcode `\%12\relax}%
\providecommand \@@startlink[1]{}%
\providecommand \@@endlink[0]{}%
\providecommand \url  [0]{\begingroup\@sanitize@url \@url }%
\providecommand \@url [1]{\endgroup\@href {#1}{\urlprefix }}%
\providecommand \urlprefix  [0]{URL }%
\providecommand \Eprint [0]{\href }%
\providecommand \doibase [0]{http://dx.doi.org/}%
\providecommand \selectlanguage [0]{\@gobble}%
\providecommand \bibinfo  [0]{\@secondoftwo}%
\providecommand \bibfield  [0]{\@secondoftwo}%
\providecommand \translation [1]{[#1]}%
\providecommand \BibitemOpen [0]{}%
\providecommand \bibitemStop [0]{}%
\providecommand \bibitemNoStop [0]{.\EOS\space}%
\providecommand \EOS [0]{\spacefactor3000\relax}%
\providecommand \BibitemShut  [1]{\csname bibitem#1\endcsname}%
\let\auto@bib@innerbib\@empty
\bibitem [{\citenamefont {Nguyen}\ \emph {et~al.}(2011)\citenamefont {Nguyen},
  \citenamefont {Zhang}, \citenamefont {Cui}, \citenamefont {Han},
  \citenamefont {Fathololoumi}, \citenamefont {Couillard}, \citenamefont
  {Botton},\ and\ \citenamefont {Mi}}]{Nguyen2011}%
  \BibitemOpen
  \bibfield  {author} {\bibinfo {author} {\bibfnamefont {H.~P.~T.}\
  \bibnamefont {Nguyen}}, \bibinfo {author} {\bibfnamefont {S.}~\bibnamefont
  {Zhang}}, \bibinfo {author} {\bibfnamefont {K.}~\bibnamefont {Cui}}, \bibinfo
  {author} {\bibfnamefont {X.}~\bibnamefont {Han}}, \bibinfo {author}
  {\bibfnamefont {S.}~\bibnamefont {Fathololoumi}}, \bibinfo {author}
  {\bibfnamefont {M.}~\bibnamefont {Couillard}}, \bibinfo {author}
  {\bibfnamefont {G.~A.}\ \bibnamefont {Botton}}, \ and\ \bibinfo {author}
  {\bibfnamefont {Z.}~\bibnamefont {Mi}},\ }\bibfield  {title} {\enquote
  {\bibinfo {title} {{p-Type Modulation Doped InGaN/GaN Dot-in-a-Wire
  White-Light-Emitting Diodes Monolithically Grown on Si(111)}},}\ }\href
  {\doibase 10.1021/nl104536x} {\bibfield  {journal} {\bibinfo  {journal} {Nano
  Lett.}\ }\textbf {\bibinfo {volume} {11}},\ \bibinfo {pages} {1919--1924}
  (\bibinfo {year} {2011})}\BibitemShut {NoStop}%
\bibitem [{\citenamefont {Puchtler}\ \emph {et~al.}(2016)\citenamefont
  {Puchtler}, \citenamefont {Wang}, \citenamefont {Ren}, \citenamefont {Tang},
  \citenamefont {Oliver}, \citenamefont {Taylor},\ and\ \citenamefont
  {Zhu}}]{Puchtler2016}%
  \BibitemOpen
  \bibfield  {author} {\bibinfo {author} {\bibfnamefont {Tim~J.}\ \bibnamefont
  {Puchtler}}, \bibinfo {author} {\bibfnamefont {Tong}\ \bibnamefont {Wang}},
  \bibinfo {author} {\bibfnamefont {Christopher~X.}\ \bibnamefont {Ren}},
  \bibinfo {author} {\bibfnamefont {Fengzai}\ \bibnamefont {Tang}}, \bibinfo
  {author} {\bibfnamefont {Rachel~A.}\ \bibnamefont {Oliver}}, \bibinfo
  {author} {\bibfnamefont {Robert~A.}\ \bibnamefont {Taylor}}, \ and\ \bibinfo
  {author} {\bibfnamefont {Tongtong}\ \bibnamefont {Zhu}},\ }\bibfield  {title}
  {\enquote {\bibinfo {title} {{Ultrafast, Polarized, Single-Photon Emission
  from m-Plane InGaN Quantum Dots on GaN Nanowires}},}\ }\href {\doibase
  10.1021/acs.nanolett.6b03980} {\bibfield  {journal} {\bibinfo  {journal}
  {Nano Lett.}\ }\textbf {\bibinfo {volume} {16}},\ \bibinfo {pages}
  {7779--7785} (\bibinfo {year} {2016})}\BibitemShut {NoStop}%
\bibitem [{\citenamefont {Kibria}\ \emph {et~al.}(2013)\citenamefont {Kibria},
  \citenamefont {Nguyen}, \citenamefont {Cui}, \citenamefont {Zhao},
  \citenamefont {Liu}, \citenamefont {Guo}, \citenamefont {Trudeau},
  \citenamefont {Paradis}, \citenamefont {Hakima},\ and\ \citenamefont
  {Mi}}]{Kibria2013}%
  \BibitemOpen
  \bibfield  {author} {\bibinfo {author} {\bibfnamefont {Md~G.}\ \bibnamefont
  {Kibria}}, \bibinfo {author} {\bibfnamefont {Hieu~P.T.}\ \bibnamefont
  {Nguyen}}, \bibinfo {author} {\bibfnamefont {Kai}\ \bibnamefont {Cui}},
  \bibinfo {author} {\bibfnamefont {Songrui}\ \bibnamefont {Zhao}}, \bibinfo
  {author} {\bibfnamefont {Dongping}\ \bibnamefont {Liu}}, \bibinfo {author}
  {\bibfnamefont {Hong}\ \bibnamefont {Guo}}, \bibinfo {author} {\bibfnamefont
  {Michel~L.}\ \bibnamefont {Trudeau}}, \bibinfo {author} {\bibfnamefont
  {Suzanne}\ \bibnamefont {Paradis}}, \bibinfo {author} {\bibfnamefont
  {Abou~Rachid}\ \bibnamefont {Hakima}}, \ and\ \bibinfo {author}
  {\bibfnamefont {Zetian}\ \bibnamefont {Mi}},\ }\bibfield  {title} {\enquote
  {\bibinfo {title} {{One-step overall water splitting under visible light
  using multiband InGaN/GaN nanowire heterostructures}},}\ }\href {\doibase
  10.1021/nn4028823} {\bibfield  {journal} {\bibinfo  {journal} {ACS Nano}\
  }\textbf {\bibinfo {volume} {7}},\ \bibinfo {pages} {7886--7893} (\bibinfo
  {year} {2013})}\BibitemShut {NoStop}%
\bibitem [{\citenamefont {Sang}\ \emph {et~al.}(2014)\citenamefont {Sang},
  \citenamefont {Liao}, \citenamefont {Liang}, \citenamefont {Takeguchi},
  \citenamefont {Dierre}, \citenamefont {Shen}, \citenamefont {Sekiguchi},
  \citenamefont {Koide},\ and\ \citenamefont {Sumiya}}]{Sang2014}%
  \BibitemOpen
  \bibfield  {author} {\bibinfo {author} {\bibfnamefont {Liwen}\ \bibnamefont
  {Sang}}, \bibinfo {author} {\bibfnamefont {Meiyong}\ \bibnamefont {Liao}},
  \bibinfo {author} {\bibfnamefont {Qifeng}\ \bibnamefont {Liang}}, \bibinfo
  {author} {\bibfnamefont {Masaki}\ \bibnamefont {Takeguchi}}, \bibinfo
  {author} {\bibfnamefont {Benjamin}\ \bibnamefont {Dierre}}, \bibinfo {author}
  {\bibfnamefont {Bo}~\bibnamefont {Shen}}, \bibinfo {author} {\bibfnamefont
  {Takashi}\ \bibnamefont {Sekiguchi}}, \bibinfo {author} {\bibfnamefont
  {Yasuo}\ \bibnamefont {Koide}}, \ and\ \bibinfo {author} {\bibfnamefont
  {Masatomo}\ \bibnamefont {Sumiya}},\ }\bibfield  {title} {\enquote {\bibinfo
  {title} {{A Multilevel Intermediate-Band Solar Cell by InGaN/GaN Quantum Dots
  with a Strain-Modulated Structure}},}\ }\href {\doibase
  10.1002/adma.201304335} {\bibfield  {journal} {\bibinfo  {journal} {Adv.
  Mater.}\ }\textbf {\bibinfo {volume} {26}},\ \bibinfo {pages} {1414--1420}
  (\bibinfo {year} {2014})}\BibitemShut {NoStop}%
\bibitem [{\citenamefont {Cheriton}\ \emph {et~al.}(2020)\citenamefont
  {Cheriton}, \citenamefont {Sadaf}, \citenamefont {Robichaud}, \citenamefont
  {Krich}, \citenamefont {Mi},\ and\ \citenamefont {Hinzer}}]{Cheriton2020}%
  \BibitemOpen
  \bibfield  {author} {\bibinfo {author} {\bibfnamefont {Ross}\ \bibnamefont
  {Cheriton}}, \bibinfo {author} {\bibfnamefont {Sharif~M.}\ \bibnamefont
  {Sadaf}}, \bibinfo {author} {\bibfnamefont {Luc}\ \bibnamefont {Robichaud}},
  \bibinfo {author} {\bibfnamefont {Jacob~J.}\ \bibnamefont {Krich}}, \bibinfo
  {author} {\bibfnamefont {Zetian}\ \bibnamefont {Mi}}, \ and\ \bibinfo
  {author} {\bibfnamefont {Karin}\ \bibnamefont {Hinzer}},\ }\bibfield  {title}
  {\enquote {\bibinfo {title} {Two-photon photocurrent in {InGaN}/{GaN}
  nanowire intermediate band solar cells},}\ }\href {\doibase
  10.1038/s43246-020-00054-6} {\bibfield  {journal} {\bibinfo  {journal}
  {Communications Materials}\ }\textbf {\bibinfo {volume} {1}},\ \bibinfo
  {pages} {63} (\bibinfo {year} {2020})}\BibitemShut {NoStop}%
\bibitem [{\citenamefont {Saito}\ and\ \citenamefont
  {Arakawa}(2002)}]{Saito2002}%
  \BibitemOpen
  \bibfield  {author} {\bibinfo {author} {\bibfnamefont {T.}~\bibnamefont
  {Saito}}\ and\ \bibinfo {author} {\bibfnamefont {Y.}~\bibnamefont
  {Arakawa}},\ }\bibfield  {title} {\enquote {\bibinfo {title} {{Electronic
  structure of piezoelectric In$_{0.2}$Ga$_{0.8}$N quantum dots in GaN
  calculated using a tight-binding method}},}\ }\href {\doibase
  10.1016/S1386-9477(02)00515-5} {\bibfield  {journal} {\bibinfo  {journal}
  {Physica E}\ }\textbf {\bibinfo {volume} {15}},\ \bibinfo {pages} {169--181}
  (\bibinfo {year} {2002})}\BibitemShut {NoStop}%
\bibitem [{\citenamefont {Winkelnkemper}\ \emph {et~al.}(2006)\citenamefont
  {Winkelnkemper}, \citenamefont {Schliwa},\ and\ \citenamefont
  {Bimberg}}]{Winkelnkemper}%
  \BibitemOpen
  \bibfield  {author} {\bibinfo {author} {\bibfnamefont {Momme}\ \bibnamefont
  {Winkelnkemper}}, \bibinfo {author} {\bibfnamefont {Andrei}\ \bibnamefont
  {Schliwa}}, \ and\ \bibinfo {author} {\bibfnamefont {Dieter}\ \bibnamefont
  {Bimberg}},\ }\bibfield  {title} {\enquote {\bibinfo {title} {{Interrelation
  of structural and electronic properties in In$_{x}$Ga$_{1-x}$N/GaN quantum
  dots using an eight-band k$\cdot$p model}},}\ }\href {\doibase
  10.1103/PhysRevB.74.155322} {\bibfield  {journal} {\bibinfo  {journal} {Phys.
  Rev. B}\ }\textbf {\bibinfo {volume} {74}},\ \bibinfo {pages} {155322}
  (\bibinfo {year} {2006})}\BibitemShut {NoStop}%
\bibitem [{\citenamefont {Andreev}\ and\ \citenamefont
  {O'Reilly}(2000)}]{Andreev2000}%
  \BibitemOpen
  \bibfield  {author} {\bibinfo {author} {\bibfnamefont {A.~D.}\ \bibnamefont
  {Andreev}}\ and\ \bibinfo {author} {\bibfnamefont {E.~P.}\ \bibnamefont
  {O'Reilly}},\ }\bibfield  {title} {\enquote {\bibinfo {title} {{Theory of the
  electronic structure of GaN/AlN hexagonal quantum dots}},}\ }\href {\doibase
  10.1103/PhysRevB.62.15851} {\bibfield  {journal} {\bibinfo  {journal} {Phys.
  Rev. B}\ }\textbf {\bibinfo {volume} {62}},\ \bibinfo {pages} {15851--15870}
  (\bibinfo {year} {2000})}\BibitemShut {NoStop}%
\bibitem [{\citenamefont {Vukmirovi{\'{c}}}\ \emph {et~al.}(2005)\citenamefont
  {Vukmirovi{\'{c}}}, \citenamefont {Indjin}, \citenamefont {Jovanovi{\'{c}}},
  \citenamefont {Ikoni{\'{c}}},\ and\ \citenamefont
  {Harrison}}]{Vukmirovic2005}%
  \BibitemOpen
  \bibfield  {author} {\bibinfo {author} {\bibfnamefont {Nenad}\ \bibnamefont
  {Vukmirovi{\'{c}}}}, \bibinfo {author} {\bibfnamefont {Dragan}\ \bibnamefont
  {Indjin}}, \bibinfo {author} {\bibfnamefont {Vladimir~D.}\ \bibnamefont
  {Jovanovi{\'{c}}}}, \bibinfo {author} {\bibfnamefont {Zoran}\ \bibnamefont
  {Ikoni{\'{c}}}}, \ and\ \bibinfo {author} {\bibfnamefont {Paul}\ \bibnamefont
  {Harrison}},\ }\bibfield  {title} {\enquote {\bibinfo {title} {Symmetry of
  k.p {Hamiltonian} in pyramidal {InAs}/{GaAs} quantum dots: Application to the
  calculation of electronic structure},}\ }\href {\doibase
  10.1103/physrevb.72.075356} {\bibfield  {journal} {\bibinfo  {journal} {Phys.
  Rev. B}\ }\textbf {\bibinfo {volume} {72}},\ \bibinfo {pages} {075356}
  (\bibinfo {year} {2005})}\BibitemShut {NoStop}%
\bibitem [{\citenamefont {Vukmirovi{\'{c}}}\ \emph {et~al.}(2006)\citenamefont
  {Vukmirovi{\'{c}}}, \citenamefont {Ikoni{\'{c}}}, \citenamefont {Indjin},\
  and\ \citenamefont {Harrison}}]{Vukmirovic2006}%
  \BibitemOpen
  \bibfield  {author} {\bibinfo {author} {\bibfnamefont {Nenad}\ \bibnamefont
  {Vukmirovi{\'{c}}}}, \bibinfo {author} {\bibfnamefont {Zoran}\ \bibnamefont
  {Ikoni{\'{c}}}}, \bibinfo {author} {\bibfnamefont {Dragan}\ \bibnamefont
  {Indjin}}, \ and\ \bibinfo {author} {\bibfnamefont {Paul}\ \bibnamefont
  {Harrison}},\ }\bibfield  {title} {\enquote {\bibinfo {title}
  {{Symmetry-based calculation of single-particle states and intraband
  absorption in hexagonal GaN/AlN quantum dot superlattices}},}\ }\href
  {\doibase 10.1088/0953-8984/18/27/008} {\bibfield  {journal} {\bibinfo
  {journal} {J. Phys. Condens. Matter}\ }\textbf {\bibinfo {volume} {18}},\
  \bibinfo {pages} {6249--6262} (\bibinfo {year} {2006})}\BibitemShut {NoStop}%
\bibitem [{\citenamefont {Vukmirov{\'{c}}}\ and\ \citenamefont
  {Tomi{\'{c}}}(2008)}]{Vukmirovc2008}%
  \BibitemOpen
  \bibfield  {author} {\bibinfo {author} {\bibfnamefont {Nenad}\ \bibnamefont
  {Vukmirov{\'{c}}}}\ and\ \bibinfo {author} {\bibfnamefont {Stanko}\
  \bibnamefont {Tomi{\'{c}}}},\ }\bibfield  {title} {\enquote {\bibinfo {title}
  {{Plane wave methodology for single quantum dot electronic structure
  calculations}},}\ }\href {\doibase 10.1063/1.2936318} {\bibfield  {journal}
  {\bibinfo  {journal} {J. Appl. Phys.}\ }\textbf {\bibinfo {volume} {103}},\
  \bibinfo {pages} {103718} (\bibinfo {year} {2008})}\BibitemShut {NoStop}%
\bibitem [{\citenamefont {Stier}\ \emph {et~al.}(1999)\citenamefont {Stier},
  \citenamefont {Grundmann},\ and\ \citenamefont {Bimberg}}]{Stier1998}%
  \BibitemOpen
  \bibfield  {author} {\bibinfo {author} {\bibfnamefont {O.}~\bibnamefont
  {Stier}}, \bibinfo {author} {\bibfnamefont {M.}~\bibnamefont {Grundmann}}, \
  and\ \bibinfo {author} {\bibfnamefont {D.}~\bibnamefont {Bimberg}},\
  }\bibfield  {title} {\enquote {\bibinfo {title} {{Electronic and optical
  properties of strained quantum dots modeled by 8-band k.p theory}},}\ }\href
  {\doibase 10.1103/PhysRevB.59.5688} {\bibfield  {journal} {\bibinfo
  {journal} {Phys. Rev. B}\ }\textbf {\bibinfo {volume} {59}},\ \bibinfo
  {pages} {5688--5701} (\bibinfo {year} {1999})}\BibitemShut {NoStop}%
\bibitem [{\citenamefont {Andreev}\ \emph {et~al.}(1999)\citenamefont
  {Andreev}, \citenamefont {Downes}, \citenamefont {Faux},\ and\ \citenamefont
  {O'Reilly}}]{Andreev1999}%
  \BibitemOpen
  \bibfield  {author} {\bibinfo {author} {\bibfnamefont {A.~D.}\ \bibnamefont
  {Andreev}}, \bibinfo {author} {\bibfnamefont {J.~R.}\ \bibnamefont {Downes}},
  \bibinfo {author} {\bibfnamefont {D.~A.}\ \bibnamefont {Faux}}, \ and\
  \bibinfo {author} {\bibfnamefont {E.~P.}\ \bibnamefont {O'Reilly}},\
  }\bibfield  {title} {\enquote {\bibinfo {title} {{Strain distributions in
  quantum dots of arbitrary shape}},}\ }\href {\doibase 10.1063/1.370728}
  {\bibfield  {journal} {\bibinfo  {journal} {J. Appl. Phys.}\ }\textbf
  {\bibinfo {volume} {86}},\ \bibinfo {pages} {297--305} (\bibinfo {year}
  {1999})}\BibitemShut {NoStop}%
\bibitem [{\citenamefont {Renard}\ \emph {et~al.}(2009)\citenamefont {Renard},
  \citenamefont {Songmuang}, \citenamefont {Tourbot}, \citenamefont {Bougerol},
  \citenamefont {Daudin},\ and\ \citenamefont {Gayral}}]{Renard2009}%
  \BibitemOpen
  \bibfield  {author} {\bibinfo {author} {\bibfnamefont {J.}~\bibnamefont
  {Renard}}, \bibinfo {author} {\bibfnamefont {R.}~\bibnamefont {Songmuang}},
  \bibinfo {author} {\bibfnamefont {G.}~\bibnamefont {Tourbot}}, \bibinfo
  {author} {\bibfnamefont {C.}~\bibnamefont {Bougerol}}, \bibinfo {author}
  {\bibfnamefont {B.}~\bibnamefont {Daudin}}, \ and\ \bibinfo {author}
  {\bibfnamefont {B.}~\bibnamefont {Gayral}},\ }\bibfield  {title} {\enquote
  {\bibinfo {title} {Evidence for quantum-confined stark effect in gan/aln
  quantum dots in nanowires},}\ }\href {\doibase 10.1103/PhysRevB.80.121305}
  {\bibfield  {journal} {\bibinfo  {journal} {Phys. Rev. B}\ }\textbf {\bibinfo
  {volume} {80}},\ \bibinfo {pages} {121305} (\bibinfo {year}
  {2009})}\BibitemShut {NoStop}%
\bibitem [{\citenamefont {Nenashev}\ \emph {et~al.}(2018)\citenamefont
  {Nenashev}, \citenamefont {Koshkarev},\ and\ \citenamefont
  {Dvurechenskii}}]{Nenashev2018}%
  \BibitemOpen
  \bibfield  {author} {\bibinfo {author} {\bibfnamefont {A.~V.}\ \bibnamefont
  {Nenashev}}, \bibinfo {author} {\bibfnamefont {A.~A.}\ \bibnamefont
  {Koshkarev}}, \ and\ \bibinfo {author} {\bibfnamefont {A.~V.}\ \bibnamefont
  {Dvurechenskii}},\ }\bibfield  {title} {\enquote {\bibinfo {title}
  {Approximate analytical description of the elastic strain field due to an
  inclusion in a continuous medium with cubic anisotropy},}\ }\href {\doibase
  10.1063/1.5019335} {\bibfield  {journal} {\bibinfo  {journal} {Journal of
  Applied Physics}\ }\textbf {\bibinfo {volume} {123}},\ \bibinfo {pages}
  {105104} (\bibinfo {year} {2018})}\BibitemShut {NoStop}%
\bibitem [{\citenamefont {Bernardini}\ \emph {et~al.}(1997)\citenamefont
  {Bernardini}, \citenamefont {Fiorentini},\ and\ \citenamefont
  {Vanderbilt}}]{Bernardini1997}%
  \BibitemOpen
  \bibfield  {author} {\bibinfo {author} {\bibfnamefont {Fabio}\ \bibnamefont
  {Bernardini}}, \bibinfo {author} {\bibfnamefont {Vincenzo}\ \bibnamefont
  {Fiorentini}}, \ and\ \bibinfo {author} {\bibfnamefont {David}\ \bibnamefont
  {Vanderbilt}},\ }\bibfield  {title} {\enquote {\bibinfo {title} {{Spontaneous
  polarization and piezoelectric constants of III-V nitrides}},}\ }\href
  {\doibase 10.1103/PhysRevB.56.R10024} {\bibfield  {journal} {\bibinfo
  {journal} {Phys. Rev. B}\ }\textbf {\bibinfo {volume} {56}},\ \bibinfo
  {pages} {R10024--R10027} (\bibinfo {year} {1997})}\BibitemShut {NoStop}%
\bibitem [{\citenamefont {Zoroddu}\ \emph {et~al.}(2001)\citenamefont
  {Zoroddu}, \citenamefont {Bernardini}, \citenamefont {Ruggerone},\ and\
  \citenamefont {Fiorentini}}]{Zoroddu2001}%
  \BibitemOpen
  \bibfield  {author} {\bibinfo {author} {\bibfnamefont {Agostino}\
  \bibnamefont {Zoroddu}}, \bibinfo {author} {\bibfnamefont {Fabio}\
  \bibnamefont {Bernardini}}, \bibinfo {author} {\bibfnamefont {Paolo}\
  \bibnamefont {Ruggerone}}, \ and\ \bibinfo {author} {\bibfnamefont
  {Vincenzo}\ \bibnamefont {Fiorentini}},\ }\bibfield  {title} {\enquote
  {\bibinfo {title} {{First-principles prediction of structure, energetics,
  formation enthalpy, elastic constants, polarization, and piezoelectric
  constants of AlN, GaN, and InN: Comparison of local and gradient-corrected
  density-functional theory}},}\ }\href {\doibase 10.1103/PhysRevB.64.045208}
  {\bibfield  {journal} {\bibinfo  {journal} {Phys. Rev. B}\ }\textbf {\bibinfo
  {volume} {64}},\ \bibinfo {pages} {045208} (\bibinfo {year}
  {2001})}\BibitemShut {NoStop}%
\bibitem [{\citenamefont {Chichibu}\ \emph {et~al.}(1998)\citenamefont
  {Chichibu}, \citenamefont {Abare}, \citenamefont {Minsky}, \citenamefont
  {Keller}, \citenamefont {Fleischer}, \citenamefont {Bowers}, \citenamefont
  {Hu}, \citenamefont {Mishra}, \citenamefont {Coldren}, \citenamefont
  {DenBaars},\ and\ \citenamefont {Sota}}]{Chichibu1998}%
  \BibitemOpen
  \bibfield  {author} {\bibinfo {author} {\bibfnamefont {S.~F.}\ \bibnamefont
  {Chichibu}}, \bibinfo {author} {\bibfnamefont {A.~C.}\ \bibnamefont {Abare}},
  \bibinfo {author} {\bibfnamefont {M.~S.}\ \bibnamefont {Minsky}}, \bibinfo
  {author} {\bibfnamefont {S.}~\bibnamefont {Keller}}, \bibinfo {author}
  {\bibfnamefont {S.~B.}\ \bibnamefont {Fleischer}}, \bibinfo {author}
  {\bibfnamefont {J.~E.}\ \bibnamefont {Bowers}}, \bibinfo {author}
  {\bibfnamefont {E.}~\bibnamefont {Hu}}, \bibinfo {author} {\bibfnamefont
  {U.~K.}\ \bibnamefont {Mishra}}, \bibinfo {author} {\bibfnamefont {L.~A.}\
  \bibnamefont {Coldren}}, \bibinfo {author} {\bibfnamefont {S.~P.}\
  \bibnamefont {DenBaars}}, \ and\ \bibinfo {author} {\bibfnamefont
  {T.}~\bibnamefont {Sota}},\ }\bibfield  {title} {\enquote {\bibinfo {title}
  {Effective band gap inhomogeneity and piezoelectric field in ingan/gan
  multiquantum well structures},}\ }\href {\doibase 10.1063/1.122350}
  {\bibfield  {journal} {\bibinfo  {journal} {Applied Physics Letters}\
  }\textbf {\bibinfo {volume} {73}},\ \bibinfo {pages} {2006--2008} (\bibinfo
  {year} {1998})}\BibitemShut {NoStop}%
\bibitem [{\citenamefont {Ibbetson}\ \emph {et~al.}(2000)\citenamefont
  {Ibbetson}, \citenamefont {Fini}, \citenamefont {Ness}, \citenamefont
  {DenBaars}, \citenamefont {Speck},\ and\ \citenamefont
  {Mishra}}]{Ibbetson2000}%
  \BibitemOpen
  \bibfield  {author} {\bibinfo {author} {\bibfnamefont {J.~P.}\ \bibnamefont
  {Ibbetson}}, \bibinfo {author} {\bibfnamefont {P.~T.}\ \bibnamefont {Fini}},
  \bibinfo {author} {\bibfnamefont {K.~D.}\ \bibnamefont {Ness}}, \bibinfo
  {author} {\bibfnamefont {S.~P.}\ \bibnamefont {DenBaars}}, \bibinfo {author}
  {\bibfnamefont {J.~S.}\ \bibnamefont {Speck}}, \ and\ \bibinfo {author}
  {\bibfnamefont {U.~K.}\ \bibnamefont {Mishra}},\ }\bibfield  {title}
  {\enquote {\bibinfo {title} {Polarization effects, surface states, and the
  source of electrons in algan/gan heterostructure field effect transistors},}\
  }\href {\doibase 10.1063/1.126940} {\bibfield  {journal} {\bibinfo  {journal}
  {Applied Physics Letters}\ }\textbf {\bibinfo {volume} {77}},\ \bibinfo
  {pages} {250--252} (\bibinfo {year} {2000})}\BibitemShut {NoStop}%
\bibitem [{\citenamefont {Kim}\ \emph {et~al.}(2004)\citenamefont {Kim},
  \citenamefont {Cho}, \citenamefont {Lee}, \citenamefont {Kim}, \citenamefont
  {Ryu}, \citenamefont {Kim}, \citenamefont {Kang},\ and\ \citenamefont
  {Chung}}]{Kim2004}%
  \BibitemOpen
  \bibfield  {author} {\bibinfo {author} {\bibfnamefont {Hwa-mok}\ \bibnamefont
  {Kim}}, \bibinfo {author} {\bibfnamefont {Yong-hoon}\ \bibnamefont {Cho}},
  \bibinfo {author} {\bibfnamefont {Hosang}\ \bibnamefont {Lee}}, \bibinfo
  {author} {\bibfnamefont {Suk~Il}\ \bibnamefont {Kim}}, \bibinfo {author}
  {\bibfnamefont {Sung~Ryong}\ \bibnamefont {Ryu}}, \bibinfo {author}
  {\bibfnamefont {Deuk~Young}\ \bibnamefont {Kim}}, \bibinfo {author}
  {\bibfnamefont {Tae~Won}\ \bibnamefont {Kang}}, \ and\ \bibinfo {author}
  {\bibfnamefont {Kwan~Soo}\ \bibnamefont {Chung}},\ }\bibfield  {title}
  {\enquote {\bibinfo {title} {{High-Brightness Light Emitting Diodes Using
  Dislocation-Free Indium Gallium Nitride/Gallium Nitride Multiquantum-Well
  Nanorod Arrays}},}\ }\href {\doibase 10.1021/nl049615a} {\bibfield  {journal}
  {\bibinfo  {journal} {Nano Letters}\ }\textbf {\bibinfo {volume} {4}},\
  \bibinfo {pages} {1059--1062} (\bibinfo {year} {2004})}\BibitemShut {NoStop}%
\bibitem [{\citenamefont {Chuang}\ and\ \citenamefont {Chang}(1996)}]{Chuang}%
  \BibitemOpen
  \bibfield  {author} {\bibinfo {author} {\bibfnamefont {S.~L.}\ \bibnamefont
  {Chuang}}\ and\ \bibinfo {author} {\bibfnamefont {C.~S.}\ \bibnamefont
  {Chang}},\ }\bibfield  {title} {\enquote {\bibinfo {title}
  {k\ensuremath{\cdot}p method for strained wurtzite semiconductors},}\ }\href
  {\doibase 10.1103/PhysRevB.54.2491} {\bibfield  {journal} {\bibinfo
  {journal} {Phys. Rev. B}\ }\textbf {\bibinfo {volume} {54}},\ \bibinfo
  {pages} {2491--2504} (\bibinfo {year} {1996})}\BibitemShut {NoStop}%
\bibitem [{\citenamefont {Joci{\'{c}}}\ and\ \citenamefont
  {Vukmirovi{\'{c}}}(2020)}]{Jocic2020}%
  \BibitemOpen
  \bibfield  {author} {\bibinfo {author} {\bibfnamefont {Milan~Joci{\'{c}}}\
  \bibnamefont {Joci{\'{c}}}}\ and\ \bibinfo {author} {\bibfnamefont {Nenad}\
  \bibnamefont {Vukmirovi{\'{c}}}},\ }\bibfield  {title} {\enquote {\bibinfo
  {title} {Ab initio construction of symmetry-adapted k.p hamiltonians for the
  electronic structure of semiconductors},}\ }\href {\doibase
  10.1103/PhysRevB.102.085121} {\bibfield  {journal} {\bibinfo  {journal}
  {Phys. Rev. B}\ }\textbf {\bibinfo {volume} {102}},\ \bibinfo {pages}
  {085121} (\bibinfo {year} {2020})}\BibitemShut {NoStop}%
\bibitem [{\citenamefont {Tomi{\'{c}}}\ \emph {et~al.}(2006)\citenamefont
  {Tomi{\'{c}}}, \citenamefont {Sunderland},\ and\ \citenamefont
  {Bush}}]{Tomic2006}%
  \BibitemOpen
  \bibfield  {author} {\bibinfo {author} {\bibfnamefont {Stanko}\ \bibnamefont
  {Tomi{\'{c}}}}, \bibinfo {author} {\bibfnamefont {Andrew~G.}\ \bibnamefont
  {Sunderland}}, \ and\ \bibinfo {author} {\bibfnamefont {Ian~J.}\ \bibnamefont
  {Bush}},\ }\bibfield  {title} {\enquote {\bibinfo {title} {{Parallel
  multi-band k{\textperiodcentered}p code for electronic structure of zinc
  blend semiconductor quantum dots}},}\ }\href {\doibase 10.1039/b600701p}
  {\bibfield  {journal} {\bibinfo  {journal} {J. Mater. Chem.}\ }\textbf
  {\bibinfo {volume} {16}},\ \bibinfo {pages} {1963--1972} (\bibinfo {year}
  {2006})}\BibitemShut {NoStop}%
\bibitem [{\citenamefont {Morrow}\ and\ \citenamefont
  {Brownstein}(1984)}]{Morrow1984}%
  \BibitemOpen
  \bibfield  {author} {\bibinfo {author} {\bibfnamefont {Richard~A.}\
  \bibnamefont {Morrow}}\ and\ \bibinfo {author} {\bibfnamefont {Kenneth~R.}\
  \bibnamefont {Brownstein}},\ }\bibfield  {title} {\enquote {\bibinfo {title}
  {{Model effective-mass Hamiltonians for abrupt heterojunctions and the
  associated wave-function-matching conditions}},}\ }\href {\doibase
  10.1103/PhysRevB.30.678} {\bibfield  {journal} {\bibinfo  {journal} {Phys.
  Rev. B}\ }\textbf {\bibinfo {volume} {30}},\ \bibinfo {pages} {678--680}
  (\bibinfo {year} {1984})}\BibitemShut {NoStop}%
\bibitem [{\citenamefont {Berkowicz}\ \emph {et~al.}(2000)\citenamefont
  {Berkowicz}, \citenamefont {Gershoni}, \citenamefont {Bahir}, \citenamefont
  {Lakin}, \citenamefont {Shilo}, \citenamefont {Zolotoyabko}, \citenamefont
  {Abare}, \citenamefont {Denbaars},\ and\ \citenamefont
  {Coldren}}]{Berkowicz2000}%
  \BibitemOpen
  \bibfield  {author} {\bibinfo {author} {\bibfnamefont {E.}~\bibnamefont
  {Berkowicz}}, \bibinfo {author} {\bibfnamefont {D.}~\bibnamefont {Gershoni}},
  \bibinfo {author} {\bibfnamefont {G.}~\bibnamefont {Bahir}}, \bibinfo
  {author} {\bibfnamefont {E.}~\bibnamefont {Lakin}}, \bibinfo {author}
  {\bibfnamefont {D.}~\bibnamefont {Shilo}}, \bibinfo {author} {\bibfnamefont
  {E.}~\bibnamefont {Zolotoyabko}}, \bibinfo {author} {\bibfnamefont {A.~C.}\
  \bibnamefont {Abare}}, \bibinfo {author} {\bibfnamefont {S.~P.}\ \bibnamefont
  {Denbaars}}, \ and\ \bibinfo {author} {\bibfnamefont {L.~A.}\ \bibnamefont
  {Coldren}},\ }\bibfield  {title} {\enquote {\bibinfo {title} {Measured and
  calculated radiative lifetime and optical absorption of
  ${\mathrm{in}}_{x}{\mathrm{ga}}_{1\ensuremath{-}x}\mathrm{N}/\mathrm{G}\mathrm{a}\mathrm{N}$
  quantum structures},}\ }\href {\doibase 10.1103/PhysRevB.61.10994} {\bibfield
   {journal} {\bibinfo  {journal} {Phys. Rev. B}\ }\textbf {\bibinfo {volume}
  {61}},\ \bibinfo {pages} {10994--11008} (\bibinfo {year} {2000})}\BibitemShut
  {NoStop}%
\end{thebibliography}%

\end{document}